\documentclass[a4paper,11pt]{article}
\pdfoutput=1 

\usepackage{jheppub} 

\usepackage[T1]{fontenc} 

\usepackage{latexsym}
\usepackage[usenames]{color}
\usepackage{fancybox}
\usepackage{simplewick}
\usepackage{comment}
\usepackage{xcolor,bbm}
\usepackage{framed}
\definecolor{shadecolor}{rgb}{0.9,0.9,0.95}
\usepackage{setspace}
\usepackage{tabularx,booktabs}
\usepackage{bm}
\definecolor{darkgreen}{rgb}{0,0.5,0}
\definecolor{darkblue}{cmyk}{0.9,0.9,0,0}
\definecolor{darkred}{rgb}{0.6,0,0.3}

\usepackage{graphicx}
\usepackage{hyperref}

\newcommand{\tr}{{\rm tr}}
\renewcommand{\Im}{{\rm Im}}
\renewcommand{\Re}{{\rm Re}}

\def\eqref#1{(\ref{#1})}

\def\beq{\begin{equation}}
\def\eeq{\end{equation}}

\title{\boldmath Correlation functions of determinant operators in conformal fishnet theory}

\author[a]{Omar Shahpo,}
\author[b,1]{Edoardo Vescovi\note{Corresponding author.}}

\affiliation[a]{Department of Mathematics, King's College London, The Strand, WC2R 2LS London, U.K.}
\affiliation[b]{Nordita, KTH Royal Institute of Technology and Stockholm University,
\\
Hannes Alfv{\'e}ns v{\"a}g 12, SE-106 91 Stockholm, Sweden
}

\emailAdd{omar.shahpo$\bullet$kcl.ac.uk}
\emailAdd{edoardo.vescovi$\bullet$su.se}

\abstract{We consider scalar local operators of the determinant type in the conformal ``fishnet'' theory that arises as a limit of gamma-deformed $\mathcal{N}=4$ super Yang-Mills theory.
We generalise a field-theory approach to expand their correlation functions to arbitrary order in the small coupling constants and apply it to the bi-scalar reduction of the model.
We explicitly analyse the two-point functions of determinants, as well as of certain deformations with the insertion of scalar fields, and describe the Feynman-graph structure of three- and four-point correlators with single-trace operators. These display the topology of globe and spiral graphs, which are known to renormalise single-trace operators, but with ``alternating'' boundary conditions. In the appendix material we further investigate a four-point function of two determinants and the shortest bi-local single trace. We resum the diagrams by the Bethe-Salpeter method and comment on the exchanged OPE states.
}

\begin{document} 
\maketitle
\flushbottom

\section{Introduction}
\label{secIntro}

In this work we initiate the study of determinant operators in the conformal field theory (CFT) emerging from the fishnet limit \cite{Gurdogan:2015csr}, combining weak coupling with strong imaginary $\gamma$-twists, of the $\gamma$-deformed $\mathcal{N} = 4$ super Yang-Mills (SYM) theory \cite{Frolov:2005dj}. The Lagrangian of the model describes the four-dimensional dynamics of three scalars and three fermions. It is controlled by three effective coupling constants and with a restricted structure of the interactions, hence the name chiral CFT, or shortly $\chi$CFT${}_4$, coined for this model in \cite{Caetano:2016ydc}. Most of the quantitative results are obtained for the bi-scalar theory, the most-studied reduction of $\chi$CFT${}_4$ with two scalars and a single non-zero coupling.

The main motivation behind our work is inspired by the application of integrability techniques to correlation functions of determinant operators in $\mathcal{N}=4$ SYM. Recently one of the authors has developed a formalism for computing the  structure constants of two \mbox{1/2-BPS} determinants and one non-BPS single trace at finite coupling \cite{Jiang:2019xdz,Jiang:2019zig}. It is formulated as a bootstrap-type programme for overlaps between a boundary state and a Bethe state in an integrable two-dimensional system (a spin chain at weak coupling and a string worldsheet at strong coupling). One can map the three-point function to such overlaps, where the boundary state corresponds to the determinant pair, solve the constraints imposed by integrability and derive a non-perturbative formula in the framework of the thermodynamic Bethe ansatz (TBA). Overlaps of integrable boundary states in relativistic \cite{Ghoshal:1993tm} and spin-chain systems \cite{Piroli:2017sei} are rare quantities, besides the spectrum, that can be calculated exactly and they appear in connection to the $g$-function \cite{Affleck:1991tk} (see also in \cite{Jiang:2019xdz}) in two-dimensional systems. Moreover, they are known to subtend a plethora of observables in higher dimensions \footnote{The range of applications extends as far to the time evolution of quenched systems; see references in the review \cite{Linardopoulos:2020jck}.}: the strategy laid out in \cite{Jiang:2019xdz,Jiang:2019zig} finds application beyond the original scope in the analogous three-point functions in ABJM theory \cite{Yang:2021hrl,komatsutoappear} and the expectation value of a single-trace operator in the presence of a domain-wall defect \cite{Komatsu:2020sup}.
In this paradigm the result comes in the form of a Fredholm determinant and it is dependent on a set of Y-functions, which obey an associated system of TBA equations.
The evaluation of Fredholm determinants as a function of the coupling and the solution of the infinitely-many non-linear integral TBA equations is feasible in principle, for example \cite{Bajnok:2013wsa} in a related context. However, they are based on numerical algorithms and they remain computationally expensive, also in spite of recent reformulations \cite{Caetano:2020dyp}.

The reason for revisiting determinant operators in the new setting of fishnet models takes its roots from the observation \cite{Jiang:2019xdz} that the three-point functions with the single trace in the $SU(2)$ sector can be written in free theory as an integral of a product of Baxter $Q$-functions. Similar integral expressions are reminiscent of the structure expected from the separation of variables (SoV) method and they are found for a growing list of observables at finite coupling, computed via supersymmetric localisation \cite{Giombi:2018qox,Giombi:2018hsx} or resummation of Feynman diagrams \cite{Cavaglia:2018lxi,McGovern:2019sdd}. The $Q$-functions, as solutions of the quantum spectral curve (QSC) equations \cite{Gromov:2013pga,Gromov:2014caa},
are well-understood at finite coupling, therefore there is an indication that SoV could lead also to study the above-mentioned three-point functions at finite coupling.
Somewhat counter-intuitively, the high symmetry content $PSU(2,2|4)$ of $\mathcal{N} = 4$ SYM brings additional complications to the construction of the basis that realises the SoV paradigm, namely the factorisation of Bethe states and other observables into products of $Q$-functions. Many important lessons came from constructing explicitly scalar products, correlation functions and form factors in lower-rank integrable spin chains \cite{Cavaglia:2019pow,Gromov:2019wmz,Gromov:2020fwh} and SoV-type expressions for field-theory observables in the bi-scalar theory \cite{Cavaglia:2021mft}. The bi-scalar theory, and more generally the $\chi$CFT$_{4}$, is the ideal starting point to advance further towards the SoV-formulation of three-point functions of \cite{Jiang:2019xdz,Jiang:2019zig}. Such formulation would ultimately contribute to borrow and develop computational methods based on the QSC equations rather than the TBA equations
\footnote{
The integrability method \cite{Jiang:2019xdz} needs to go through few alterations to deliver results valid for fishnets. At the level of Bethe ansatz equations it requires going to a spin chain where the deformation parameters enter the boundary conditions as twists. This suggests that the arguments in favour of the integrability of the boundary state may be revisited. The weak-coupling analysis relies on the perturbative eigenstates of the dilatation operator, which is altered by the phase deformations. In the undeformed theory  the interesting outcome is to find a compact determinant formula for asymptotic overlaps and a pairing condition on the Bethe roots. There has been limited success in retaining these properties in the case of defect one-point functions \cite{Widen:2018nnu}. Moreover, the deformation breaks the superconformal symmetry, which is crucial to bootstrap the fundamental (two-particle) overlap in the TBA approach. In conclusion, the avenue appears to need some effort for finite twists, unless some simplification occurs in smaller sub-sectors like $SU(2)$ or in the double-scaling limit.
\\
There is work in progress for boundary states in twisted theories. A step was already made for the overlap between the CFT wave-function and a fixed boundary state within SoV \cite{Cavaglia:2021mft} and for the TBA for conformal dimensions. The latter can be recovered from the TBA of twisted $\mathcal{N}=4$ SYM \cite{Caetano:2016ydc,Gromov:2017cja,Ahn:2011xq} in the double-scaling limit or that of the $d$-dimensional fishnet \cite{Basso:2019xay}.
}.

In this paper we explore the subject of determinant operators with an approach based on Feynman perturbation theory, conformal symmetry and the Bethe-Salpeter operatorial method. Several reasons favour this methodology in $\chi$CFT${}_4$. The ``chiral'' form of the Lagrangian brings massive simplifications: Feynman diagrams obey conformal properties and display a regular bulk topology (a square fishnet in the bi-scalar model \cite{Zamolodchikov:1980mb,Gurdogan:2015csr} and a ``dynamical'' fishnet in $\chi$CFT$_{4}$ \cite{Kazakov:2018gcy}), whose boundary is determined by the observable under consideration. One can in turn exploit the iterative structure of Feynman graphs to resum the perturbative series exactly and extract the non-perturbative and explicit operator product expansion (OPE) data of the exchanged operators \cite{Grabner:2017pgm,Gromov:2018hut,Kazakov:2018gcy}. This is a remarkable achievement in an interacting CFT in more than two dimensions. The main drawbacks of this setting are the loss of supersymmetry (SUSY) \cite{Frolov:2005dj} (due to the non-zero $\gamma$-twists \footnote{The case of three equal twists is the $\beta$-deformed SYM \cite{Leigh:1995ep,Lunin:2005jy} with one unbroken supersymmetry.}) and of unitarity \cite{Gurdogan:2015csr} (due to their imaginary values), the presence of Lagrangian counter-terms \cite{Fokken:2014soa,Sieg:2016vap,Grabner:2017pgm,Kazakov:2018gcy} (for UV consistency of the quantum theory) and the restriction to the planar limit (as it is unknown if quantum conformal symmetry persists at finite $N$). The first and third points affect our calculations the most: determinants cease to be super-conformal primary operators and the extra interaction vertices appear as lattice defects in the graphs.

To present few more motivations, we define the determinant operators and sketch the approach to the perturbative expansion at large $N$. In $\mathcal{N}=4$ SYM they are gauge-invariant local operators made up of $N$ scalar fields
 \cite{Balasubramanian:2001nh}
\begin{flalign}
\label{eq11}
\mathcal{D}(x)
=
\det(y \cdot \Phi(x))
=\frac{\varepsilon_{i_{1}...i_{N}}\varepsilon^{j_{1}\dots j_{N}}}{N!}
\left(y\cdot\Phi\left(x\right)\right)_{~j_{1}}^{i_{1}}\left(y\cdot\Phi\left(x\right)\right)_{~j_{2}}^{i_{2}}\dots\left(y\cdot\Phi\left(x\right)\right)_{~j_{N}}^{i_{N}}
\,.
\end{flalign}
The argument is a linear combination $y \cdot \Phi=\sum_{I=1}^6 y^I\, \Phi^I$ of the six real adjoint-valued scalars $\Phi^I$. The colour indices ($i,j=1,2,\dots,N$) are contracted with Levi-Civita symbols. With the polarisation vector $y^I$ being a null vector ($y\cdot y=0$), they are invariant under 1/2 of the SUSY transformations and they have protected conformal dimension $\Delta=N$. In the AdS/CFT correspondence it was proposed in \cite{Balasubramanian:2001nh} and confirmed in \cite{Corley:2001zk} that determinants create macroscopic (maximal) giant gravitons D3-branes in $AdS_5\times S^5$ \cite{McGreevy:2000cw} placed at the center of $AdS_5$, wrapping a (maximum size) three-sphere inside $S^5$ and dynamically prevented from collapsing by their (maximal) angular momentum.
\\
The matter content of $\chi$CFT${}_4$ consists of three complex scalars and fermions in the adjoint representation of global $SU(N)$
\begin{gather}
\label{eq13}
\tilde{\chi}_i\in\{\phi^j,{\phi}^\dagger_j,\psi^j_\alpha,\bar{\psi}_j^{\dot{\alpha}}\}\,,
\qquad
j=1,2,3\,,
\qquad
\alpha,\dot{\alpha}=1,2\,,
\end{gather}
with a simple relation to the scalars of the ``mother'' theory
\begin{gather}
\label{eq20}
\phi^1=\frac{\Phi^{1}+i\Phi^{2}}{\sqrt{2}}\,,
\qquad
\phi^2=\frac{\Phi^{3}+i\Phi^{4}}{\sqrt{2}}\,,
\qquad
\phi^3=\frac{\Phi^{5}+i\Phi^{6}}{\sqrt{2}}\,.
\end{gather}
It is natural to borrow the definition of determinant operators in $\chi$CFT${}_4$ from \eqref{eq11}. We still work with null polarisations, although the SUSY breaking puts all choices on equal footing.
\\
Perturbative computations of heavy operators 
are a non-trivial task (see introduction of \cite{Jiang:2019xdz}). The factorial growth of the number of non-planar diagrams can overwhelm the suppression by powers of $1/N^2$, thus planarity fails to be an attribute of the dominant diagrams at large $N$ \cite{Witten:1998xy,Balasubramanian:2001nh} \footnote{Notice however that the methods in sections \ref{secETwhy} and \ref{secETderivation} reduce to consider trace operators of length of order $N^0$, for which planar diagrams are the dominant contributions at large $N$.}. For this reason we generalise the semi-classical approach of \cite{Jiang:2019xdz,Vescovi:2021fjf} \footnote{Other generalisation were pushed forward in free theory: for correlators of Schur polynomial operators in $\mathcal{N}=4$ SYM \cite{Chen:2019gsb} and for (sub-)determinants and permanents in ABJ(M) \cite{Chen:2019kgc,Yang:2021hrl}.}, dubbed effective theory, to write all dominant Feynman diagrams with an arbitrary number of vertices. The strategy is to represent the correlation function
\begin{equation}
\label{eq10}
G_m
=
\left\langle \mathcal{D}_1(x_1) \dots \mathcal{D}_m(x_m)\, \mathcal{O}(\{x_{i,j}\}) \right\rangle
\end{equation}
of $m$ determinant operators \eqref{eq11}
\begin{equation}
\label{eq91}
\mathcal{D}_k(x_k)
=
\det(y_k \cdot \Phi(x_k))\,,
\qquad
k=1,2,\dots,m
\end{equation}
and one multi-trace operator, made up of any field of the theory \eqref{eq13},
\begin{gather}
\label{eq12}
\mathcal{O}(\{x_{i,j}\})=\prod_{i}\textrm{tr}\left(\tilde{\chi}_{i,1}(x_{i,1})\tilde{\chi}_{i,2}(x_{i,2})...\tilde{\chi}_{i,L_{i}}(x_{i,L_i})\right)
\end{gather}
as a zero-dimensional integral over an auxiliary matrix $\rho$. The dominant diagrams can be read off solving the integral in saddle-point approximation at large $N$. The computational labour of obtaining more complex diagrams correlates with the number of vertices, but it can be implemented in an algorithmic way and it skips over a lot of combinatorial dexterity required by traditional perturbative techniques.

The second motivation behind our work comes from open strings in holography. A lot of progress has been made in the first-principle formulation of holography for the bi-scalar model: a (fish)chain of quantum particles from the conformal properties of fishnet integrals \cite{Gromov:2019aku,Gromov:2019bsj,Gromov:2019jfh,Gromov:2021ahm} and a string sigma-model from the ``continuum'' limit of fishnet graphs \cite{Basso:2019xay,Basso:2021omx}. The determinants/giant gravitons are interesting objects to gain further insights.
In a series of works \cite{Berenstein:2002ke,Balasubramanian:2002sa} determinant-like operators
\begin{gather}
\label{eq90}
\varepsilon_{i_{1}...i_{N}}\varepsilon^{j_{1}\dots j_{N}}\left(\phi^1(x)\right)_{~j_{1}}^{i_{1}}\dots\left(\phi^1(x)\right)_{~j_{N-1}}^{i_{N-1}}\left(\mathcal{W}(x)\right)_{~j_{N}}^{i_{N}}
\end{gather}
were constructed by inserting an open chain, or word, $\mathcal{W}$ of fields and covariant derivatives, similar to the closed chains in single-trace operators. Such deformations are mostly non-BPS \cite{Das:2000st} and their spectrum has been the subject of studies using various approaches \cite{Berenstein:2003ah,Berenstein:2004kk,Balasubramanian:2004nb,Berenstein:2005vf,deMelloKoch:2007rqf,deMelloKoch:2007nbd,deMelloKoch:2010zrl,DeComarmond:2010ie,Carlson:2011hy}, including integrability \cite{Hofman:2007xp,Mann:2006rh,Berenstein:2005fa,Berenstein:2006qk,Bajnok:2013wsa} (see also in \cite{Jiang:2019xdz}). What is more interesting occurs in AdS/CFT: while a determinant creates a single giant graviton D-brane on its ground state, words add open-string excitations on the giant \cite{Balasubramanian:2002sa} and in the presence of multiple giants one can also assign Chan-Paton factors to open strings by intertwining the indices of multiple words. Fluctuations of giants contain more information on the local bulk physics of $AdS_5$ than a point-like giant moving in time; moreover they are not too heavy, like giants, to deform the $AdS$ geometry. In \cite{Balasubramanian:2002sa} open strings were shown to emerge from field-theory degrees of freedom by quantising the fluctuations and constructing explicitly their world-sheets. Since the fishnet theory are fully under control at quantum level, one could use it to approach a non-perturbative description of open strings.
\\
In this paper we make a first step considering a family of operators similar to \eqref{eq90} with multiple insertions of one-letter words. We play with the polarisation vectors~\footnote{The authors of \cite{Basso:2017khq} exploits the trick to suppress wrapping corrections and control the flow of R-charge in \emph{asymptotic} four-point functions of single traces.} in such a way to replace some of the scalars $y\cdot \Phi$ in \eqref{eq11} with $\ell_1$ insertions of the type $\tilde{\chi}_1$, $\ell_2$ insertions of the type $\tilde{\chi}_2$ and so on up to $\ell_n$ insertions of the type $\tilde{\chi}_n$:
\begin{flalign}
\label{eq32}
&
\left.\prod_{i=1}^{n}\left(\frac{\partial}{\partial a_{i}}\right)^{\ell_{i}}\det\left(y\cdot\Phi+\sum_{i=1}^{n} a_{i}\,\tilde{\chi}_{i}\right)\right|_{ a_{1}=\dots= a_{n}=0}=\frac{\varepsilon_{i_{1}\dotsi_{N}}\varepsilon^{j_{1}\dots j_{N}}}{\left(N-\sum_{i=1}^{n}\ell_{i}\right)!}
\underbrace{\left(\tilde{\chi}_{1}\right)_{~j_{1}}^{i_{1}}\dots\left(\tilde{\chi}_{1}\right)_{~j_{\ell_{1}}}^{i_{\ell_{1}}}}_{\ell_{1}\textrm{ insertions of the type }\tilde{\chi}_{1}}
\nonumber
\\
&
\times
\underbrace{\left(\tilde{\chi}_{2}\right)_{~j_{\ell_{1}+1}}^{i_{\ell_{1}+1}}\dots\left(\tilde{\chi}_{2}\right)_{~j_{\ell_{1}+\ell_{2}}}^{i_{\ell_{1}+\ell_{2}}}}_{\ell_{2}\textrm{ insertions of the type }\tilde{\chi}_{2}}\dots\underbrace{\left(\tilde{\chi}_{n}\right)_{~j_{\ell_{1}+\dots+\ell_{n-1}+1}}^{i_{\ell_{1}+\dots+\ell_{n-1}+1}}\dots\left(\tilde{\chi}_{n}\right)_{~j_{\ell_{1}+\dots+\ell_{n}}}^{i_{\ell_{1}+\dots+\ell_{n}}}}_{\ell_{n}\textrm{ insertions of the type }\tilde{\chi}_{n}}\left(y\cdot\Phi\right)_{~j_{\ell_{1}+\dots+\ell_{n}+1}}^{i_{\ell_{1}+\dots+\ell_{n}+1}}\dots\left(y\cdot\Phi\right)_{~j_{N}}^{i_{N}}\,.
\end{flalign}
The formula generally comes with the constraint $\ell_i\ll N$ in order to tell apart the insertion scalars $\tilde{\chi}_i$ from the more numerous scalars $y\cdot \Phi$ of the undeformed determinant. This way of packing up determinant-like operators into determinants helps study their correlators with our tools. All we need to do is to crank up the effective theory with suitable $a_i$-dependent polarisations and take $a_{i}$-derivatives of the output.
We are then able to consider insertions $\tilde{\chi}_i$ that are ``{elementary}'', if picked from the alphabet $\bar{\chi}_i\in\{\phi^1,\phi^2,\phi^3,\phi^\dagger_1,\phi^\dagger_2,\phi^\dagger_3\}$, or ``{composite}'', if linear combinations of elementary ones, e.g. ${\phi^1-2\phi_2^\dagger}$. The distinction comes in handy in the classification of section \ref{sec2pt} and it is unambiguous in $\chi$CFT${}_4$, as the breaking of the $SO(6)$ R-symmetry leaves no transformation able to rotate elementary into composite insertions and vice versa.

The third motivation comes from the similarity between colour singlets made up of $N$ fields in theories with adjoint matter (determinants of scalars) and in QCD (baryons of quarks), for which one can see three-point functions of \cite{Jiang:2019xdz,Jiang:2019zig} as the translation of the baryon-baryon-meson vertex into $\mathcal{N}=4$ SYM. Resummation techniques for determinants are an important tool for the study of more complicated operators in large-$N$ quantum field theories. In combination with integrability, they have the potential to render fishnet models an ideal toy model to achieve a non-perturbative solution of baryon-like operators.

This paper is organised as follows. In section \ref{secET} we show that correlation functions of determinants in $\chi$CFT${}_4$ can be systematically expanded at weak coupling by means of the effective theory. We put this strategy to work in the bi-scalar reduction in the rest of the paper. In section \ref{sec2pt} we carry out the perturbative analysis of two-point functions. Under some working assumptions we can identify a class of operators that have exact dimension $\Delta=N$ in the planar limit. In section \ref{sec3pt} we conduct a survey on three- and four-point functions of a determinant pair and a (local or bi-local) single-trace operator made up of $L$ scalars. The minimal-length cases ($L=2$) generally connect back to correlators of single traces, both protected and non-protected from quantum corrections, previously known in the literature. The description of the cases with $L>2$ is mostly qualitative and limited to their Feynman graph expansions. The graphs have the same topology of the globe and spiral graphs of 
\cite{Gurdogan:2015csr,Caetano:2016ydc} but with different, ``alternating'' boundary conditions. The section \ref{secConclusion} has a summary of continuations of this work. Details about $\chi$CFT${}_4$ are in appendix \ref{secConventions}. In the appendices \ref{sec4pt}-\ref{secShift} we start the exact study of a four-point correlator of two determinants and a short, bi-local scalar operator. It is one observable found in section \ref{sec3pt} and characterised by a simple, yet non-trivial, graph content. We resum the graphs using the Bethe-Salpeter method with the help of conformal symmetry and write the result in the form of a conformal partial wave expansion. We match the lowest orders against a direct perturbative calculation and comment on the spectrum of exchanged operator in the $s$-channel of the four-point function.

\section{Effective theory for the correlators of determinant operators}
\label{secET}
The effective theory is a rewriting of the (infinite-dimensional) path-integral  \eqref{eq10} in $\chi$CFT${}_4$ into a (zero-dimensional) matrix integral \cite{Jiang:2019xdz} in a form amenable to the large-$N$ expansion. The formula \eqref{eq5} below is a straightforward generalisation of the effective theory in $\mathcal{N}=4$ SYM at leading \cite{Jiang:2019xdz} and sub-leading order \cite{Vescovi:2021fjf} at weak coupling. We motivate the need of such reformulation in section \ref{secETwhy}, explain how to accommodate the dependence on the couplings at all loop orders in section \ref{secETderivation} and comment on the output in section \ref{secETfiniteness}.

\subsection{Perturbation theory in the fishnet theory}
\label{secETwhy}

The diagrammatic strategy pursued in \cite{Jiang:2019xdz,Vescovi:2021fjf} for the correlators \eqref{eq10} with $m=2$ consists of two steps. First, one operates free Wick contractions within the determinant pair
\begin{gather}
\label{eq27}
\left.\det(y_1 \cdot \Phi(x_1))\det(y_2 \cdot \Phi(x_2))
\right|_{\rm partial~contraction}
=
\sum_{L=0}^{N}\left(\frac{y_1\cdot y_2}{N}I_{x_1x_2}\right)^{N-L}\,\mathcal{G}_{L}(x_{1},x_{2})\,,
\end{gather}
where the ``scalar propagator'' $I_{x_1x_2}=(4\pi^2 (x_1-x_2)^2)^{-1}$ is defined below \eqref{eq4}. The result is a sum of bi-local multi-traces of length $2L=0,2,\dots, 2N$, called partially-contracted giant graviton (PCGG),
\begingroup \allowdisplaybreaks
\begin{flalign}
\label{eq29}
\mathcal{G}_{L}(x_{1},x_{2})
&=
\left(-\right)^{L}\left(N-L\right)!
\sum_{k_{1},k_{2},\dots,k_{L}=0}^{L}\prod_{n=1}^{L}\frac{1}{k_{n}!}\left(-\frac{1}{n}\textrm{tr}\left(\left((y_{1}\cdot\Phi(x_{1}))(y_{2}\cdot\Phi(x_{2}))\right)^{n}\right)\right)^{k_{n}}
\nonumber
\\
&
{\rm with}~~~\sum_{n=1}^N n\,k_{n}=L\,.
\end{flalign}
\endgroup
Second, one plugs \eqref{eq27} into \eqref{eq10} and takes into account the contractions with $\mathcal{O}$ and the interaction vertices. At this stage contractions between $\Phi^I(x_1)$ and $\Phi^I(x_2)$ in \eqref{eq29} are not to be considered, as all of them are supposed to take place in the first step. Only in this way the two-step strategy becomes equivalent to the standard contraction of the scalars in \eqref{eq10} at once via the Wick's theorem.
\\
The key observation is that $\mathcal{O}$ contracts only with the PCGGs with length comparable to the length $O(N^0)$ of $\mathcal{O}$. Moreover, the multi-trace terms in \eqref{eq29} come with the same power of $N$ of the single-trace term, preventing the competition between planar and non-planar diagrams. These facts mark the difference between the PCGG approach and the naive perturbation theory mentioned in section \ref{secIntro}.

The PCGG approach puts correlation functions of determinant operators, which lack the standard notion of planar approximation  \cite{Witten:1998xy,Balasubramanian:2001nh}, on a solid ground in $\chi$CFT${}_4$, where finiteness and conformality are proven only for planar fishnet graphs. The (single-trace \eqref{eq9} and double-trace \eqref{eq2}) interaction vertices produce the same effect of an external trace $\mathcal{O}$, so correlators of determinants decompose into correlators of short traces of the kind described by fishnet graphs. An exciting corollary is that determinant correlators inherit the simple graph description and the quantum integrability properties of it.

However, the double-trace counter-terms put an obstacle to the PCGG approach in practice. Correlators are described by fishnet graphs when all external operators and intermediate states in the OPEs have length greater than two \cite{Sieg:2016vap,Caetano:2016ydc}. The fact that the PCGG contains such operators (the terms with $k_1\neq 0$ in \eqref{eq29}) leaves a generic correlator \eqref{eq10} potentially affected by the counter-terms, which break the regular structure of the fishnet. By contrast, in the works on multi-trace correlators \cite{Gurdogan:2015csr,Gromov:2017cja,Caetano:2016ydc,Grabner:2017pgm,Gromov:2018hut,Kazakov:2018gcy,Basso:2018cvy,Derkachov:2020zvv} there is control over the external operators' lengths and it is possible to track down or exclude the existence of minimal-length operators from the get-go. Moving on to determinants in a systematic way, one would have to spell out the relevant multi-traces \eqref{eq29} for different values of $L$ and to distribute the scalars if $y_k \cdot \Phi(x_k)$ is a combination of two or more scalars. These preliminary operations become cumbersome as the length of $\mathcal{O}$ increases, thus driving up the maximal value of $L$ and the complexity of the multi-traces \eqref{eq29}. In the next section we present an approach equivalent to Feynman diagrams which bypasses this technical limitation.

\subsection{Effective theory}
\label{secETderivation}
The main idea \cite{Jiang:2019xdz} behind the effective theory is to express each determinant in \eqref{eq10}
\footnote{The derivation holds for $U(N)$ rather than for the $SU(N)$ gauge group of $\chi$CFT${}_4$. The existence of the $U(1)$ mode is immaterial at large $N$ for our practical purposes. Moreover, each $y_k$ is a null vector $y_k \cdot y_k=0$.}
as a Berezin integral of the zero-dimensional fermions $\chi^{a}_k$ and their conjugates $\bar{\chi}_{k,a}$ 
\footnote{Adjusting a notation established since \cite{Caetano:2016ydc}, we denote a physical field with $\tilde{\chi}$ in \eqref{eq13} and the non-physical fermions here with $\chi$.}. They have indices $a=1,2,\dots, N$ in the (anti-)fundamental representation of $SU(N)$ and there is pair of them for each determinant labeled by $k=1,2,\dots, m$. What follows is a judicious sequence of operations (the steps 1-4 in section IV of \cite{Vescovi:2021fjf}) to integrate out the physical fields \eqref{eq13} and integrate in a position-independent matrix $\rho_{kl}$ with $k,l=1,2,\dots, m$, which obeys $\rho_{kl}=\rho_{lk}^*$ and $\rho_{kk}=0$ ($k$ not summed). The result is the effective theory for \eqref{eq10} 
\begin{gather}
\label{eq5}
G_{m}=\frac{1}{Z_\rho}\int d\rho \, \left\langle \mathcal{O}^{(S)}\right\rangle _{\chi} \,e^{-NS_{\textrm{eff}}\left[\rho\right]}
\end{gather}
with measure
\begin{gather}
\label{eq14}
d\rho=\prod_{k<l}d\textrm{Re}\left(\rho_{kl}\right)\,d\textrm{Im}\left(\rho_{kl}\right)\,.
\end{gather}
While the effective theory is valid at finite coupling and finite $N$, in the rest of the section we also show its practical use at weak coupling and large $N$. Moreover, the current form of the counter-term Lagrangian \eqref{eq2} guarantees that only planar correlators are conformal, therefore our interest is limited to the leading orders of $N$.
\paragraph{The effective action.}
The action of the matrix $\rho$
\begin{gather}
\label{eq6}
S_{\textrm{eff}}\left[\rho\right]=8\pi^{2}\,\textrm{tr}\left(\rho^2\right)-\log\textrm{det}\left(-2\,i\hat{\rho}\right)
\end{gather}
is non-polynomial and depends on it also through the ``rescaled'' matrix $\hat{\rho}_{kl}=\sqrt{d_{kl}}\,\rho_{kl}$ (indices not summed). The factor ${d_{kl}=y_{kl}/x_{kl}^2 = y_k \cdot y_l/(x_k-x_l)^2}$ is basically the scalar propagator between the determinants labeled by $k$ and $l$ in \eqref{eq91} and it knows about their positions $x_k$ and polarisations $y_k$. The fact that the latter are null vectors ensures that the ambiguous quantity $d_{kk}$, as well as $\rho_{kk}$ and $\hat{\rho}_{kk}$, does not appear at any step of the derivation of \eqref{eq6} and in the rest of this section \footnote{This should also be true without the null condition. The operators in \eqref{eq10} are normal-ordered and one would have to exclude self-contractions of scalar pairs sitting in the same point $x_k$, which generate the ``scalar propagator'' $d_{kk}$.}. The determinant-independent part of the action fixes the normalisation
\begin{gather}
\label{eq15}
Z_\rho
=\int d\rho \,e^{-8\pi^{2}N\,\textrm{tr}\left(\rho^2\right)}
=\left(16\pi N\right)^{-m\left(m-1\right)/2}\,.
\end{gather}

\paragraph{The integrand.}
The prefactor in \eqref{eq5} is the average over fermions (to be defined in \eqref{eq7} below) of the effective operator $\mathcal{O}^{(S)}$. This can be thought of as the external operator \eqref{eq12} ``dressed up'' with the interactions of the theory. The analogous expression in $\mathcal{N}=4$ SYM is given in free theory in \cite{Jiang:2019xdz} (for a \emph{single-trace} $\mathcal{O}$ made up of scalars and derivatives only) and up to one loop in \cite{Vescovi:2021fjf} (\emph{without} the operator, namely $\mathcal{O}=1$). Uplifting the derivation of the latter to finite coupling \footnote{In \cite{Vescovi:2021fjf}  we take note of the footnote 26 to allow for the presence of $\mathcal{O}$ and we do not expand the relevant exponentials in step 2.}, we easily obtain
\begingroup \allowdisplaybreaks
\begin{flalign}
\label{eq16}
\mathcal{O}^{(S)}
&=
\frac{1}{Z}\int D\phi^j\,D{\phi}^\dagger_j\,D\psi^j_\alpha\,D\bar{\psi}_j^{\dot{\alpha}} \,\,
\mathcal{O}(\phi^j+S^j,\phi_j^\dagger+S_j^\dagger,\psi_\alpha^j,\bar{\psi}^{\dot{\alpha}}_j)
\\
&
\times
e^{\int d^4x\, [\mathcal{L}_{\rm free}(\phi^j,\phi_j^\dagger,\psi_\alpha^j,\bar{\psi}^{\dot{\alpha}}_j)
+
(\mathcal{L}_{\rm int}+\mathcal{L}_{\rm dt})(\phi^j+S^j,\phi_j^\dagger+S_j^\dagger,\psi_\alpha^j,\bar{\psi}^{\dot{\alpha}}_j)]}
\nonumber
\\
{\rm with}~~~Z
&=
\int D\phi^j\,D{\phi}^\dagger_j\,D\psi^j_\alpha\,D\bar{\psi}_j^{\dot{\alpha}} \,\,
e^{\int d^4x\,
(\mathcal{L}_{\rm free}+\mathcal{L}_{\rm int}+\mathcal{L}_{\rm dt})(\phi^j,\phi_j^\dagger,\psi_\alpha^j,\bar{\psi}^{\dot{\alpha}}_j)}\,.
\end{flalign}
\endgroup
The scalars in the multi-trace and in the interacting Lagrangian \eqref{eq9}-\eqref{eq2}, unlike those in the free Lagrangian \eqref{eq1}, are shifted by the ``classical backgrounds''
\begin{flalign}
\label{eq94}
(S^{j}\left(x\right))^a_{~b}&=-\frac{1}{\sqrt{2}N}\sum_{k=1}^mI_{xx_{k}}(y_{k}^{2j-1}+i y_{k}^{2j})\chi_{k}^a\bar{\chi}_{k,b}
\\
\label{eq95}
(S^\dagger_{j}\left(x\right))^a_{~b}&=-\frac{1}{\sqrt{2}N}\sum_{k=1}^mI_{xx_{k}}(y_{k}^{2j-1}-i y_{k}^{2j})\chi_{k}^a\bar{\chi}_{k,b}
~~~
{\rm with}
~~~
j=1,2,3\,,
\end{flalign}
where we remind the definition $I_{xx_k}=(4\pi^2 (x-x_k)^2)^{-1}$ given below \eqref{eq4}. The path integral over physical fields \eqref{eq16} returns a function of the fermion bilinears $S^j$ and $S_j^\dagger$
\footnote{The origin of the shifts can be so understood. In the steps 1-2 in section IV of \cite{Vescovi:2021fjf}, when we express the determinants as Berezin integrals
\begin{flalign}
\label{eq115}
G_{m}
&=
\frac{1}{Z}\int D\phi^{j}\,D{\phi}_{j}^{\dagger}\,D\psi_{\alpha}^{j}\,D\bar{\psi}_{j}^{\dot{\alpha}}\,d\chi_{k}\,d\bar{\chi}_{k}\,\,\mathcal{O}(\phi^{j},\phi_{j}^{\dagger},\psi_{\alpha}^{j},\bar{\psi}_{j}^{\dot{\alpha}})
\\
&
\times
\exp\left(\int d^{4}x\,(\mathcal{L}_{{\rm free}}+\mathcal{L}_{{\rm int}}+\mathcal{L}_{{\rm dt}})(\phi^{j},\phi_{j}^{\dagger},\psi_{\alpha}^{j},\bar{\psi}_{j}^{\dot{\alpha}})+\sum_{k=1}^{m}\bar{\chi}_{k,a}\left(y_{k}\cdot\Phi(x_{k})\right)_{~b}^{a}\chi_{k}^{b}\right)\,,
\nonumber
\end{flalign}
we generate the last term, linear in the scalars, in the exponent. The proof requires to cancel it with a shift of the real scalars $\Phi^I$ by ${S^{I}\left(x\right)=-\frac{1}{N}\sum_{k=1}^mI_{xx_{k}}y_{k}^{I}\chi_{k}\bar{\chi}_{k}}$, or equivalently $\phi^j\to \phi^j+S^j$ and $\phi_j^\dagger\to \phi_j^\dagger+S_j^\dagger$,
\begin{flalign}
\label{eq116}
&
\frac{1}{Z}\int D\phi^{j}\,D{\phi}_{j}^{\dagger}\,D\psi_{\alpha}^{j}\,D\bar{\psi}_{j}^{\dot{\alpha}}\,d\chi_{k}\,d\bar{\chi}_{k}\,\,\mathcal{O}(\phi^{j}+S^j,\phi_{j}^{\dagger}+S_j^\dagger,\psi_{\alpha}^{j},\bar{\psi}_{j}^{\dot{\alpha}})
\,
\exp\left(-\frac{N}{2}\int d^{4}x\,\textrm{tr}\left(S^{I}\square S^{I}\right)\right)
\\
&
\times\exp\left(\int d^{4}x\,[\mathcal{L}_{{\rm free}}(\phi^{j},\phi_{j}^{\dagger},\psi_{\alpha}^{j},\bar{\psi}_{j}^{\dot{\alpha}})+(\mathcal{L}_{{\rm int}}+\mathcal{L}_{{\rm dt}})(\phi^{j}+S^{j},\phi_{j}^{\dagger}+S_{j}^{\dagger},\psi_{\alpha}^{j},\bar{\psi}_{j}^{\dot{\alpha}})]\right)\,.
\nonumber
\end{flalign}
The shifts in the free Lagrangian $\mathcal{L}_{{\rm free}}$ cancel the last term in \eqref{eq115} and produce a quartic fermion interaction in the first line of \eqref{eq116}. The integrand can be compared to \eqref{eq16}. The rest of the proof (steps 3-4) trades the quartic interaction for the auxiliary boson $\rho_{kl}$ via a Hubbard-Stratonovich transformation.}. It is easy to compute at weak coupling: keeping only terms with $V$ interaction vertices at most in the expansion of the exponent and applying the Wick's theorem with the Feynman rules \eqref{eq3}-\eqref{eq4}. We can understand this logic with an example:
\begin{gather}
\label{eq109}
\mathcal{O}(y_1,y_2,y_3,y_4)=\tr\left(\phi^1(y_1)\phi_1^\dagger(y_2)\phi^3(y_3)\phi_3^\dagger(y_4)\right)
\end{gather}
corresponds to
\begingroup \allowdisplaybreaks
\begin{flalign}
\label{eq102}
&\mathcal{O}^{(S)}
=
\left\langle
\mathcal{O}(y_1,y_2,y_3,y_4)
\right\rangle_{\rm tree~level}
+
\left\langle
\tr\left(\phi^1(y_1)\phi_1^\dagger(y_2)S^3(y_3)S_3^\dagger(y_4)\right)
\right\rangle_{\rm tree~level}
\\
&
+
\left\langle
\tr\left(S^1(y_1)S_1^\dagger(y_2)\phi^3(y_3)\phi_3^\dagger(y_4)\right)
\right\rangle_{\rm tree~level}
+
\tr\left(S^1(y_1)S_1^\dagger(y_2)S^3(y_3)S_3^\dagger(y_4)\right)
+O(\xi_j^2,\alpha_j^2)\,,
\nonumber
\end{flalign}
\endgroup 
where the interactions in $\mathcal{L}_{\rm int}+\mathcal{L}_{\rm dt}$ in \eqref{eq16} are ignored at the lowest order, the bilinears are constants in the expectation values and only the non-zero correlators \footnote{Non-vanishing correlators have zero overall R-charges, see section 2.3 of \cite{Kazakov:2018gcy}. The charge assignments are in table 4 therein while the auxiliary fermions are charge-less.} are reported. We drop the brackets $\left\langle \dots \right\rangle$, defined by \eqref{eq17}, in the last term, since this does not depend on the physical fields. A basic implementation on the calculator is to pair-wise contract scalars in \eqref{eq102} by repeated application of \eqref{eq3}.
\\
The counterpart of \eqref{eq16} in $\mathcal{N}=4$ SYM, which we do not write here, would reproduce formula (3.10) of \cite{Jiang:2019xdz} at leading order and formula (23) of \cite{Vescovi:2021fjf} at sub-leading order in the SYM coupling. Furthermore, in a fishnet theory there is the hope that the weak-coupling expansion of $\mathcal{O}^{(S)}$ is such that the perturbative series of \eqref{eq5} can be resummed into an analytical function of the couplings, thus opening up the way to the finite- and strong-coupling regime. We verify and exploit this expectation in the case of the four-point function of appendix \ref{sec4pt}.

\begin{figure}[t]
\centering
\includegraphics[scale=0.3]{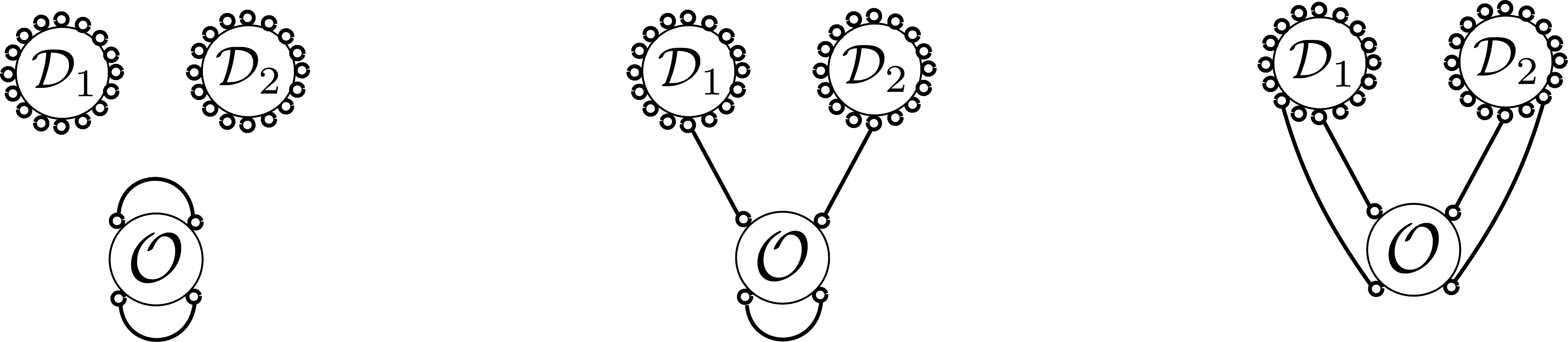}
\caption{Feynman diagrams in free theory for $G_2=\left\langle\mathcal{D}_1\mathcal{D}_2\,\mathcal{O}\right\rangle$ in \eqref{eq10} with $\mathcal{O}$ in \eqref{eq109}. In the effective theory they are generated by the first term (left), second and third term (middle) and last term (right) in \eqref{eq102}. Dots are the external fields and propagators between determinants are not depicted.}
\label{fig6}
\end{figure}

The last object to define is the average over the fermions
\begin{flalign}
\label{eq7}
\left\langle \mathcal{O}^{(S)}\right\rangle _{\chi}&=\frac{\int d\chi_k \, d\bar{\chi}_k \, \mathcal{O}^{(S)}\,e^{2\,i\sum_{k,l=1}^m\hat{\rho}_{kl}\,\bar{\chi}_{l,a}\chi_{k}^a}}{\int d\chi_k \, d\bar{\chi}_k\,e^{2\,i\sum_{k,l=1}^m\hat{\rho}_{kl}\,\bar{\chi}_{l,a}\chi_{k}^a}}\,,
\end{flalign}
which is calculable for fixed $\rho$ with the Feynman rule
$\left\langle \bar{\chi}_{k,a}\chi_{l}^{b}\right\rangle _{\chi}=-\frac{i}{2}\delta_{a}^{b}\left(\hat{\rho}^{-1}\right)_{kl}$. In the case of \eqref{eq102} the average acts trivially on the first term because this does not depend on bilinears, hence its contribution to \eqref{eq5} is a product of disconnected correlators (see figure \ref{fig6}, left panel)
\begin{gather}
\left(\frac{1}{Z_\rho}\int d\rho \,
 \,e^{-NS_{\textrm{eff}}\left[\rho\right]}
 \right)
  \left\langle
 \mathcal{O}
 \right\rangle_{\rm tree~level}
 =
  \left\langle
\mathcal{D}_1\dots \mathcal{D}_m
 \right\rangle_{\rm tree~level}
   \left\langle
 \mathcal{O}
 \right\rangle_{\rm tree~level}\,.
\end{gather}
The other terms in \eqref{eq102} contribute to the connected component of \eqref{eq5} (see figure \ref{fig6}, middle and right panel).

\paragraph{Large-$N$ limit.}
The advantage of the effective theory \eqref{eq5} is to make the $N$-dependence explicit. Because a factor of $N$ multiplies the exponent, we can saddle-point approximate the $\rho$-integral to leading order in ${1/N\to 0}$. The saddle points extremise $S_{\rm eff}[\rho]$ irrespective of $\left\langle\mathcal{O}^{(S)}\right\rangle$, so they are completely determined by the determinants' positions and polarisations. The relevant case in this paper is $m=2$: the saddle point is
\begin{gather}
\label{eq18}
\rho=
\frac{1}{4\pi}
\left(\begin{array}{cc}
0 & e^{i\theta} \\
e^{-i\theta} & 0
\end{array}\right)
\end{gather}
and it is parametrised by the zero mode $\theta\in[0,2\pi)$ \cite{Jiang:2019xdz}.
The planar contribution to \eqref{eq5} is a Gaussian integral over the fluctuations around the saddle point (see details in \cite{Vescovi:2021fjf}): we expand the effective action to quadratic order in the fluctuations and evaluate \eqref{eq7} and the Jacobian determinant, due to the change from Cartesian $(\Re (\rho_{12}),\Im (\rho_{12}))$ to polar coordinates, on the saddle point \eqref{eq18}.

We also comment on a subtlety in the algorithmic calculation of \eqref{eq7}. Taking up the example \eqref{eq102}
\begin{flalign}
\label{eq103}
&\left\langle \tr\left(S^{1}(y_{1})S_{1}^{\dagger}(y_{2})S^{3}(y_{3})S_{3}^{\dagger}(y_{4})\right)\right\rangle _{\chi}
\propto \frac{1}{N^4}\sum_{k_{1},k_{2},k_{3},k_{4}=1}^{m}I_{y_1x_{k_{1}}}I_{y_2x_{k_{2}}}I_{y_3x_{k_{3}}}I_{y_4x_{k_{4}}}
\\
&\times(y_{k_{1}}^{1}+iy_{k_{1}}^{2})(y_{k_{2}}^{1}-iy_{k_{2}}^{2})(y_{k_{3}}^{5}+iy_{k_{3}}^{6})(y_{k_{4}}^{5}-iy_{k_{4}}^{6})
\left\langle \chi_{k_{1}}^{a}\bar{\chi}_{k_{1},b}\chi_{k_{2}}^{b}\bar{\chi}_{k_{2},c}\chi_{k_{3}}^{c}\bar{\chi}_{k_{3},d}\chi_{k_{4}}^{d}\bar{\chi}_{k_{4},a}\right\rangle _{\chi}\,,
\nonumber
\end{flalign}
we find convenient to perform the fermionic average for generic $k_i$'s before the sums. While the highest power of $N$ is produced by the contractions of fermions sharing the same color index \cite{Jiang:2019xdz}
\begin{flalign}
\label{eq104}
&\left\langle \bar{\chi}_{k_{4},a}\chi_{k_{1}}^{a}\right\rangle _{\chi}\left\langle \bar{\chi}_{k_{1},b}\chi_{k_{2}}^{b}\right\rangle _{\chi}\left\langle \bar{\chi}_{k_{2},c}\chi_{k_{3}}^{c}\right\rangle _{\chi}\left\langle \bar{\chi}_{k_{3},d}\chi_{k_{4}}^{d}\right\rangle _{\chi}
\\
&\propto
N^4
\left(\hat{\rho}^{-1}\right)_{k_4k_1} \left(\hat{\rho}^{-1}\right)_{k_1k_2}\left(\hat{\rho}^{-1}\right)_{k_2k_3}\left(\hat{\rho}^{-1}\right)_{k_3k_4}\,,
\nonumber
\end{flalign}
there is the possibility that all terms in the sums \eqref{eq103} are zero because of some vanishing components of the polarisation vectors ($y^I_k$ and inside $\hat{\rho}$ given below \eqref{eq6}) and the vanishing entries of $\rho$ \eqref{eq18} (inside $\hat{\rho}$). A work-around is to calculate all $4!$ pair-wise contractions and discard terms that are too suppressed in $1/N$ to contribute to the planar correlator.

\subsection{Structure of the result and regularisation}
\label{secETfiniteness}

The previous section glosses over two important questions: what the effective theory really computes and how it deals with the usual divergences in loop calculations.

We address the first point with the case of the four-point function in the bi-scalar theory taken from appendix \ref{sec4pt}. At the heart of the output of the effective theory there are Feynman integrands in position space: in this example they read up to two loops \footnote{The form of the prefactors comes from the Stirling's approximation ${N!\,N^{-N}\sim\sqrt{2\pi N}e^{-N}}$.}
\begingroup \allowdisplaybreaks
\begin{flalign}
\label{eq96}
&\left\langle \mathcal{D}_{1}(x_1)\mathcal{D}_{2}(x_2)\mathcal{O}(x_3,x_4)\right\rangle
=\frac{e^{-N}}{N}\sqrt{\frac{2\pi}{N}}\left(I_{x_{1}x_{2}}\right)^{N-1}
\left[I_{x_{1}x_{3}}I_{x_{2}x_{4}}+\right.
\\
&
~~~~+2\left(4\pi\alpha_{1}\right)^{2}\int d^{4}x_{1'}\,I_{x_{1}x_{1'}}I_{x_{2}x_{1'}}I_{x_{3}x_{1'}}I_{x_{4}x_{1'}}+
\nonumber
\\
&
~~~~+4\left(4\pi\alpha_{1}\right)^{4}\int d^{4}x_{1'}d^{4}x_{2'}\,I_{x_{1}x_{1'}}I_{x_{2}x_{1'}}\left(I_{x_{1'}x_{2'}}\right)^{2}I_{x_{3}x_{2'}}I_{x_{4}x_{2'}}\,+
\nonumber
\\
&~~~~+\left(4\pi\xi\right)^{4}\int d^{4}x_{1'}d^{4}x_{2'}\,I_{x_{1}x_{1'}}I_{x_{2}x_{2'}}\left(I_{x_{1'}x_{2'}}\right)^{2}I_{x_{3}x_{1'}}I_{x_{4}x_{2'}}\,+
\nonumber
\\
&
~~~~\left.-\frac{\left(4\pi\xi\right)^{4}}{I_{x_{1}x_{2}}}\int d^{4}x_{1'}d^{4}x_{2'}\,I_{x_{1}x_{1'}}I_{x_{1}x_{2'}}I_{x_{2}x_{1'}}I_{x_{2}x_{2'}}I_{x_{1'}x_{2'}}I_{x_{3}x_{1'}}I_{x_{4}x_{2'}}+\left(x_{3}\leftrightarrow x_{4}\right)\right]
\nonumber
+\,\dots\,\,.
\end{flalign}
\endgroup
\\
The scalar propagators $I_{x_k x_{k'}}=(4\pi^2 (x_k-x_{k'})^2)^{-1}$ that end on one determinant come from \eqref{eq94}, whereas all others from the Wick contractions in \eqref{eq16}
Although one goal of appendix \ref{sec4pt} is to resum the series \eqref{eq96}, such ambition is not within the scope of the effective theory.

The perspective on the second issue is agnostic: the effective theory is not equipped with a regularisation scheme, nor does it require quantum conformal symmetry. The situation is not different from the works on single-trace correlators \cite{Grabner:2017pgm,Gromov:2018hut,Kazakov:2018gcy}, where UV divergences can be treated in dimensional regularisation and then removed by fine tuning the double-trace couplings to one RG fixed point, or the study of determinant correlators in $\mathcal{N}=4$ SYM, where at one loop the choice can fall on point-splitting regularisation \cite{Jiang:2019xdz,Vescovi:2021fjf}. Similarly to \cite{Grabner:2017pgm} in this paper we work at a fixed point and ignore the choice of regularisation both in section \ref{sec4ptBS}, because the result of the Bethe-Salpeter resummation is well-defined, and in the perturbative test of section \ref{sec4ptweak}, because the Feynman integrals therein are finite.

\section{Two-point functions}
\label{sec2pt}
Determinant operators are 1/2-BPS in $\mathcal{N}=4$ SYM and their two- and three-point functions are tree-level exact. Their correlation functions depend on the relative orientations $y_k\cdot y_l$ of the polarisations in \eqref{eq91} by virtue of the R-symmetry $SO(6)$.
Two reasons render the situation more complex in $\chi$CFT${}_4$ \footnote{The same problems affect the three-point functions in the hexagon framework \cite{Basso:2018cvy}.}.
First, the $\gamma$-twisting breaks $SO(6)$ down to the Cartan group $U(1)_1\times U(1)_2\times U(1)_3$. The complex scalars $\phi^j$ and $\phi_j^\dagger$ have charge $+1$ and $-1$ respectively under $U(1)_j$ and they are neutral under the other $U(1)$'s. In lack of the $SO(6)$ group, operators charged differently are distinct at quantum level and correlators become sensitive to the individual polarisations.
Second, the R-symmetry breaking causes the complete loss of SUSY. No other manifest symmetry can prevent determinants from developing anomalous dimensions quantum mechanically, similarly to the case of the BMN ``vacuum'' $\tr(\phi_1)^L$ \cite{Gurdogan:2015csr}.

In this section we aim to study the renormalisation properties of determinants \eqref{eq11} and their deformations \eqref{eq32}. The extent of the analysis is limited by the maximal number of vertices that our implementation of the effective theory can handle on a standard calculator. We are assisted in our survey by working exclusively in the bi-scalar reduction thanks to the smaller set of building blocks (see appendix \ref{secConventions}).
In what follows we mod out operators related by the $\mathbb{Z}_4$ discrete symmetry \cite{Cavaglia:2021mft} (see also \cite{Gromov:2018hut})
\begin{gather}
\label{eq112}
\phi^1\rightarrow \phi^{2}\,,
\qquad
\phi^2\rightarrow \phi_{1}^\dagger\,,
\qquad
\phi_{1}^\dagger\rightarrow \phi_{2}^\dagger\,,
\qquad
\phi_{2}^\dagger\rightarrow \phi^1\,.
\end{gather}

The perturbative way to find conformal primary operators is by diagonalising the mixing matrix. This procedure is complicated by the extent of operator mixing \footnote{See appendix F of \cite{Kazakov:2018gcy} for an overview in (non-unitary) fishnets and section 5 of \cite{Caetano:2016ydc} for an application.}. This allows in general the operators \eqref{eq11} and \eqref{eq32} to mix with other operators that share the same bare dimension, Cartan charges and spin. Resolving the mixing in this sector is hard because it entails a generalisation of the tools in section \ref{secET} \footnote{
One could recast the multi-trace operators into a generating function. In the matrix-model literature a trick for single traces is accomplished by the resolvent $\textrm{tr}(z-\phi^1)^{-1}=\sum_{i=0}^\infty z^{-i-1}\textrm{tr}(\phi^1)^i$. In this approach one could extract multi-traces from a product of resolvents (cf. \cite{Benvenuti:2006qr}). In the matrix-model literature the replica method is used to relate a single resolvent to another generating function that takes a determinant form: $\textrm{tr}(x-\phi^1)^{-1}=\lim_{n\to 0}\frac{1}{n}\frac{\partial}{\partial z}(\det(z-\phi^1))^n$. Alternatively, one could express $\textrm{tr}(z-\phi^1)^{-1}=\frac{\partial}{\partial z}\left.\frac{\det(z-\phi^1)}{\det(z'-\phi^1)}\right|_{z'\to z}$. These determinants can be included in the effective theory with an appropriate number of bosonic and fermionic degrees of freedom (see section 3.5.4 \cite{Jiang:2019xdz}).
The effective theory and its modifications, based on field-theory arguments, do not pose any conceptual difficulty to other non-supersymmetric theories. One should also notice that the fishnet is much nicer since the conformality plays the role of a regulator.} and possibly the diagonalisation of a mixing matrix whose dimension grows with $N$. In what follows we take a first look into two-point correlators under the assumption that the interaction vertices do not induce transitions from the operators \eqref{eq11} and \eqref{eq32} to operators of other type. On this premise, open to refinement in future works, we find two possible behaviours.

In the first case determinants made up of one type of scalar
\begin{gather}
\label{eq22}
\mathcal{D}(x)=\det(\phi^1(x))
=\frac{\varepsilon_{i_{1}\dots i_{N}}\varepsilon^{j_{1}\dots j_{N}}}{N!}
\left(\phi^1(x)\right)_{~j_{1}}^{i_{1}}
\dots\left(\phi^1(x)\right)_{~j_{N}}^{i_{N}}
\end{gather}
or decorated with elementary insertions in total number $\sum_{i=1}^3\ell_i\ll N$
\begin{flalign}
\label{eq23}
\mathcal{D}(x)&=\frac{\varepsilon_{i_{1}\dots i_{N}}\,\varepsilon^{j_{1}\dots j_{N}}}{\sqrt{N!(N-\sum_{i=1}^3\ell_i)!\prod_{i=1}^3\ell_i!}}
\left[
\left(\phi^{2}\right)_{~j_{1}}^{i_{1}}\dots\left(\phi^{2}\right)_{~j_{\ell_{1}}}^{i_{\ell_{1}}}
\left(\phi_{1}^{\dagger}\right)_{~j_{\ell_{1}+1}}^{i_{\ell_{1}+1}}\dots\left(\phi_{1}^{\dagger}\right)_{~j_{\ell_{1}+\ell_{2}}}^{i_{\ell_{1}+\ell_{2}}}\right.
\\
&
\left.
\times\left(\phi_{2}^{\dagger}\right)_{~j_{\ell_{1}+\ell_{2}+1}}^{i_{\ell_{1}+\ell_{2}+1}}\dots\left(\phi_{2}^{\dagger}\right)_{~j_{\ell_{1}+\ell_{2}+\ell_{3}}}^{i_{\ell_{1}+\ell_{2}+\ell_{3}}}
\left(\phi^{1}\right)_{~j_{\ell_{1}+\ell_{2}+\ell_{3}+1}}^{i_{\ell_{1}+\ell_{2}+\ell_{3}+1}}\dots\left(\phi^{1}\right)_{~j_{N}}^{i_{N}}\right](x)
\nonumber
\end{flalign}
have two-point functions that are protected from quantum corrections
\begin{gather}
\label{eq21}
\left\langle \mathcal{D}(x) \mathcal{D}^\dagger(0) \right\rangle=\frac{\mathcal{N}}{x^{2\Delta}}
~~~~~~~{\rm with}~~~
\mathcal{N}=\frac{\sqrt{2\pi N}}{(4\pi^2e)^N}\,,
~~~
\Delta=N\,.
\end{gather}
To see why it is so, let us have a look at some examples. Let us begin with \eqref{eq22}. We read off the tree-level coefficient from the ``full contraction'' (i.e. the term $L=0$) in \eqref{eq27}. Going to the loop corrections, the scalars contract both within themselves and with the vertices \eqref{eq1}-\eqref{eq2}. We know from \eqref{eq27} that the former contractions produce a single trace $\langle\tr(\phi^1(x_1)\phi_1^\dagger(x_2))^L\rangle$ (with $L>0$ and coefficient $O(\sqrt{N}e^{-N})/(x_{12})^{2N-2L}$) and multi-traces, while the latter attach vertices and try to retain terms of the same magnitude of the tree level $\mathcal{N}=O(\sqrt{N}e^{-N})$. One can verify that this attempt yields only non-planar terms (suppressed by $N^{-2}$ at least) in the case of the single trace, as well as for the multi-traces.
The same logic applies to \eqref{eq23}. Let us inspect an interesting example
\begin{flalign}
\label{eq117}
\mathcal{D}(x)=\left.\frac{1}{\sqrt{N}}\frac{d}{da_{1}}\det(\phi^{1}+a_{1}\,\phi_{1}^{\dagger})(x)\right|_{a_{1}=0}=\frac{\sqrt{N}}{N!}\varepsilon_{i_{1}\dots i_{N}}\,\varepsilon^{j_{1}\dots j_{N}}\left[(\phi_{1}^{\dagger})_{~j_{1}}^{i_{1}}(\phi^{1})_{~j_{2}}^{i_{2}}\dots(\phi^{1})_{~j_{N}}^{i_{N}}\right](x),
\end{flalign}
where we name $a_2$ the parameter in $\mathcal{D}^\dagger(0)$. Here we can easily point at a class of diagrams. Plugging \eqref{eq117} into \eqref{eq27} and taking the derivatives in $a_1$ and $a_2$, one finds also $\langle\tr(\phi^{1}(x_{1})\phi^{1}(x_{2}))\tr(\phi_{1}^{\dagger}(x_{1})\phi_{1}^{\dagger}(x_{2}))\rangle$. This correlator is known to be quantum corrected \cite{Grabner:2017pgm} (see a ``point-split'' version \eqref{eq77} in section \ref{sec3pt}), but the problem is that the combinatorial coefficient that comes along is sub-leading with respect to the tree-level one $\mathcal{N}$. This can be ascribed to different combinatorics if one works with the definition \eqref{eq117} in terms of $\varepsilon$ and operates Wick contractions. The tree level contracts $N-1$ pairs of $\phi^1$'s from $\mathcal{D}$ and $\phi_1^\dagger$'s from $\mathcal{D}^\dagger$, whereas the correlator above appears after $N-2$ pairs are contracted. Putting together all factors, the latter pairing brings a lessened power of $N$.
\\
%
%
%
\\
Under the aforementioned assumption, we extract the conformal data from perturbation theory as follows. The two-point function of a (bare) conformal primary operator $\mathcal{O}_\Delta$, ignoring the operator mixing, is expected to have the expansion
\begin{gather}
\label{eq110}
\left\langle \mathcal{O}_\Delta(x) \mathcal{O}_\Delta^\dagger(0) \right\rangle =
\frac{\mathcal{N}(\xi^2)}{x^{2\Delta_0}\left(x/\epsilon\right)^{2\gamma(\xi^2)}}
=\frac{\mathcal{N}(0)}{x^{2\Delta_0}}\left[1+\xi^{2}\left(\frac{\mathcal{N}'(0)}{\mathcal{N}(0)}-\gamma'(0)\,\log\frac{x^2}{\epsilon^2}\right)+\dots\right]
\end{gather}
where one splits the dimension $\Delta=\Delta_0+\gamma(\xi^2)$ into the classical and anomalous part and $\epsilon\ll 1$ is a UV cutoff used to regulate Feynman diagrams. On the other hand, as in the two examples above, we empirically find that the calculation of two-point functions fits the form \eqref{eq110} if one sets $\mathcal{N}'(0)/\mathcal{N}(0)=\mathcal{N}''(0)/\mathcal{N}(0)=\dots=O(N^{-1})$, $\Delta_0=N$ and $\gamma'(0)=\gamma''(0)=\dots=O(N^{-1})$. This conclusion is tantamount to ignore the terms $\xi^{2n}$ in \eqref{eq110} and write \eqref{eq21}. On a side note, the operators \eqref{eq22} and \eqref{eq23} are reminiscent of the non-chiral single traces \cite{Caetano:2016ydc}
\begin{gather}
\tr\!
\left(
(\phi^1)^{L_1}\,
\phi_1^\dagger\,
(\phi^1)^{L_2}\,
\phi_1^\dagger\,
\dots
\right)(x)\,,
\qquad~~~
L_1,L_2,\dots \neq0,
\end{gather}
whose anomalous dimension vanishes too in the planar limit due to diagrammatical arguments.
\\
Determinants with composite insertions \eqref{eq32} with $y\cdot \Phi=\phi^1$ and with the $\tilde{\chi}_i$'s equal to linear combinations of the scalars, for example $\phi^1-\phi^2+2\phi_3^\dagger$, have dimension $\Delta=N$. Their two-point functions are linear combinations of those in \eqref{eq21}.

The second case includes determinants of more than one scalar, e.g. ${\mathcal{D}=\det(\phi^1-2\,\phi_2^\dagger)}$, and the deformations by a small number of insertions. They are combinations of operators of the type \eqref{eq23} but with large number of insertions $\sum_{i=1}^5\ell_i\sim N$. The perturbative expansion of two-point functions fails to exponentiate in the form \eqref{eq110}. For example one can use the effective theory to quantify the bare two-point function of $\mathcal{D}=\det(\phi^1+\phi^2)$:
\begin{flalign}
\frac{
\left\langle 
\mathcal{D}(x)\mathcal{D}^\dagger(0)
\right\rangle
}
{
\left\langle 
\mathcal{D}(x)\mathcal{D}^\dagger(0)
\right\rangle_{\rm tree~level}
}
&=
1+8\pi^{4}\xi^{4}I_{x_{1}x_{2}}^{-4}X_{x_{1}x_{1}x_{2}x_{2}}^{2}+\frac{64}{3}\pi^{6}\xi^{6}I_{x_{1}x_{2}}^{-6}X_{x_{1}x_{1}x_{2}x_{2}}^{3}+O\!\left(\xi^{8}\right)\,.
\end{flalign}
At the fixed point \eqref{eq87} we ignore Feynman integrals multiplied by powers of $\alpha_2^2+\xi^2$ and write the resulting sum in terms of the box master integral
$X_{x_{1}x_{2}x_{3}x_{4}}=\int d^{4}y\,I_{x_{1}y}I_{x_{2}y}I_{x_{3}y}I_{x_{4}y}$. While this is finite for distinct points \cite{Usyukina:1992jd}, we use point-splitting regularisation to regulate the divergence at coincident points \cite{Drukker:2008pi}
\begin{gather}
X_{x_{1}x_{1}x_{2}x_{2}}=\frac{I_{x_{1}x_{2}}^{2}}{8\pi^{2}}\left(1-\log\frac{\epsilon^{2}}{x_{12}^{2}}\right)\,,
\end{gather}
after which the result reads
\begin{gather}
\frac{
\left\langle 
\mathcal{D}(x)\mathcal{D}^\dagger(0)
\right\rangle
}
{
\left\langle 
\mathcal{D}(x)\mathcal{D}^\dagger(0)
\right\rangle_{\rm tree~level}
}
=1
+\frac{\xi^4}{8}\left(1-\log\frac{\epsilon^2}{x_{12}^2}\right)^2
+\frac{\xi^6}{24}\left(1-\log\frac{\epsilon^2}{x_{12}^2}\right)^3
+O(\xi^8)\,.
\end{gather}
The pattern fits a multi-exponential behaviour and the operators should be rather viewed as combinations of primaries.
\\
We can tweak \eqref{eq117} into the operator $\mathcal{D}(x)=\det(\phi^{1}+\phi_{1}^{\dagger})(x)$ and see what changes in the counting of the relative powers of $N$ between tree and loop level. Repeating the argument below \eqref{eq117}, there are the diagrams that describe $\langle\tr(\phi^{1}(x_{1})\phi^{1}(x_{2}))\tr(\phi_{1}^{\dagger}(x_{1})\phi_{1}^{\dagger}(x_{2}))\rangle$, but the overall coefficient is enhanced and scales as that of the tree level. A new combinatorics takes the place of the previous argument: the tree level contracts all scalars $\phi^{1}+\phi_{1}^{\dagger}$ of $\mathcal{D}$ with those $\phi^{1}+\phi_{1}^{\dagger}$ of $\mathcal{D}^\dagger$ in $N$ pairs, whereas $N-2$ pairings are necessary to build the said correlator. Overall this evens out the crucial factors and elevates loop diagrams to planar level.

\section{Three- and four-point functions}
\label{sec3pt}

In this section we assemble the basic correlation functions of two determinant-like operators \eqref{eq22}-\eqref{eq23}, which are identified as protected primaries in relation to the working assumption below \eqref{eq112}, and one single trace of the type considered in \cite{Gurdogan:2015csr,Caetano:2016ydc,Grabner:2017pgm}.

\paragraph{Length-2 traces.} We begin with various examples of three-point functions of two determinants with a single insertion each
\begingroup \allowdisplaybreaks
\begin{flalign}
\!\!\!\!\mathcal{D}_1(x_1)
&
=\left.\frac{1}{N}\frac{d}{da_1}\det(\phi^2+a_1\, \tilde{\chi}_1)(x_1)\right|_{a_1=0}
=\frac{\varepsilon_{i_{1}\dots i_{N}}\varepsilon^{j_{1}\dots j_{N}}}{N!}
\left[\left(\tilde{\chi}_1\right)_{~j_{1}}^{i_{1}}
\left(\phi^2\right)_{~j_{2}}^{i_{2}}
\dots\left(\phi^2\right)_{~j_{N}}^{i_{N}}\right](x_1)\,,
\nonumber
\\
\!\!\!\!\mathcal{D}_2(x_2)
&=\left.\frac{1}{N}\frac{d}{da_2}\det(\phi_2^\dagger+a_2\, \tilde{\chi}_2)(x_2)\right|_{a_2=0}
=\frac{\varepsilon_{i_{1}\dots i_{N}}\varepsilon^{j_{1}\dots j_{N}}}{N!}
\left[\left(\tilde{\chi}_2\right)_{~j_{1}}^{i_{1}}
\left(\phi_2^\dagger\right)_{~j_{2}}^{i_{2}}
\dots\left(\phi_2^\dagger\right)_{~j_{N}}^{i_{N}}\right](x_2)
\nonumber
\\
\end{flalign}
\endgroup
and one single trace of minimal length
\begin{gather}
\mathcal{O}(x_3,x_4)
=
\tr
\left(
\tilde{\chi}_1^\dagger(x_3)
\,
\tilde{\chi}_2^\dagger(x_4)
\right)\,.
\end{gather}
The latter projects the determinant pair onto a single trace of the same length via \eqref{eq27}, so all cases are either tree-level exact or reduce to the four-point functions of \cite{Grabner:2017pgm}. An exception is the last case below. For the sake of completeness, let us mention that four-point functions generally receive disconnected (d) and connected (c) contributions
\begin{gather}
\label{eq88}
\left\langle \mathcal{D}_1\mathcal{D}_2\,\mathcal{O} \right\rangle
=
\left\langle \mathcal{D}_1\mathcal{D}_2\,\mathcal{O} \right\rangle_{\rm d}
+
\left\langle \mathcal{D}_1\mathcal{D}_2\,\mathcal{O} \right\rangle_{\rm c}\,.
\end{gather}
The former, defined as $\left\langle \mathcal{D}_1\mathcal{D}_2\right\rangle \left\langle\mathcal{O} \right\rangle$ by the factorisation of the coordinate dependence, can vanish, if the two-point functions carry non-zero R-charges, or it is dominant in powers of $N$, but it is usually the latter to harbour all the interesting physics.

The simplest choice $(\tilde{\chi}_1,\tilde{\chi}_2)=(\phi^2,\phi_2^\dagger)$ corresponds to no insertion at all and generates only a tree-level contribution
\begin{gather}
\label{eq98}
\left\langle \mathcal{D}_1(x_1)\mathcal{D}_2(x_2)\,\mathcal{O}(x_3,x_4) \right\rangle
=
\frac{C_2}{x_{12}^{2N-2}x_{13}^{2}x_{24}^{2}}
\qquad
{\rm with}
\qquad
C_L=\frac{\left(-\right)^{L/2+1}\sqrt{2\pi N}}{\left(4\pi^{2}\right)^{N+L/2}e^{N}}\,.
\end{gather}
The same is true for $(\tilde{\chi}_1,\tilde{\chi}_2)=(\phi^1,\phi_1^\dagger)$ with a rescaled constant due to the different combinatorics:
\begin{gather}
\left\langle \mathcal{D}_1(x_1)\mathcal{D}_2(x_2)\,\mathcal{O}(x_3,x_4) \right\rangle
=
\frac{C_2}{N^2\,x_{12}^{2N-2}x_{13}^{2}x_{24}^{2}}\,.
\end{gather}
The next cases
$(\tilde{\chi}_1,\tilde{\chi}_2)=
(\phi^2,\phi^1)
$
and
$\tilde{\chi}_1,\tilde{\chi}_2
=
(\phi^2,\phi_1^\dagger)
$
are the tree-level exact correlators of \cite{Grabner:2017pgm}. The vanishing of the $\beta$-functions of the couplings $\alpha_2^2$ and $\alpha_3^2$ forces loop corrections to vanish. More explicitly, $\ell$-loop diagrams come with the powers $(\alpha_2^2+\xi^2)^\ell$ or $(\alpha_3^2+\xi^2)^\ell$ and only the tree level survives at the conformal fixed points \eqref{eq87}:
\begin{gather}
\left\langle \mathcal{D}_1(x_1)\mathcal{D}_2(x_2)\,\mathcal{O}(x_3,x_4) \right\rangle
=
\frac{C_2}{N\,x_{12}^{2N-2}x_{13}^{2}x_{24}^{2}}\,.
\end{gather}

The case with $(\tilde{\chi}_1,\tilde{\chi}_2)=
(\phi^2,\phi^2)$ projects onto the simplest non-trivial four-point function:
\begin{gather}
\label{eq97}
\left\langle \mathcal{D}_1(x_1)\mathcal{D}_2(x_2)\,\mathcal{O}(x_3,x_4) \right\rangle
=
\frac{C_{2}}{x_{12}^{2N-2}}\frac{\left(4\pi^{2}\right)^{2}}{N}
\,
\left\langle \tr \left(
\phi^2(x_1) \phi^2(x_2)
\right)
\tr \left(
\phi_2^\dagger(x_3) \phi_2^\dagger(x_4)
\right)
\right\rangle\,.
\end{gather}
This was already solved in the bi-scalar theory \cite{Grabner:2017pgm,Gromov:2018hut} and in $\chi$CFT${}_4$ \cite{Kazakov:2018gcy}, where in the latter paper it is extensively studied in many limits of the model. The conformal partial wave expansion of it reads in our conventions (see appendix \ref{sec4pt})
\begin{flalign}
\label{eq77}
\left\langle \textrm{tr}\left(\phi^{2}\left(x_{1}\right)\phi^{2}\left(x_{2}\right)\right)
\tr \left(
\phi_2^\dagger(x_3) \phi_2^\dagger(x_4)
\right)
\right\rangle 
&=\frac{1}{x_{12}^{2}x_{34}^{2}}\sum_{S=0,2,\dots}\sum_{\Delta=\Delta_{-},\Delta_{+}}
C^2_{\mathcal{O}_{\Delta,S}\mathcal{O}}
\,g_{\Delta,S}^{0,0}\left(u,v\right)\,.
\end{flalign}
We denote the four-dimensional conformal block \cite{Dolan:2000ut} as
\begingroup \allowdisplaybreaks
\begin{flalign}
\label{eq68}
&g^{\Delta_1-\Delta_2,\Delta_3-\Delta_4}_{\Delta,S}=(-1)^S \frac{z\bar  z}{
 z - \bar z} \left[ k(\Delta + S, z) k(\Delta - S - 2, \bar z) - 
   k(\Delta + S, \bar z) k(\Delta - S - 2, z)\right],
\nonumber   
   \\
&\text{with} \quad k(\beta, x) = x^{\beta/2}
  \,\,_2F_1\left(\frac{\beta - (\Delta_{1} -\Delta_{2})}{2}, \frac{\beta + (\Delta_{3} - \Delta_{4})}{2}; \beta; 
   x\right)\,,
\end{flalign}
\endgroup
which is a function of the cross-ratios
\begin{flalign}
\label{eq67}
u=z\bar{z}&=\frac{x_{12}^2 x_{34}^2}{x_{13}^2 x_{24}^2}\,,
\qquad\qquad
v=(1-z)(1-\bar{z})=\frac{x_{14}^2 x_{23}^2}{x_{13}^2 x_{24}^2}\,.
\end{flalign}
The $s$-channel expansion runs over the exchanged states $\mathcal{O}_{\Delta,S}$, with Lorentz spin $S$ and scaling dimensions $\Delta$ given by
\begin{gather}
\label{eq113}
\Delta_{\pm}=2+\sqrt{\left(S+1\right)^{2}+1\pm2\sqrt{\left(S+1\right)^{2}+4\xi^{4}}}\,.
\end{gather}
The square of the OPE coefficients at finite coupling is
\begin{gather}
\label{eq106}
C^2_{\mathcal{O}_{\Delta,S}\mathcal{O}}
=
\frac{(S+1)\,\Gamma^{2}\left(\frac{S+\Delta}{2}\right)\Gamma\left(S-\Delta+4\right)}{4\pi^{4}\left[\left(4-\Delta\right)\Delta+S\left(S+2\right)-2\right]\Gamma^{2}\left(\frac{S-\Delta}{2}+2\right)\Gamma\left(S+\Delta-1\right)}\,.
\end{gather}
At weak coupling the partial waves labeled by $-$ and $+$ correspond respectively to twist-$2$ states $\textrm{tr}(\phi^1 (n\cdot \partial)^S \phi^1+\dots)$ and twist-$4$ states $\textrm{tr}(\square \phi^1 (n\cdot \partial)^S \phi^1+\dots)$, where the dots stand for similar terms with light-cone derivatives distributed among the fields.

The last case 
$(\tilde{\chi}_1,\tilde{\chi}_2)=
(\phi^1,\phi^1)$ corresponds to a correlator with a richer graph structure. We evaluate it and attempt to explain the resulting OPE content in appendix \ref{sec4pt}.

\paragraph{Higher-length traces.} The non-chiral local operators of the type
$\tr\!
\left(
(\phi^2)^{L_1}\,
\phi_2^\dagger\,
(\phi^2)^{L_2}\,
\phi_2^\dagger
\dots
\right)$
with $L_1,L_2,\dots \neq0$ are protected \cite{Caetano:2016ydc}. Since the single-trace part of the PCGG \eqref{eq29} contains an alternating sequence of scalars, it can only overlap a non-chiral trace with the same pattern:
\begin{gather}
\left\langle
\det(\phi^2(x_1))
\det(\phi_2^\dagger(x_2))
\,
\tr\!
\left(
\phi^2(x_3)
\phi_2^\dagger(x_3)
\right)^{L/2}
\right\rangle
=
\frac{C_L}{x_{12}^{2N-L}x_{13}^{L}x_{23}^{L}}\,.
\end{gather}

More general correlators have even more complicated graphs. We qualitatively describe the planar diagrammatics of three-point functions of two determinant-like operators and a multi-magnon chiral operator \cite{Caetano:2016ydc}
\begingroup \allowdisplaybreaks
\begin{gather}
\label{eq101}
\tr\!
\left(
(\phi^1)^{L_1}\,
\phi^2\,
(\phi^1)^{L_2}\,
\phi^2\,
\dots
\right)(x_3)\,,
\qquad~~~
M+\sum_i L_i=L\,,
\qquad~~~
M\leq \sum_i L_i\,,
\end{gather}
\endgroup
where $M$ is the number of magnons $\phi^2$. In the bi-scalar theory the operators \eqref{eq101} develop anomalous dimensions \cite{Gurdogan:2015csr,Caetano:2016ydc}, therefore the relevant observable is the (finite) structure constant of the renormalised operator.

\begin{figure}[t]
\centering
\includegraphics[scale=0.19]{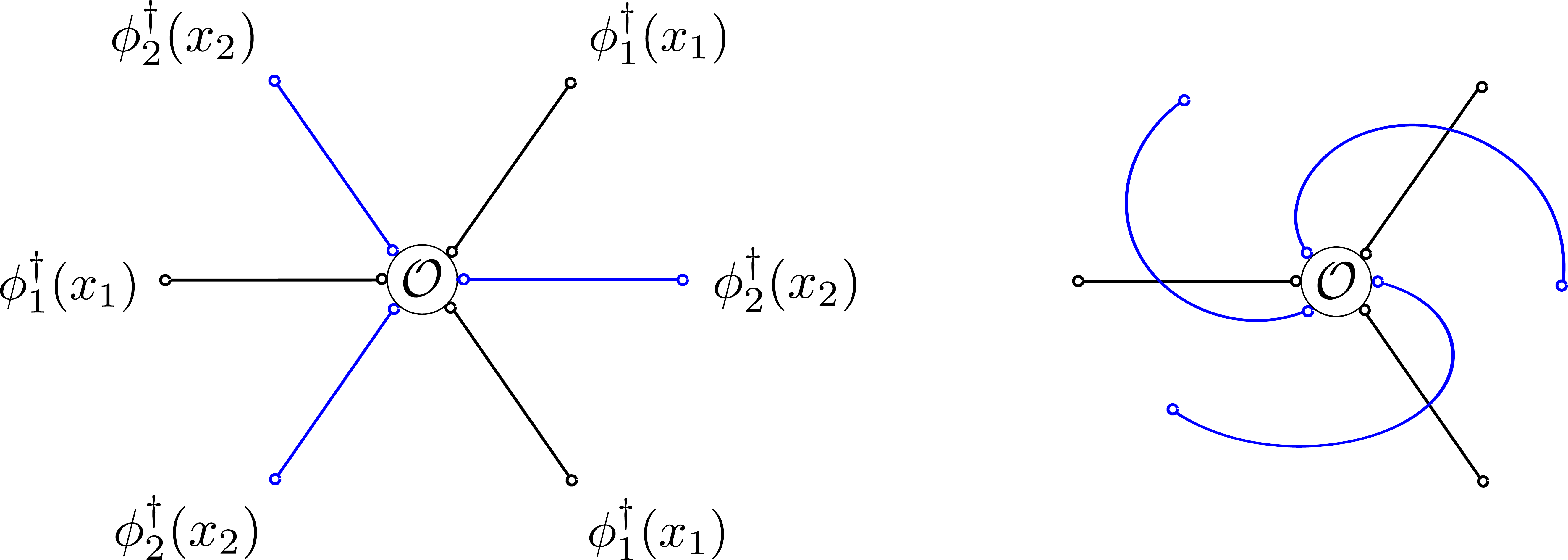}
\caption{Feynman graphs in the expansion of \eqref{eq99}. For visual clarity the single-trace contribution of the determinant pair is ``smeared'' in spacetime around $\mathcal{O} $ at the center. Colours stand for the ``flavour'' of the scalars (black for $\phi^1$ and blue for $\phi^2$) and the dots for the external fields. The intersections between the lines are $\xi^2$-vertices. We represent the tree level (left) and an unwrapped spiral graph (right) for the three-magnon $\mathcal{O}=\tr(\phi^1\phi^2)^3$.
}
\label{fig4}
\end{figure}

We can construct a non-zero three-point function with \eqref{eq101} by adding the impurity $\phi_1^\dagger$ into one determinant
\begingroup \allowdisplaybreaks
\begin{flalign}
\mathcal{D}_1(x_1)
&=\frac{\varepsilon_{i_{1}\dots i_{N}}\varepsilon^{j_{1}\dots j_{N}}}{N!}
\left[
\left(\phi^2+\phi_1^\dagger\right)_{~j_{1}}^{i_{1}}
\dots 
\left(\phi^2+\phi_1^\dagger\right)_{~j_{L/2}}^{i_{L/2}}
\left(\phi^2\right)_{~j_{L/2+1}}^{i_{L/2+1}}
\dots\left(\phi^2\right)_{~j_{N}}^{i_{N}}\right](x_1)\,,
\\
\mathcal{D}_2(x_2)
&=\frac{\varepsilon_{i_{1}\dots i_{N}}\varepsilon^{j_{1}\dots j_{N}}}{N!}
\left[
\left(\phi_2^\dagger\right)_{~j_{1}}^{i_{1}}
\dots
\left(\phi_2^\dagger\right)_{~j_{N}}^{i_{N}}\right](x_2)
\end{flalign}
\endgroup
and alternating the scalars in the trace
\begin{gather}
\mathcal{O}(x_3)
=
\tr\left(\phi^1(x_3) \phi^2(x_3)\right)^{L/2}\,.
\end{gather}
The reason for this choice can be understood as follows. We operate $N-L$ Wick contractions among $\phi^2(x_1)$'s and $\phi^\dagger_2(x_2)$'s and create a length-$L$ PCGG. Planarity implies that the single trace $\mathcal{O}$ overlaps with the single-trace term of the PCGG:
\begingroup \allowdisplaybreaks
\begin{gather}
\label{eq99}
\left\langle \mathcal{D}_1(x_1)\mathcal{D}_2(x_2)\,\mathcal{O}(x_3) \right\rangle
\propto
\left\langle 
\tr\left(\phi_1^\dagger(x_1) \phi_2^\dagger(x_2)\right)^{L/2}
\,\mathcal{O}(x_3) \right\rangle\,.
\end{gather}
\endgroup
The alternating sequence of scalars in the PCGG forces the same pattern upon $\mathcal{O}$ in order to have a non-zero tree level. The resulting three-point function \eqref{eq99} is a point-split version of the two-point function of the same operators with $x_1=x_2$, so the Feynman graphs in figure \ref{fig4} resemble the (wrapped and unwrapped) spiral graphs of \cite{Caetano:2016ydc}. However, two differences prevent a straightforward adaptation of spectral methods to the calculation of the renormalised structure constant. First, the boundary conditions of the bulk lattice are different: the outer propagators end in $x_1$ and $x_2$ rather than converging to a single point. Second, if we were interested only in the divergent part of \eqref{eq99}, as in the computation of the anomalous dimension when $x_1=x_2$, we could amputate the outer propagators and reduce the computation to the spiral graphs of \cite{Caetano:2016ydc}. This operation is no longer permitted for the purpose of extracting the finite parts of \eqref{eq99}. The knowledge of the structure constant would be equivalent to the knowledge of the overlap between the CFT wave-function \cite{Gromov:2019jfh} and a ``Dirac-delta'' state that anchors the outer propagators to the points $x_1$ and $x_2$.

\begin{figure}[t]
\centering
\includegraphics[scale=0.19]{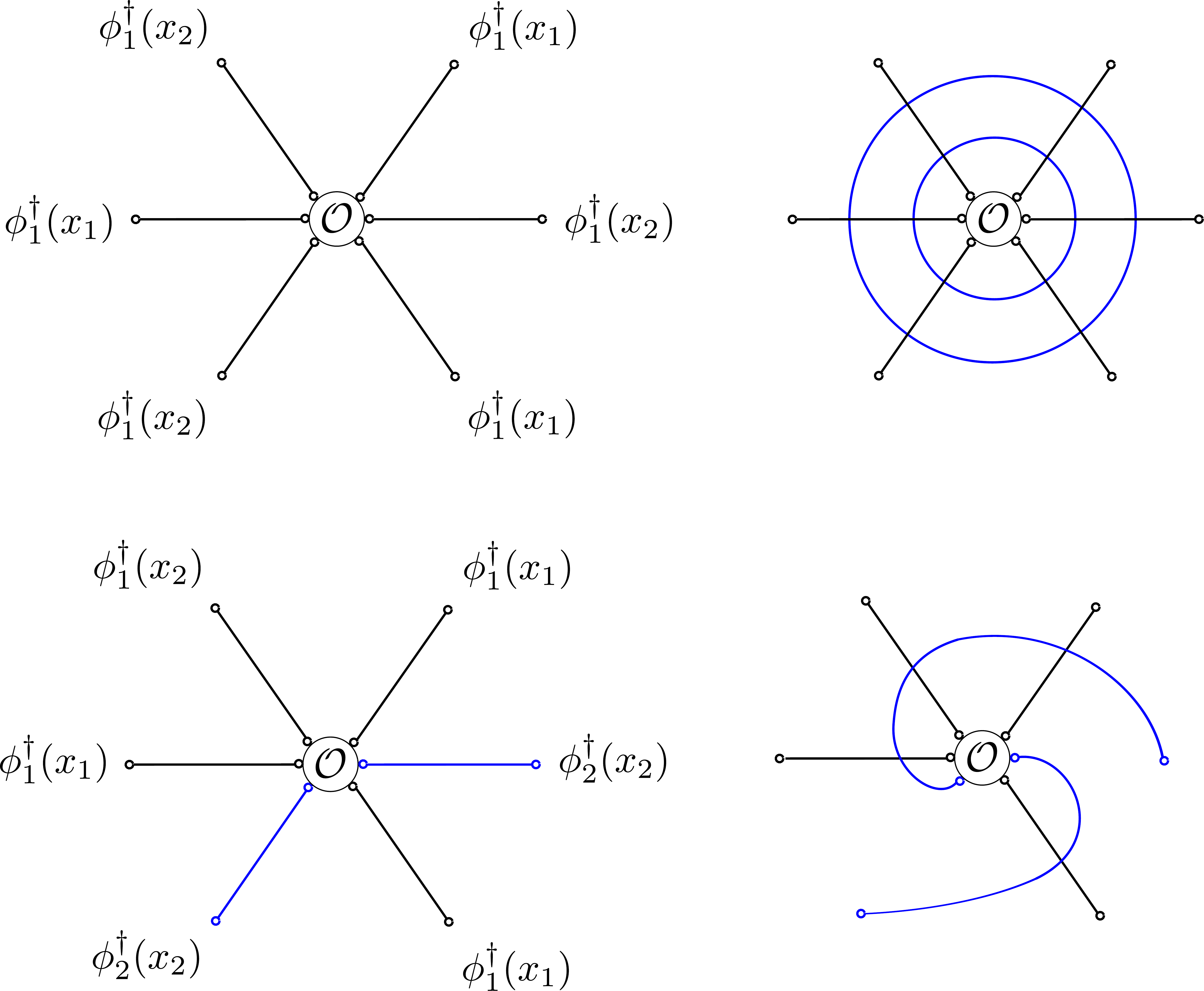}
\caption{
A class of Feynman graphs in the expansion of the three-point function of \eqref{eq100}-\eqref{eq111}. For the BMN ``vacuum'' $\mathcal{O}=\tr({\phi^1})^6$: tree level (top left) and a globe graph (top right).
For the two-magnon trace $\mathcal{O}=\tr((\phi^1)^3\phi^2\phi^1\phi^2)$: tree level (bottom left) and an unwrapped spiral graph (bottom right).
}
\label{fig5}
\end{figure}

Similarly, we can place the same impurity in both determinants
\begin{flalign}
\label{eq100}
\mathcal{D}_1(x_1)
&=\frac{\varepsilon_{i_{1}\dots i_{N}}\varepsilon^{j_{1}\dots j_{N}}}{N!}
\left[
\left(\phi_1^\dagger\right)_{~j_{1}}^{i_{1}}
\dots 
\left(\phi_1^\dagger\right)_{~j_{L/2-M}}^{i_{L/2-M}}
\left(\phi^2\right)_{~j_{L/2-M+1}}^{i_{L/2-M+1}}
\dots\left(\phi^2\right)_{~j_{N}}^{i_{N}}\right](x_1)
\\
\mathcal{D}_2(x_2)
&=\frac{\varepsilon_{i_{1}\dots i_{N}}\varepsilon^{j_{1}\dots j_{N}}}{N!}
\left[
\left(\phi_1^\dagger\right)_{~j_{1}}^{i_{1}}
\dots 
\left(\phi_1^\dagger\right)_{~j_{L/2}}^{i_{L/2}}
\left(\phi_2^\dagger\right)_{~j_{L/2+1}}^{i_{L/2+1}}
\dots\left(\phi_2^\dagger\right)_{~j_{N}}^{i_{N}}\right](x_2)
\end{flalign}
and look for the operators \eqref{eq101} with a non-zero overlap at tree level. They turn out to be of the type
\begingroup \allowdisplaybreaks
\begin{gather}
\label{eq111}
\mathcal{O}(x_3)
=
\tr\left(
(\phi^{1}\phi^{2})^{{L_{1}/2}}
(\phi^{1}\phi^{1})^{{L_{2}/2}}
(\phi^{1}\phi^{2})^{{L_{3}/2}}
(\phi^{1}\phi^{1})^{{L_{4}/2}}
\dots
\right)(x_3)\,,
\\
L=\sum_i L_i\,,
\qquad\qquad
M=\sum_{i~\rm odd} L_i
\nonumber
\end{gather}
\endgroup
where magnons $\phi^2$ are separated by an odd number of $\phi^1$'s. Some of the loop diagrams are in figure \ref{fig5} and compare to the globe and spiral graphs of \cite{Caetano:2016ydc} with the caveat below \eqref{eq99}.

\section{Conclusion}
\label{secConclusion}

This paper represents the first study of operators of R-charge of order $N$ in the planar $\gamma$-deformation of $\mathcal{N}=4$ SYM, in the limit combining the weak coupling and the strong imaginary deformation parameter \cite{Gurdogan:2015csr}. The investigation of two-, three- and four-point functions of determinant(-like) and single trace operators walks the fine line between tree-level exact and quantum-corrected correlation functions, which are in principle solvable by the whole arsenal of operatorial and integrability methods.

Let us put together the loose ends in the main text.
\\
The effective theory in section \ref{secET} is constrained by our implementation of the Wick's theorem on physical (below \eqref{eq102}) and auxiliary fields (below \eqref{eq104}). The former operation hits a bottleneck when contractions occur among many interaction vertices of the same type, in particular the $\alpha_1^2$-vertices \eqref{eq2} which have a single type of scalars. One could set up a more ingenious algorithm: list all adjacency matrices of Feynman graphs with given sites (vertices and external operators), mod out permutations of vertices of the same type and compensate for them with a symmetry factor. Another benefit would come from a compiled programming language. The optimisation is a prerequisite to refine the investigation of section \ref{sec2pt} and move on to fishnets with a richer graph content \cite{Caetano:2016ydc,Kazakov:2018gcy}.
\\
Our glance into the spectrum in section \ref{sec2pt} is far from exhaustive. The interesting questions revolve around the extent of operatorial mixing and the identification of operators with anomalous dimensions. In light of the analogy in section \ref{secIntro}, the latter challenge compares to the measurement of the effective mass of ``baryons'' from the two-point functions of these ``particle states''. The fishnet theories represent a setting where such goal could be completed
without approximation in the coupling.
\\
The three- and four-point functions in section \ref{sec3pt} show a great deal of variety in the graph content. A complete understanding would pave the way to access new conformal data. The Bethe-Salpeter kernels for an overlap with a length-$L$ trace (with $L=2$ in appendix \ref{sec4pt}) need the diagonalisation of a $L$-site spin chain in an infinite-dimensional representation of the conformal group $SU(2,2)$ \cite{Gromov:2017cja}. The solution is readily provided by the conformal triangles basis in the case $L=2$, whereas for $L>2$ quantum spin chain methods have been developed only in recent times \cite{Derkachov:2019tzo,Derkachov:2021rrf}.

There are several further directions worth pursuing out of the scope of this paper.
\\
There is an obvious generalisation of the effective model to a plethora of more complicated composite operators: permanents/dual giant gravitons \cite{Grisaru:2000zn,Hashimoto:2000zp}, sub-determinants/non-maximal giant gravitons \cite{McGreevy:2000cw,Balasubramanian:2001nh}, Schur polynomials \cite{Corley:2001zk} and restricted Schur polynomials/excited giants \cite{Balasubramanian:2004nb,deMelloKoch:2007rqf,deMelloKoch:2007nbd,Bekker:2007ea}. A motivation is to investigate the existence of heavy operators providing integrable boundary states \cite{Cavaglia:2021mft}.
\\
An interesting setting is obviously the $\chi$CFT${}_4$. The search for integrability may lead only to a handful of positive results since only determinants showed evidence in favour of integrability in $\mathcal{N}=4$ SYM \cite{Chen:2019gsb}. However, such conclusion could be hasty when one takes into account open spin chains/open strings. Open strings ending on maximal giant gravitons have an integrable dynamics \cite{Berenstein:2005vf,Hofman:2007xp}, whereas integrability is less certain on less-than-maximal giants \cite{Berenstein:2006qk,Ciavarella:2010tp,deMelloKoch:2016mhc,deMelloKoch:2018tlb}. The chiral theory should be a good testing ground for the spectrum of the spin chain attached to heavy states via Bethe ansatz.
\\
Another interesting direction would be towards other fishnet theories. The analysis of \cite{Chen:2019kgc,Yang:2021hrl} suggests that (sub-)determinants should preserve integrability in ABJM, making in turn the doubly-scaled $\chi$CFT${}_3$ of \cite{Caetano:2016ydc} a testing ground for similar ideas \cite{Chen:2018sbp,Bai:2019soy}. Little is known in the $\gamma$-deformation of (non-integrable) $\mathcal{N}=2$ theories in four dimensions, although the authors of \cite{Pittelli:2019ceq} singled out an integrable sector in a double-scaling limit of the quiver.
\\
A motivation in section \ref{secIntro} is to move towards a SoV approach for the observables in this paper.
%
%
%
%
A latent reason behind fishnets is to move away from the double-scaling limit and recover the undeformed $\mathcal{N}=4$ SYM as a perturbation in the twists.
\\
Another direction of study could be holography. In \cite{Balasubramanian:2002sa} open strings are shown to originate from the quantisation of fluctuations around states of large R-charge/momentum. The energies are mapped to dimensions of operators in the field theory and the world-sheet vibration spectrum is reproduced in gauge theory. It would be interesting to set up a similar investigation in the bi-scalar model and make connections to the holographic descriptions \cite{Gromov:2019aku,Basso:2019xay}. The priority is to establish evidence of an exponential scaling typical for a semiclassical description. In the absence of data we venture to outline what could change in the fishchain. The classical model emerges directly from a mapping between boundary (length-$J$ single traces) and bulk ($J$ particles in $AdS_5$) degrees of freedom. The map goes through the identification of a single graph-building operator with a Hamiltonian constraint on the particles. Insofar as determinants cannot be projected onto a (finite number of) single traces, the new model may well support infinitely-many degrees of freedom or multiple constraints. At the same time, as much as the locality of the fishchain correlates with a Polyakov-type action for a ``discretised world-sheet'', one expects that a similar mechanism generates a DBI-like action for a ``discretised brane'', whose equations of motion reproduce the scaling at large coupling. Moreover in the fishnet, for a given single trace, there exists a special correlator (the CFT boundary wave-function) that allows to express any other correlator. A feature of the quantum fishchain is to raise this to the bulk wave-function of the dual Hilbert space. An indication towards a fishchain-like description perhaps addresses whether such special correlator exists for determinants in the first place.
\\
The integrating in-and-out procedure in section \ref{secETderivation} has an interpretation of graph duality and a relation to the open-closed-open duality in the AdS${}_5$/CFT${}_4$ system \cite{Jiang:2019xdz} (later extended to dual giants \cite{Chen:2019gsb} and to AdS${}_4$/CFT${}_3$ \cite{Chen:2019kgc}), as first discussed in \cite{Gopakumar}. It might be interesting to investigate a connection to the recent progress in holography.

\appendix

\section{Conventions}
\label{secConventions}

The action $S_{\phi\psi}=\int d^4x\, (\mathcal{L}_{\rm free}+\mathcal{L}_{\rm int}+\mathcal{L}_{\rm dt})$ of $\chi$CFT$_{4}$ is the double scaling limit \cite{Gurdogan:2015csr} of the $\gamma$-deformed SYM action \cite{Frolov:2005dj}
\begingroup \allowdisplaybreaks
\begin{flalign}
\label{eq1}
\mathcal{L}_{\rm free}&=N\,\textrm{tr}\left(-\partial_{\mu}\phi_{j}^{\dagger}\partial_{\mu}\phi^{j}+i\bar{\psi}_{j}^{\dot{\alpha}}\left(\tilde{\sigma}^{\mu}\right)_{\dot{\alpha}}^{\:\alpha}\partial_{\mu}\psi_{\alpha}^{j}\right)
\\
\label{eq9}
\mathcal{L}_{\rm int}&
=N\,\textrm{tr}\left(
\left(4\pi\right)^{2}\xi_{j}^{2}\phi_{j+}^{\dagger}\phi_{j-}^{\dagger}\phi^{j+}\phi^{j-}+4\pi i\sqrt{\xi_{j}\xi_{j+}}\left(\psi^{j+}\phi^{j-}\psi^{j}+\bar{\psi}_{j+}\phi_{j-}^{\dagger}\bar{\psi}_{j}\right)\right)
\end{flalign}
\endgroup
supplemented with the double-trace counter-terms \cite{Fokken:2014soa,Sieg:2016vap} valid in the planar limit
\begin{flalign}
\label{eq2}
\frac{\mathcal{L}_{\rm dt}}{\left(4\pi\right)^{2}}&
=\alpha_{1}^{2}\,\textrm{tr}\left(\phi_{j}\phi_{j}\right)\textrm{tr}\left(\phi_{j}^{\dagger}\phi_{j}^{\dagger}\right)
+\alpha_{2}^{2}\,\textrm{tr}\left(\phi_{j}\phi_{j+}^{\dagger}\right)\textrm{tr}\left(\phi_{j}^{\dagger}\phi_{j+}\right)
+\alpha_{3}^{2}\,\textrm{tr}\left(\phi_{j}\phi_{j+}\right)\textrm{tr}\left(\phi_{j}^{\dagger}\phi_{j+}^{\dagger}\right)\,.
\end{flalign}
We abbreviate $j\pm = (j \pm 1)~\textrm{mod}~3$ in the sums over $j=1,2,3$ and set the values of the double-trace couplings on the lines(s) of conformal fixed points \cite{Kazakov:2018gcy}
\begin{gather}
\label{eq26}
\alpha_1^2=\alpha_{1\star}^2=\mp i \frac{\xi^2_2-\xi^2_3}{2}+\dots\,,
\qquad\qquad
\alpha_3^2=\alpha_{3\star}^2=-\xi^2_3\,,
\\
\alpha_2^2=\alpha_{2\star}^2=-\xi^2_3+i\sqrt{\xi_1\xi_2}\sqrt{\xi_1\xi_2+2\xi_3^2}+\dots\,.
\nonumber
\end{gather}
In the conventions of \cite{Gurdogan:2015csr,Kazakov:2018gcy} the Euclidean path integral reads
\begin{gather}
\label{eq17}
\left\langle\dots\right\rangle=\int 
D\phi^j\,D{\phi}^\dagger_j\,D\psi^j_\alpha\,D\bar{\psi}_j^{\dot{\alpha}} \,(\dots)\, e^{S_{\phi\psi}}
\end{gather}
and the $SU(N)$ generators are normalised as (with $A,B=1,\dots,N^2-1$ and $i,j,k,l=1,\dots,N$)
\begin{gather}
\textrm{tr}(T^A T^B)=\delta^{AB}\,,
\qquad
(T^{A})^i_{~j}(T^{A})^k_{~l}=\delta^i_{l}\delta^k_{j}-\frac{1}{N}\delta^i_{j}\delta^k_{l}\,.
\end{gather}
The normalisation explains the factors of $\sqrt{2}$ in section \ref{secET} when compared to the effective theory in \cite{Vescovi:2021fjf,Jiang:2019xdz}. The propagators are
\begin{flalign}
\label{eq3}
\left\langle \phi_{i}^{A \dagger}\left(x\right)\phi_{j}^{B}\left(y\right)\right\rangle
&=
\frac{1}{N}\delta_{ij}\delta^{AB}I_{xy}\,,
\\
\label{eq4}
\left\langle \bar{\psi}_{i}^{\dot{\alpha}A_{1}}\left(x\right)\psi_{\alpha}^{jA_{2}}\left(y\right)\right\rangle 
&=
\frac{i}{N}\delta_{i}^{j}\delta^{AB}\left(\sigma^{\mu}\right)_{\alpha}^{\:\dot{\alpha}}\frac{\partial}{\partial x^\mu}I_{xy}
\end{flalign}
with $I_{xy}=(4\pi^2 (x-y)^2)^{-1}$, $x_{12}^\mu=x_1^\mu-x_2^\mu$, $x_{12}^2=(x_1-x_2)^2$ and
\begin{gather}
\label{eq8}
\left(\tilde{\sigma}^{\left\{\mu\right.}\right)_{\dot{\alpha}}^{\:\alpha}\left(\sigma^{\left.\nu\right\}}\right)_{\alpha}^{\:\dot{\beta}}=\left(\sigma^{\left\{\mu\right.}\right)_{\dot{\alpha}}^{\:\alpha}\left(\tilde{\sigma}^{\left.\nu\right\}}\right)_{\alpha}^{\:\dot{\beta}}=2\,\delta^{\mu\nu}\,\delta_{\dot{\alpha}}^{\dot{\beta}}\,.
\end{gather}
Most of the quantitative results in this paper are derived in the conformal bi-scalar theory, a single-coupling reduction of the $\chi$CFT$_{4}$ that descends from setting \cite{Gurdogan:2015csr, Sieg:2016vap, Grabner:2017pgm}
\begin{gather}
\label{eq87}
\xi_1=\xi_2=0\,,
\qquad
\xi_3=\xi\,,
\qquad
\alpha_1^2=\pm\frac{i}{2}\xi^{2}+ O(\xi^4)\,,
\qquad
\alpha_2^2=\alpha_3^2=-\xi^2 
\end{gather}
and dropping all counter-terms that involve fields other than $\phi_1$, $\phi_2$ and their conjugates.

\section{An exactly solvable four-point function}
\label{sec4pt}
In this appendix we consider a four-point function mentioned in section \ref{sec3pt}, which involves two determinants 
\begingroup \allowdisplaybreaks
\begin{flalign}
\label{eq30}
\mathcal{D}_1(x_1)&
=\left.\frac{1}{N}\frac{d}{da_1}\det(\phi^2+a_1 \phi^1)(x_1)\right|_{a_1=0}
=\frac{\varepsilon_{i_{1}\dots i_{N}}\varepsilon^{j_{1}\dots j_{N}}}{N!}\left[\left(\phi^1\right)_{~j_{1}}^{i_{1}}
\left(\phi^2\right)_{~j_{2}}^{i_{2}}\dots\left(\phi^2\right)_{~j_{N}}^{i_{N}}\right](x_1)\,,
\\
\label{eq35}
\mathcal{D}_2(x_2)&
=\left.\frac{1}{N}\frac{d}{da_2}\det(\phi_2^\dagger+a_2 \phi^1)(x_2)\right|_{a_2=0}
=\frac{\varepsilon_{i_{1}\dots i_{N}}\varepsilon^{j_{1}\dots j_{N}}}{N!}\left[\left(\phi^1\right)_{~j_{1}}^{i_{1}}
\left(\phi_2^\dagger\right)_{~j_{2}}^{i_{2}}\dots\left(\phi_2^\dagger\right)_{~j_{N}}^{i_{N}}\right](x_2)\,,
\end{flalign}
and a bi-local single trace of minimal length
\begin{gather}
\label{eq31}
\mathcal{O}(x_3,x_4)={\rm tr}\left(\phi^\dagger_1(x_3)\phi^\dagger_1(x_4)\right)\,.
\end{gather}
\endgroup
The two scalars in \eqref{eq31} are primary operators and, together with the working assumption below \eqref{eq112} that leads to identify the operators \eqref{eq30}-\eqref{eq35} as primaries in section \ref{sec2pt}, one should expect the correlator to admit an OPE representation. We make important remarks on this point in appendix \ref{sec4ptops}.

The simplicity of the graph content is behind the choice of \eqref{eq30}-\eqref{eq31}. The operators \eqref{eq30}-\eqref{eq35} (see section \ref{sec2pt}) and the matrix field $\phi_1^\dagger$ (e.g. see section 3 in \cite{Kazakov:2018gcy}) are protected in the planar theory, with dimensions $N$ and $1$ respectively. The R-charge assignments play a crucial role too, allowing for a number $O(N)$ of Wick contractions between the determinants and giving no other option to the two insertions $\phi^1$ than annihilating an equal number of the conjugate fields $\phi_1^\dagger$ in \eqref{eq31}. To understand this argument, we examine the lowest orders of the weak coupling expansion of the effective theory. The diagrams match those that enter the combination of two single-trace correlators \footnote{Free Wick contractions and vertex insertions involving only the determinants' scalars in $x_1$ and $x_2$ are forbidden at this stage, see below \eqref{eq29}. They are already taken into account in the coefficients $c_1$ and $c_2$ by the partial contraction \eqref{eq86}.}
\begingroup \allowdisplaybreaks
\begin{flalign}
\label{eq36}
\left\langle \mathcal{D}_1\mathcal{D}_2\mathcal{O} \right\rangle
&=
c_1 \left\langle\textrm{tr}\left(\phi^{1}\left(x_{1}\right)\phi^{1}\left(x_{2}\right)\right)\mathcal{O}\right\rangle 
+
c_2 \left\langle \textrm{tr}\left(\phi^{2}\left(x_{1}\right)\phi^{1}\left(x_{2}\right)\phi^{1}\left(x_{1}\right)\phi_{2}^\dagger\left(x_{2}\right)\right)\mathcal{O}\right\rangle
\\
&
\!\!\!{\rm with}~~~
c_1=
\sqrt{\frac{2\pi}{N^3}}e^{-N}\left(I_{x_{1}x_{2}}\right)^{N-1}
\,,
\qquad
c_2=
-\sqrt{\frac{2\pi}{N^3}}e^{-N}\left(I_{x_{1}x_{2}}\right)^{N-2}\,.
\nonumber
\end{flalign}
\endgroup
In hindsight this suggests that the determinant pair can be projected onto two single traces of length $2L=2$ and $2L=4$ via the decomposition formula \eqref{eq27}
\begin{flalign}
\label{eq86}
\left.
\mathcal{D}_1\mathcal{D}_2
\right|_{\rm partial~contraction}
&=
c_1\, \textrm{tr}\left(\phi^{1}\left(x_{1}\right)\phi^{1}\left(x_{2}\right)\right)
\\
&
+c_2\, \textrm{tr}\left(\phi^{2}\left(x_{1}\right)\phi^{1}\left(x_{2}\right)\phi^{1}\left(x_{1}\right)\phi_{2}^\dagger\left(x_{2}\right)\right)+\dots\,.
\nonumber
\end{flalign}
The multi-traces omitted in the dots, with higher length or with different matter content such as~$\textrm{tr}(\phi_{2}^{\dagger}\left(x_{2}\right)\phi^{1}\left(x_{1}\right)\phi^{1}\left(x_{2}\right)\phi^{2}\left(x_{1}\right))$, have vanishing overlap with \eqref{eq31}, so they drop out of the right-hand side of \eqref{eq36}. Some of these would contribute beyond planar limit, or also in the planar limit under examination in the presence of an operator more complicated than \eqref{eq31}. Motivated by the perturbative analysis, we focus on the correlators on the right-hand side of \eqref{eq36}. The first correlator is known \cite{Grabner:2017pgm,Gromov:2018hut} and equivalent to \eqref{eq77} due to the symmetry \eqref{eq112}, whereas the second one is calculated below in \eqref{eq80} in appendix \ref{sec4ptBS}. We test this expression at weak coupling in appendix \ref{sec4ptweak} and comment on the exchanged states in the OPE in appendix \ref{sec4ptops}.

\begin{figure}[t]
\centering
\includegraphics[scale=0.16]{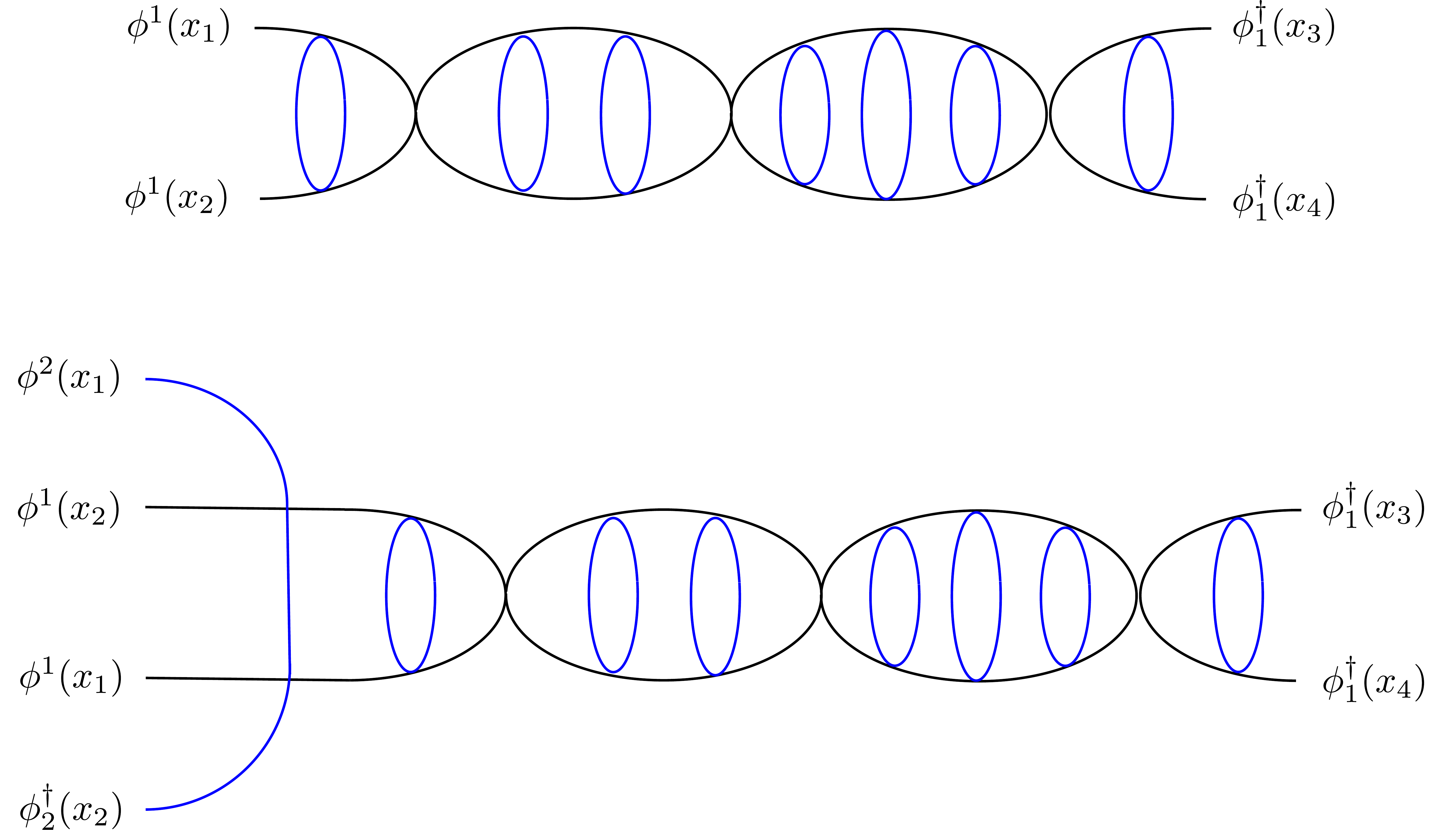}
\caption{Typical Feynman diagrams contributing to the perturbative expansions of \eqref{eq42} (top) and \eqref{eq55} (bottom). The top diagram \cite{Grabner:2017pgm,Gromov:2018hut} is a linear chain of scalar loops, built using a pair of $\xi^2$-vertices and interspersed with $\alpha_1^2$-vertices. The propagators are curved to hint at their cylindrical topology. The bottom diagram ends with two $\xi^2$-vertices. The colours represent the ``flavours'' of the scalars: black for $\phi^1$ and blue for $\phi^2$. The topologies are related to those in figure 1 of \cite{Gromov:2019bsj}.}
\label{fig1}
\end{figure}

\subsection{The Bethe-Salpeter method}
\label{sec4ptBS}
\begin{figure}[t]
\centering
\includegraphics[scale=0.22]{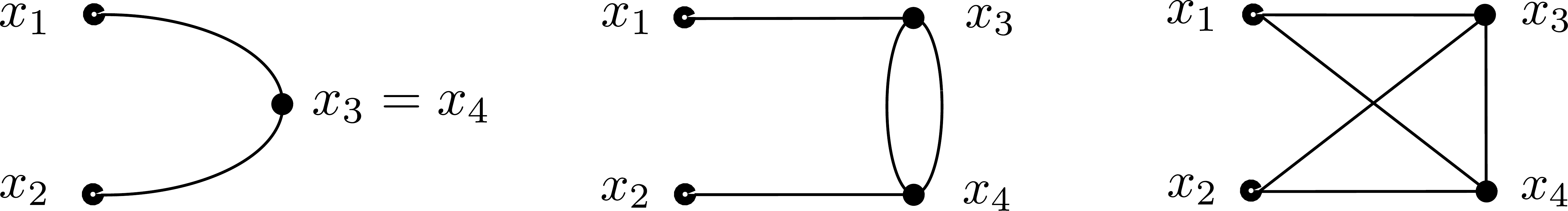}
\caption{The kernels associated to the graph-building operators $\hat{\mathcal{V}}$ (left), $\hat{\mathcal{H}}_{B}$ (center) and $\hat{\mathcal{H}}_{\rm basis}$ (right). White dots are the external points and black dots the points integrated over $\mathbb{R}^4$.
}
\label{fig2}
\end{figure}

The Feynman graphs that describe the two correlators in \eqref{eq36} display an iterative structure in position space, see figure \ref{fig1}. Each graph can be obtained from one at the previous perturbative order by the action of the integral graph-building operators in figure \ref{fig2}. This observation enables to exploit conformal symmetry and the Bethe-Salpeter method in order to obtain the exact expression of the correlator. We briefly fit the derivation \cite{Grabner:2017pgm} of the first correlator \eqref{eq77} in the scheme of this appendix, in order to emphasise the peculiar aspects of the second correlator, in the notation of \cite{Gromov:2018hut,Kazakov:2018gcy}.

\paragraph{Graph-building operators.}
The perturbative expansion of the relevant correlators in figure \ref{fig1} can be written in the form \footnote{\label{foot}The permutations account for the four ways the graphs are attached to the external points. The normalisations in \eqref{eq42} guarantee that the leading term $\hat{G}\approx\hat{\mathcal{H}}_B$ reproduces the first Feynman integral in \eqref{eq96}. Similarly, the leading order of \eqref{eq55}, obtained by convoluting \eqref{eq56} with $\hat{G}\approx\hat{\mathcal{H}}_B$, coincides with the last integral in \eqref{eq96}.
}
\begingroup \allowdisplaybreaks
\begin{flalign}
\label{eq42}
&\left\langle\textrm{tr}\left(\phi^{1}\left(x_{1}\right)\phi^{1}\left(x_{2}\right)\right)\mathcal{O}\right\rangle
=
\frac{1}{2I_{x_3x_4}^{2}}\left[\left\langle x_{1},x_{2}|\hat{G}|x_{3},x_{4}\right\rangle +\left(x_{1}\leftrightarrow x_{2}\right)\right]+\left(x_{3}\leftrightarrow x_{4}\right)\,,
\\
\label{eq55}
&\left\langle \textrm{tr}\left(\phi^{2}\left(x_{1}\right)\phi^{1}\left(x_{2}\right)\phi^{1}\left(x_{1}\right)\phi^\dagger_{2}\left(x_{2}\right)\right)\mathcal{O}\right\rangle
=
\frac{\left(4\pi\xi\right)^{4}}{2I_{x_3x_4}^{2}}
\left[\left\langle x_{1},x_{2}\right|\hat{\mathcal{H}}_{\rm basis}\,\hat{G}\left|x_{3},x_{4}\right\rangle +\left(x_{1}\leftrightarrow x_{2}\right)\right]
\nonumber
\\
&\qquad\qquad\qquad\qquad\qquad\qquad\qquad\qquad~~~~~~+\left(x_{3}\leftrightarrow x_{4}\right)\,.
\end{flalign}
\endgroup
The expressions contain the ``chain''-building operator \cite{Grabner:2017pgm}
\begin{gather}
\label{eq38}
\hat{G}
=\sum_{n=0}^\infty \left(\chi_{V}\hat{\mathcal{V}}+\chi_{B}\hat{\mathcal{H}}_{B}\right)^n\,\hat{\mathcal{H}}_{B}
=\frac{\hat{\mathcal{H}}_{B}}{1-\chi_{V}\hat{\mathcal{V}}-\chi_{B}\hat{\mathcal{H}}_{B}}\,,
\end{gather}
which is the geometric series of two commuting graph-building operators $\hat{\mathcal{V}}$ and $\hat{\mathcal{H}}_B$. A generic term in the series is depicted in figure \ref{fig1}, top panel. The operators are represented in position space by the integral kernels
\begin{gather}
\label{eq39}
\left\langle x_{1},x_{2}|\hat{\mathcal{V}}|x_{3},x_{4}\right\rangle =\frac{2\,\delta^{\left(4\right)}\left(x_{3}-x_{4}\right)}{\left(4\pi^{2}\right)^{2}x_{13}^{2}x_{23}^{2}}\,,
\qquad
\left\langle x_{1},x_{2}|\hat{\mathcal{H}}_{B}|x_{3},x_{4}\right\rangle =\frac{1}{\left(4\pi^{2}\right)^{4}x_{13}^{2}x_{24}^{2}x_{34}^{4}}\,.
\end{gather}
They act as convolutions on functions of $x_3$ and $x_4$, for example on a test function
\begin{flalign}
\label{eq114}
\left\langle x_{1},x_{2}|
\hat{\mathcal{H}}_B\,f
|x_{3},x_{4}\right\rangle
&=\int 
\frac{1}{\left(4\pi^{2}\right)^{4}}
\frac{d^4x_{3}\,d^4x_{4}\,}{x_{13}^{2}x_{24}^{2}x_{34}^{4}}
f(x_{3},x_{4})\,.
\end{flalign}
The normalisations are such that the couplings enter \eqref{eq38} via the constants ${\chi_{V}=\left(4\pi\alpha_{1}\right)^{2}}$ and $\chi_{B}=\left(4\pi \xi\right)^{4}$. To build the diagrams in figure \ref{fig1}, bottom panel, we define a new graph-generating operator, whose action is to fasten one end of the chain to the positions $x_1$ and $x_2$ of the length-4 trace in \eqref{eq55}:
\begin{gather}
\label{eq56}
\left\langle x_{1},x_{2}|\hat{\mathcal{H}}_{\rm basis}|x_{3},x_{4}\right\rangle =\frac{1}{\left(4\pi^{2}\right)^{5}x_{13}^{2}x_{14}^{2}x_{23}^{2}x_{24}^{2}x_{34}^{2}}\,.
\end{gather}

\paragraph{Spectral decomposition of $\hat{G}$.} We recollect the approach of \cite{Grabner:2017pgm} to compute \eqref{eq42} via the spectral decomposition of $\hat{G}$. The graph-building operators mutually commute and can be simultaneously diagonalised:
\begin{flalign}
\label{eq43}
\left[\hat{\mathcal{V}}\,\Phi_{\Delta,S,x_{0}}^{\Delta_{\mathcal{O}_{1}},\Delta_{\mathcal{O}_{2}}}\right]\left(x_{1},x_{2}\right)&=h_{\mathcal{V}_{\Delta,S}}^{\Delta_{\mathcal{O}_{1}},\Delta_{\mathcal{O}_{2}}}\,\Phi_{\Delta,S,x_{0}}^{\Delta_{\mathcal{O}_{1}},\Delta_{\mathcal{O}_{2}}}\,,
\\
\label{eq44}
\left[\hat{\mathcal{H}}_{B}\,\Phi_{\Delta,S,x_{0}}^{\Delta_{\mathcal{O}_{1}},\Delta_{\mathcal{O}_{2}}}\right]\left(x_{1},x_{2}\right)&=h_{B_{\Delta,S}}^{\Delta_{\mathcal{O}_{1}},\Delta_{\mathcal{O}_{2}}}\,\Phi_{\Delta,S,x_{0}}^{\Delta_{\mathcal{O}_{1}},\Delta_{\mathcal{O}_{2}}}
\,.
\end{flalign}
Since they commute with the generators of the conformal group, conformal symmetry fixes the eigenfunctions to be the conformal triangles $\Phi^{\Delta_{\mathcal{O}_1},\Delta_{\mathcal{O}_2}}_{\Delta,S,x_0}(x_{1},x_{2})$, namely the three-point functions of two scalar operators, with dimensions $\Delta_{\mathcal{O}_{1}}$ and $\Delta_{\mathcal{O}_{2}}$ and at the positions $x_1$ and $x_2$, and an operator $\mathcal{O}_{\Delta,S}(x_0)$ with dimension $\Delta$, Lorentz spin $S$ and at $x_0$:
\begin{flalign}
\label{eq65}
&\Phi^{\Delta_{\mathcal{O}_1},\Delta_{\mathcal{O}_2}}_{\Delta,S,x_0}(x_{1},x_{2})
= \langle{{\rm tr}[\mathcal{O}_1(x_1) \mathcal{O}_2(x_2)] \mathcal{O}_{\Delta,S}(x_0)}\rangle
\\
&= (x_{12}^2)^{p-\tfrac{\Delta_{\mathcal{O}_1}+\Delta_{\mathcal{O}_2}}{2}}
(x_{10}^2)^{\tfrac{\Delta_{\mathcal{O}_2}-\Delta_{\mathcal{O}_1}}{2}-p}
(x_{20}^2)^{\tfrac{\Delta_{\mathcal{O}_1}-\Delta_{\mathcal{O}_2}}{2}-p}
\left(\frac{2(n\cdot x_{02})}{x_{02}^2}-\frac{2(n\cdot x_{01})}{x_{01}^2}\right)^S\,.
\nonumber
\end{flalign}
The operator $\mathcal{O}_{\Delta,S}$ has bare twist $2p=\Delta-S$, the dimension is parametrised by $\Delta=2+2i\nu$ with $\nu\geq 0$ \cite{Dobrev:1977qv}, $S$ spans the non-negative integers and all Lorentz indices are projected onto the auxiliary null vector $n^\mu$. The eigenvalues in \eqref{eq43}-\eqref{eq44} can be calculated via the star-triangle relations \cite{DEramo:1971hnd,Vasiliev:1981yc}. Here we need the eigenbasis in the sector with ${\Delta_{\mathcal{O}_1}=\Delta_{\mathcal{O}_2}=1}$:
\begin{flalign}
\label{eq46}
h_{\mathcal{V}_{\Delta,S}}^{1,1}&=\frac{\delta\!\left(\nu\right)\delta_{S,0}}{\left(4\pi\right)^{2}}\,,
\\
\label{eq47}
h_{B_{\Delta,S}}^{1,1}&=\frac{1}{\left(2\pi\right)^{4}\left(\Delta+S\right)\left(\Delta+S-2\right)\left(\Delta-S-2\right)\left(\Delta-S-4\right)}\,.
\end{flalign}

The matrix operators in \eqref{eq42} have ${\Delta_{\mathcal{O}_1}=\Delta_{\mathcal{O}_2}=1}$, since the chain has one propagator attached to each external point. We can then write the spectral decomposition of $\hat{G}$
\begin{flalign}
\label{eq51}
&I_{x_3x_4}^{-2}\left\langle x_{1},x_{2}|\hat{G}|x_{3},x_{4}\right\rangle
\\
&=
I_{x_3x_4}^{-2}\sum_{S\in\mathbb{N}}\frac{\left(-1\right)^{S}}{x_{34}^{4}}\int_{0}^{\infty}\frac{d\nu}{c_{1}\left(\nu,S\right)}h_{\Delta,S}^{1,1}\int d^{4}x_{0}\,\Phi_{\Delta,S,x_{0}}^{1,1}\left(x_{1},x_{2}\right)\overline{\Phi_{\Delta,S,x_{0}}^{1,1}}\left(x_{3},x_{4}\right)
\nonumber
\end{flalign}
in terms of its eigenvalue $h_{\Delta,S}^{1,1}$ and the coefficient $c_{1}\left(\nu,S\right)$
\begin{gather}
\label{eq50}
h_{\Delta,S}^{1,1}=\frac{h_{B_{\Delta,S}}^{1,1}}{1-\chi_{\mathcal{V}}\,h_{\mathcal{V}_{\Delta,S}}^{1,1}-\chi_{B}\,h_{B_{\Delta,S}}^{1,1}}\,,
\qquad
c_{1}\left(\nu,S\right)=
  \frac{ 2^{S-1}\, \pi ^7}{ (S+1)\nu ^2\left(4 \nu ^2+(S+1)^2\right)}\,.
\end{gather}
We use an identity \cite{Dobrev:1977qv,Dolan:2000ut,Dolan:2011dv} to trade the $x_0$-integral for the conformal blocks \eqref{eq68} \footnote{See section 3 and appendix A of \cite{Gromov:2018hut} and section 3 of \cite{Kazakov:2018gcy}.}
\begingroup \allowdisplaybreaks
\begin{flalign}
\label{eq69}
&I_{x_3x_4}^{-2}\left\langle x_{1},x_{2}|\hat{G}|x_{3},x_{4}\right\rangle
\\
&
=I_{x_3x_4}^{-2}\sum_{S=0}^{\infty}\frac{\left(-1\right)^{S}}{x_{12}^{2}x_{34}^{6}}\int_{0}^{\infty}\frac{d\nu}{c_{1}\left(\nu,S\right)}h_{\Delta,S}^{1,1}\left[\frac{c_{1}\left(\nu,S\right)}{c_{2}\left(\nu,S\right)}g_{\Delta,S}^{0,0}\left(u,v\right)+\frac{c_{1}\left(-\nu,S\right)}{c_{2}\left(-\nu,S\right)}g_{4-\Delta,S}^{0,0}\left(u,v\right)\right]
\nonumber
\\
&{\rm with}~~
c_{2}\left(\nu,S\right)=-\frac{i\pi^{5}\left(-1\right)^{S}\Gamma^{2}\left(\frac{S}{2}-i\nu+1\right)\Gamma\left(S+2i\nu+1\right)}{\nu\left(S+1\right)\Gamma^{2}\left(\frac{S}{2}+i\nu+1\right)\Gamma\left(S-2i\nu+2\right)}\,.
\end{flalign}
\endgroup
The symmetry of the integrand under $\nu\rightarrow-\nu$ combines the two terms in \eqref{eq69} and extends the $\nu$-integral to the real axis using the property $h_{\Delta,S}^{1,1}=h_{4-\Delta,S}^{1,1}$
\begin{gather}
\label{eq54}
I_{x_3x_4}^{-2}\left\langle x_{1},x_{2}|\hat{G}|x_{3},x_{4}\right\rangle
=\left(4\pi^{2}\right)^{2}\sum_{S\in\mathbb{N}}\frac{\left(-1\right)^{S}}{x_{12}^{2}x_{34}^{2}}\int_{-\infty}^{\infty}\frac{d\nu}{c_{2}\left(\nu,S\right)}h_{\Delta,S}^{1,1}\,g^{0,0}_{\Delta,S}\left(u,v\right)\,.
\end{gather}

\paragraph{Shift relation.}
The strategy above carries over to \eqref{eq55}. First, we notice that the action of the basis-building operator on the relevant conformal triangles is to increase their dimensions by one
\begin{gather}
\label{eq57}
\left[\hat{\mathcal{H}}_{\rm basis}\Phi_{\Delta,S,x_{0}}^{1,1}\right]\left(x_{1},x_{2}\right)
=h_{\Delta,S}\,\Phi_{\Delta,S,x_{0}}^{2,2}\left(x_{1},x_{2}\right)
\end{gather}
at the cost of introducing a complex factor
\begin{flalign}
\label{eq58}
h_{\Delta,S}&=\frac{1+\left(-1\right)^{S}}{2}\frac{i\pi^{4}}{2\left(4\pi^{2}\right)^{5}\left(S+1\right)\nu}\left[\psi^{\left(1\right)}\left(\frac{\frac{S}{2}+i\nu+1}{2}\right)+\right.
\\
&\left.-\psi^{\left(1\right)}\left(\frac{\frac{S}{2}+i\nu+2}{2}\right)-\psi^{\left(1\right)}\left(\frac{\frac{S}{2}-i\nu+1}{2}\right)+\psi^{\left(1\right)}\left(\frac{\frac{S}{2}-i\nu+2}{2}\right)\right]\,.
\nonumber
\end{flalign}
This is a non-trivial function of $\Delta=2+2i\nu$ and $S$ through the derivative of the digamma function $\psi^{(1)}(z)=\frac{d^2}{dz^2}\log\Gamma(z)$ and it is zero for odd spins. In appendix \ref{secShift} we prove the shift relation and connect it with the spectrum of the two-magnon graph-building operator of \cite{Gromov:2018hut}. This observation justifies the name of pseudo-eigenvalue that we reserve for $h_{\Delta,S}$.

Second, we employ the shift relation to ``glue'' the basis to the rest of the chain \eqref{eq51}:
\begin{flalign}
\label{eq59}
&I_{x_3x_4}^{-2}\left\langle x_{1},x_{2}|\hat{\mathcal{H}}_{\rm basis}\hat{G}|x_{3},x_{4}\right\rangle = I_{x_3x_4}^{-2} \int d^{4}x_{1'}d^{4}x_{2'}\,\left\langle x_{1},x_{2}|\hat{\mathcal{H}}_{\rm basis}|x_{1'},x_{2'}\right\rangle
\\
&~~~\times \sum_{S\in2\mathbb{N}}\frac{1}{x_{34}^{4}}\int_{0}^{\infty}\frac{d\nu}{c_{1}\left(\nu,S\right)}h_{\Delta,S}\int d^{4}x_{0}\Phi_{\Delta,S,x_{0}}^{1,1}\left(x_{1'},x_{2'}\right)\overline{\Phi_{\Delta,S,x_{0}}^{1,1}}\left(x_{3},x_{4}\right)
\nonumber
\\
&=I_{x_3x_4}^{-2}\sum_{S\in2\mathbb{N}}\frac{1}{x_{12}^{2}x_{34}^{4}}\int_{0}^{\infty}\frac{d\nu}{c_{1}\left(\nu,S\right)}h_{\Delta,S}^{1,1}\,h_{\Delta,S}\int d^{4}x_{0}\Phi_{\Delta,S,x_{0}}^{1,1}\left(x_{1},x_{2}\right)\overline{\Phi_{\Delta,S,x_{0}}^{1,1}}\left(x_{3},x_{4}\right)\,.
\nonumber
\end{flalign}
Here we use \eqref{eq57} to replace the basis-building operator with the pseudo-eigenvalue and the formula \eqref{eq82} to restore a pair of $\Phi_{\Delta,S,x_{0}}^{1,1}$ as in \eqref{eq51}. The relation \eqref{eq59} shows that the ``gluing'' is equivalent to the insertion of the pseudo-eigenvalue \eqref{eq58} in the spectral decomposition of the chain-building operator. We notice that the pseudo-eigenvalue restricts the spin to even integers and that the non-conformal factor $I_{x_3x_4}^{-2}x_{12}^{-2}x_{34}^{-4}$ is affected by the power of $x_{12}^{-2}$ in \eqref{eq82} in order to reproduce the correct non-conformal factor in the result \eqref{eq80} below.

Finally the reflection symmetry of pseudo-eigenvalue \eqref{eq84} allows to extend the $\nu$-integration to the full real axis again:
\begin{gather}
\label{eq61}
I_{x_3x_4}^{-2}\left\langle x_{1},x_{2}|\hat{\mathcal{H}}_{\rm basis}\hat{G}|x_{3},x_{4}\right\rangle
=
\left(4\pi^{2}\right)^{2}\sum_{S\in2\mathbb{N}}\frac{1}{x_{12}^{4}x_{34}^{2}}\int_{-\infty}^{\infty}\frac{d\nu}{c_{2}\left(\nu,S\right)}h_{\Delta,S}^{1,1}\,h_{\Delta,S}\,g_{\Delta,S}^{0,0}\left(u,v\right)\,.
\end{gather}

\paragraph{Conformal partial wave expansion.}

The next step is to cast \eqref{eq54} into the OPE form of a four-point function. In the $s$-channel ($x_{12}\to 0$, namely $u\to 0$ and $v\to 1$), we have ${g^{0,0}_{2+2i\nu,S}(u,v)\sim u^{1+i\nu -S/2} (1-v)^S}$, so the integrand decays exponentially for $\Re (i\nu)\to \infty$. We close integration contour in the lower half-plane and solve \eqref{eq54} by the residue theorem. One neglects the Dirac delta in \eqref{eq50}, which comes from the double-trace operator \eqref{eq46}, and later takes it into account by a correct treatment of the singularity in the final result for $S=0$ when $\xi\to 0$ \cite{Grabner:2017pgm}.

The three factors in the integrand of \eqref{eq54} have simples poles. The measure is singular at $2i\nu=S+k+1$ (with $S,k=0,1,\dots$) with residues
\begin{gather}
\label{eq73}
r^{\rm (m)}_{k,S}=-\frac{\left(-\right)^{k}ik\Gamma^{2}\left(\frac{k+1}{2}\right)}{2\Gamma^{2}\left(k+1\right)\Gamma^{2}\left(\frac{-k+1}{2}\right)}\frac{1}{c_{2}\left(\frac{S+1}{2i},S+k\right)}\,.
\end{gather}
The eigenvalue has poles at $2i\nu=\Delta_{\pm}-2$, each labeled by $S=0,1,\dots$ and $\pm$, with
\begingroup \allowdisplaybreaks
\begin{flalign}
\label{eq74}
\Delta_{\pm}&=2+\sqrt{\left(S+1\right)^{2}+1\pm2\sqrt{\left(S+1\right)^{2}+4\xi^{4}}}\,,
\\
r^{\rm (e)}_{S,\pm}
&
=\frac{1}{128\,\pi^4\left(\Delta_\pm-2\right)\left[\left(4-\Delta_\pm\right)\Delta_\pm+S\left(S+2\right)-2\right]}
\,.
\nonumber
\end{flalign}
\endgroup
The conformal block is singular at $2i\nu=S-k$ (with $S=1,2,\dots$ and $k=0,\dots, S-1$) with residues
\begin{gather}
\label{eq75}
r^{\rm (c)}_{k,S}=-\frac{\left(-\right)^{k}i\left(k+1\right)\Gamma^{2}\left(\frac{k}{2}+1\right)}{2\Gamma^{2}\left(k+2\right)\Gamma^{2}\left(-\frac{k}{2}\right)}g_{S+3,S-k-1}^{0,0}\left(u,v\right)\,.
\end{gather}
One can prove that the so-called ``spurious'' poles cancel out
\begin{gather}
\label{eq76}
\sum_{S=0}^{\infty} \sum_{k=0}^{\infty} \left(-1\right)^{S}r^{\rm (m)}_{k,S}\,g_{S+k+3,S}
+
\sum_{S=1}^{\infty} \sum_{k=0}^{S-1} \left(-1\right)^{S}r^{\rm (c)}_{k,S}\, \frac{1}{c_2\left(\frac{S-k}{2},S\right)}
=
0\,,
\end{gather}
due to the relation $h_{S+3+k,S}^{1,1}=h_{S+3,S+k}^{1,1}$ (with $S,k=0,1,\dots$), therefore \eqref{eq54} is determined by the ``physical'' poles of $h^{1,1}_{\Delta,S}$. The proper definition of \eqref{eq42} takes into account a symmetrisation. Considering the definition of the cross-ratios \eqref{eq67} and the property ${g^{0,0}_{\Delta,S}(u/v,1/u)=(-1)^S g^{0,0}_{\Delta,S}(u,v)}$ under the exchange $x_3\leftrightarrow x_4$,  the terms with odd $S$ cancel out and the contribution of those with even $S$ gets quadrupled. The final result reads as in \eqref{eq77}. 

The same analysis carries over to \eqref{eq61} with important differences. The scaling of $h_{2+2i\nu,S}\sim\nu^{-4}$ does not alter the exponential decay. The residue analysis of the eigenvalue factor \eqref{eq74} and the conformal block \eqref{eq75} is unaffected, whereas half of the simple poles of the measure overlap with those of $h_{\Delta,S}$. The combination $h_{\Delta,S}/c_{2}\left(\nu,S\right)$ is singular at $2i\nu=S+k+1$ (with $S,k=0,1\dots $): for even $k$ the measure develops simple poles with residue
\begingroup \allowdisplaybreaks
\begin{gather}
\label{eq78}
r^{\rm (m,pe)}_{k,S}=-\frac{(-)^k i k\Gamma^{2}\left(\frac{k+1}{2}\right)}{2\Gamma^{2}\left(k+1\right)\Gamma^{2}\left(\frac{-k+1}{2}\right)}\frac{h_{S+k+3,S}}{c_{2}\left(\frac{S+1}{2i},S+k\right)}\,,
\end{gather}
\endgroup
whereas for odd $k$ both the measure and the pseudo-eigenvalue are singular with residue
\begin{gather}
\label{eq79}
r^{\rm (m,pe)}_{k,S}=\frac{\left(-\right)^{\frac{k-1}{2}}i\,\Gamma\left(\frac{k+1}{2}\right)\Gamma\left(S+\frac{k+3}{2}\right)}{\left(4\pi^{2}\right)^{5}4^{S+k}\Gamma\left(\frac{k}{2}\right)\Gamma\left(S+\frac{k}{2}+1\right)}\,.
\end{gather}
The factors $h_{\Delta,S}^{1,1}$ and $h_{\Delta,S}$ carry the dynamical data of the chain and we can call ``physical'' the poles of their product. The cancellation of the spurious poles of the measure \eqref{eq78} and of the conformal blocks \eqref{eq75} with $k$ odd is not spoiled:
\begin{gather}
\sum_{S=0,2,\dots} \sum_{k=0,2,\dots}r^{\rm (m,pe)}_{k,S} g_{S+k+3,S}
+
\sum_{S=2,4,\dots} \sum_{k=1,3,\dots,S-1} r^{\rm (c)}_{k,S}\, \frac{h_{S-k+2,S}}{c_{2}\left(\frac{S-k}{2i},S\right)}
=
0\,.
\end{gather}
The cancellation takes note of the second formula in \eqref{eq84}. Likewise the residues of the full integrand at the poles of the measure \eqref{eq75} with $k$ even vanish. The correlator \eqref{eq55} is thus determined by the sum of residues at the physical poles \eqref{eq74} and \eqref{eq79}. Similarly to the first correlator, due to the symmetry under the exchange of $x_3$ and $x_4$, the residues get multiplied by a factor of $4$. We can finally write \eqref{eq55} in the OPE form
\begingroup \allowdisplaybreaks
\begin{flalign}
\label{eq80}
& \left\langle \textrm{tr}\left(\phi^{2}\left(x_{1}\right)\phi^{1}\left(x_{2}\right)\phi^{1}\left(x_{1}\right)\phi_{2}^{\dagger}\left(x_{2}\right)\right)\mathcal{O}\right\rangle 
=
\frac{1}{x_{12}^{4}x_{34}^{2}}\sum_{S=0,2,\dots}\sum_{k=1,3,\dots}\frac{\left(-\right)^{\frac{k-1}{2}}\xi^4}{4^{k+S}\pi^{5}}
\\
&
\times\frac{1}{\left(k^{2}-1\right)\left(2S+k+1\right)\left(2S+k+3\right)-16\xi^{4}}\frac{\Gamma\left(\frac{k+1}{2}\right)\Gamma\left(S+\frac{k+3}{2}\right)}{\Gamma\left(\frac{k}{2}\right)\Gamma\left(S+\frac{k}{2}+1\right)}
\,g_{S+k+3,S}^{0,0}\left(u,v\right)
\nonumber
\\
&
+\frac{1}{x_{12}^{4}x_{34}^{2}}\sum_{S=0,2,\dots}\sum_{\Delta=\Delta_{-},\Delta_{+}}\frac{-\,\xi^{4}}{16\pi^{6}}
\frac{\Gamma^{2}\left(\frac{S+\Delta}{2}\right)\Gamma\left(S-\Delta+4\right)}{\left(\Delta-2\right)\left[\left(4-\Delta\right)\Delta+S\left(S+2\right)-2\right]\Gamma^{2}\left(\frac{S-\Delta}{2}+2\right)\Gamma\left(S+\Delta-1\right)}
\nonumber
\\
&
\times\left[\psi^{\left(1\right)}\left(\frac{S+\Delta}{4}\right)-\psi^{\left(1\right)}\left(\frac{S+\Delta+2}{4}\right)-\psi^{\left(1\right)}\left(\frac{S-\Delta+4}{4}\right)+\psi^{\left(1\right)}\left(\frac{S-\Delta+6}{2}\right)\right]g_{\Delta,S}^{0,0}\left(u,v\right)
\,.
\nonumber
\end{flalign}
\endgroup

\subsection{Diagrams at weak coupling from Feynman diagrams}
\label{sec4ptweak}

We prove that the double sums in \eqref{eq80} in the weak-coupling limit agree with the Feynman-diagram expansion of the four point-function.

The expansion of \eqref{eq80} contains the powers $\xi^{2n}$ with $n=2,3,\dots$ in agreement with perturbation theory (see \eqref{eq87} and figure \ref{fig1}, bottom panel). Since some poles $\Delta_{+}\sim S+4$ and $\Delta_{-}\sim S+2$ overlap near $\xi=0$, it is safer to extract the order $\xi^4$ and check the cancellation of the lower powers in different ways. In the first approach we expand the integrand \eqref{eq61} and then calculate the residues. Going through the calculation we come to a rather cumbersome sum of residues
\begingroup \allowdisplaybreaks
\begin{flalign}
\label{eq107}
& \left\langle \textrm{tr}\left(\phi^{2}\left(x_{1}\right)\phi^{1}\left(x_{2}\right)\phi^{1}\left(x_{1}\right)\phi_{2}^{\dagger}\left(x_{2}\right)\right)\mathcal{O}\right\rangle 
\\
=& \frac{\xi^{4}}{x_{12}^{4}x_{34}^{2}}\left\{ \sum_{S=2,4,\dots}\sum_{k=1}^{S-1}\frac{\left(-\right)^{k+1}2^{13}\pi^{9}\left(k+1\right)\Gamma^{2}\left(\frac{k}{2}+1\right)}{\Gamma^{2}\left(-\frac{k}{2}\right)\Gamma^{2}\left(k+2\right)}\frac{h_{S-k+2,S}\,h_{S-k+2,S}^{1,1}}{c_{2}\left(\frac{S-k}{2i},S\right)}\,g_{S+3,S-k-1}^{0,0}\left(u,v\right)\right.
\nonumber
\\
&
+\sum_{S=0,2,\dots}\sum_{k=0,2,\dots}-\frac{2^{13}\pi^{9}k\Gamma^{2}\left(\frac{k+1}{2}\right)}{\Gamma^{2}\left(\frac{1-k}{2}\right)\Gamma^{2}\left(k+1\right)}\frac{h_{S+k+3,S}\,h_{S+k+3,S}^{1,1}}{c_{2}\left(\frac{S+1}{2i},S+k\right)}\,g_{S+k+3,S}^{0,0}\left(u,v\right)
\nonumber
\\
&
+\sum_{S=0,2,\dots}\sum_{k=3,5,\dots}\frac{\left(-\right)^{\frac{k-1}{2}}\Gamma\left(\frac{k+1}{2}\right)\Gamma\left(S+\frac{k+3}{2}\right)}{4^{S+k-2}\pi\Gamma\left(\frac{k}{2}\right)\Gamma\left(S+\frac{k}{2}+1\right)}h_{S+k+3,S}^{1,1}\,g_{S+k+3,S}^{0,0}\left(u,v\right)
\nonumber
\\
&
+\sum_{S=0,2,\dots}\frac{64\pi^{5}}{S\left(S+1\right)}\frac{h_{S+2,S}}{c_{2}\left(\frac{S}{2i},S\right)}g_{S+2,S}^{0,0}\left(u,v\right)+\sum_{S=0,2,\dots}\frac{-\Gamma^{2}\left(S+1\right)}{16\pi^{6}\left(S+2\right)^{2}\Gamma\left(2S+3\right)}
\nonumber
\\
&
\times\left[\left(S^{2}+5S+5+2\left(S+1\right)\left(S+2\right)\left(H_{2S+2}-H_{S+1}\right)\right)g_{S+4,S}^{0,0}\left(u,v\right)\right.
\nonumber
\\
&
\left.\left.-2\left(S+1\right)\left(S+2\right)\left.\frac{\partial}{\partial{\Delta}}g_{\Delta,S}^{0,0}\left(u,v\right)\right|_{\Delta=S+4}\right]\right\}_{\xi=0}+O\left(\xi^{6}\right)\,.
\nonumber
\end{flalign}
\endgroup
The first and second sums cancel and ensure the cancellation of the spurious poles. The exception is the pole at $2i\nu=S+2$, which has to be treated separately and remains in the last double sum. Alternatively, we expand directly \eqref{eq80} and verify that it matches the last three sums in \eqref{eq107}. In particular, we notice that the order $\xi^0$ in the summands with $k=1$ and with $\Delta=\Delta_+$ cancel out in \eqref{eq80}.

\begin{figure}[t]
\centering
\includegraphics[scale=0.25]{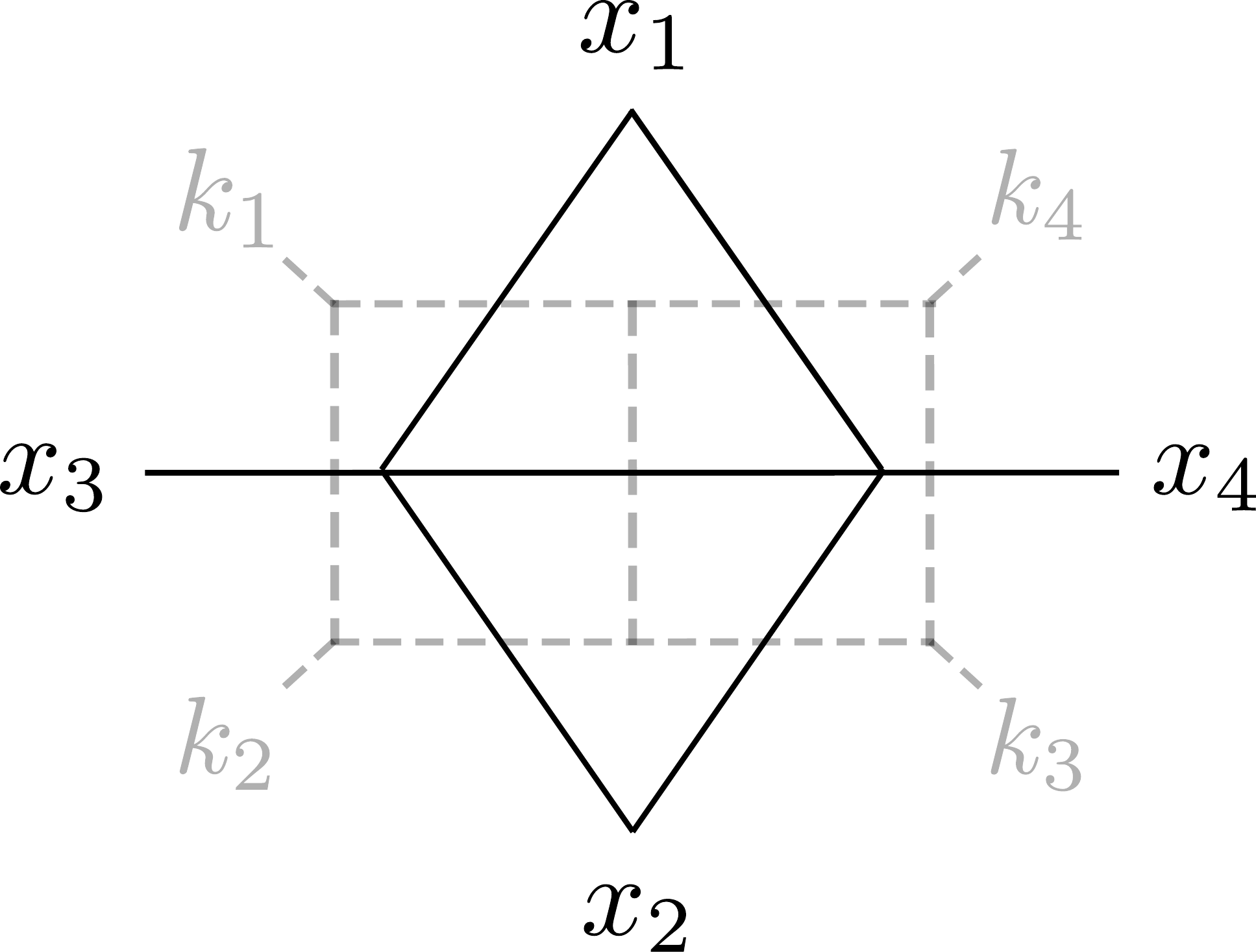}
\caption{Leading contribution to the four-point functions \eqref{eq55} in position-space variables (solid) and dualised momenta (dotted).
}
\label{fig3}
\end{figure}

The Feynman expansion starts at order $\xi^4$ with the diagram written in the last term of \eqref{eq96} and shown in figure \ref{fig3}:
\begin{flalign}
\label{eq71}
&\left\langle \textrm{tr}\left(\phi^{2}\left(x_{1}\right)\phi^{1}\left(x_{2}\right)\phi^{1}\left(x_{1}\right)\phi^\dagger_{2}\left(x_{2}\right)\right)\mathcal{O}\right\rangle
=
2\,\frac{(4\pi\xi)^4}{(4\pi^2)^7} \,\, \mathcal{I}(x_1,x_2,x_3,x_4)+\dots
\\
&\text{with} \quad
\mathcal{I}(x_1,x_2,x_3,x_4)=\int \frac{d^4 x_{1'}d^4 x_{2'}}{x_{11'}^2 x_{12'}^2 x_{21'}^2 x_{22'}^2 x_{1'2'}^2
 x_{31'}^2 x_{42'}^2}\,.
\nonumber
\end{flalign}
The integral is easily computed as
\begin{gather}
\mathcal{I}(x_1,x_2,x_3,x_4)=\!
\int\frac{d^{4}k\,d^{4}r}{\left(k+k_{1}\right)^{2}\left(k_{4}-r\right)^{2}\left(k-k_{2}\right)^{2}\left(k_{3}+r\right)^{2}\left(k_{1}+k_{4}+k-r\right)^{2}k^{2}r^{2}}
\end{gather}
by passing to the dual momentum space via the change of variables
\begin{gather}
k_{1}=x_{13}\,,\qquad k_{2}=x_{32}\,,\qquad k_{3}=x_{24}
\,,
\qquad
k_{4}=x_{41}\,,\qquad k=x_{31'}\,,\qquad r=x_{42'}\,.
\nonumber
\end{gather}
The external momenta obey the momentum conservation $\sum_{i=1}^4 k_i=0$, so the dual integral is the master double-box integral $D^{\left(2\right)}$ of \cite{Usyukina:1992jd}, whose expression was linked to the two-loop ladder integral $C^{\left(2\right)}$ therein:
\begingroup \allowdisplaybreaks
\begin{flalign}
\label{eq72}
&\mathcal{I}(x_1,x_2,x_3,x_4)
=
-D^{\left(2\right)}\!\left(k_{1}^{2},k_{2}^{2},k_{3}^{2},k_{4}^{2},s,t\right)
=
-t\,C^{\left(2\right)}\!\left(k_{1}^{2}k_{3}^{2},k_{2}^{2}k_{4}^{2},s\,t\right)
\\
&=
-t\left(\frac{\pi^{2}}{s\,t}\right)^{2}\int_{0}^{1}\frac{d\xi}{y\xi^{2}+\left(1-x-y\right)\xi+x}\left(\log^{3}\xi+\frac{3}{2}\log\frac{y}{x}\log^{2}\xi+\frac{1}{2}\log^{2}\frac{y}{x}\log\xi\right)
\nonumber
\end{flalign}
\endgroup
with the definitions
\begin{gather}
s=\left(k_{1}+k_{2}\right)^{2}\,,
\qquad
t=\left(k_{2}+k_{3}\right)^{2}\,,
\qquad
x=\frac{k_{1}^{2}\,k_{3}^{2}}{s\,t}\,,
\qquad
y=\frac{k_{2}^{2}\,k_{4}^{2}}{s\,t}\,.
\nonumber
\end{gather}
The integration delivers a combination of polylogarithms ${\rm Li}_{n}$, written here in a form numerically equivalent to that provided in \cite{Usyukina:1992jd}:
\begin{flalign}
&\mathcal{I}(x_1,x_2,x_3,x_4)
=
\frac{\pi^{4}z\bar{z}}{2x_{12}^{4}x_{34}^{2}\left(z-\bar{z}\right)}\left\{ \log^{2}\left[\left(1-z\right)\left(1-\bar{z}\right)\right]\left(\textrm{Li}_{2}\left(1-\bar{z}\right)-\textrm{Li}_{2}\left(1-z\right)\right)\right.
\\
&
\left.+6\log\left[\left(1-z\right)\left(1-\bar{z}\right)\right]\left(\textrm{Li}_{3}\left(1-z\right)-\textrm{Li}_{3}\left(1-\bar{z}\right)\right)+12\left(\textrm{Li}_{4}\left(1-\bar{z}\right)-\textrm{Li}_{4}\left(1-z\right)\right)\right\}\,.
\nonumber
\end{flalign}
\\
Once we plug this into \eqref{eq71}, it is easy to verify that this expression agrees numerically with \eqref{eq107}, provided that a sufficiently high number of terms is kept in the infinite sums, for arbitrary values of the cross-ratios. We emphasise that the check tests only the leading order, in particular it does not probe the double-trace vertices. It would be desirable to make a prediction at the next-to-leading order $\xi^6$ and verify the need of counter-terms for restoring the finiteness of the Feynman expansion at order $\xi^4\alpha_1^2$, in the spirit of what is achieved in \cite{Gromov:2018hut,Kazakov:2018gcy}.

We also find that it is possible to expand \eqref{eq71} over a special class of iterated integrals called harmonic polylogarithms (HPLs) \cite{Remiddi:1999ew,Maitre:2005uu,HPL} of weight up to 4:
\begingroup \allowdisplaybreaks
\begin{flalign}
&\mathcal{I}(x_{1},x_{2},x_{3},x_{4})
=\frac{\pi^{4}z\bar{z}}{2x_{12}^{4}x_{34}^{2}\left(z-\bar{z}\right)}\left\{
-2H_{1,0,1,1}+2\bar{H}_{1,0,1,1}+2H_{1,1,0,1}-2\bar{H}_{1,1,0,1}
\right.
\\
&\left.
-2H_{1,0}\bar{H}_{1,1}+2\bar{H}_{1,0}H_{1,1}
+2H_{1}\left(\bar{H}_{1,0,1}-\bar{H}_{1,1,0}+6\zeta_{3}\right)-2\bar{H}_{1}\left(H_{1,0,1}-H_{1,1,0}+6\zeta_{3}\right)
\right\} \,,
\nonumber
\end{flalign}
\endgroup
where $\zeta_n$ is the Riemann zeta function and we use the abbreviation $H_{a_1,a_2,\dots}=H_{a_1,a_2,\dots}(z)$ and $\bar{H}_{a_1,a_2,\dots}=H_{a_1,a_2,\dots}(\bar{z})$. A similar property was proved in \cite{Gromov:2018hut} for the first correlator in \eqref{eq36}, whose $\ell$-th perturbative order (with~$\ell\geq 1$) can be expanded as
\begin{gather}
\sum_{a_1,a_2,\dots,a_n=0,1}
\sum_{b_1,b_2,\dots,b_m=0,1} 
C_{a_1,a_2,\dots,a_n, b_1,b_2,\dots,b_m}
H_{a_1,a_2,\dots, a_n}(z)\, H_{b_1,b_2,\dots, b_m}(\bar{z})\,,
\end{gather}
where each term has weight $n+m\leq\ell+1$.

\subsection{Comment on the spectrum of exchanged operators}
\label{sec4ptops}
In the previous paragraphs we show that the four-point function of \eqref{eq30}-\eqref{eq31} receives two contributions in \eqref{eq36}, both of which are expressible in the form of an $s$-channel expansion in \eqref{eq77} and \eqref{eq80} and in agreement with conformal symmetry. The spectrum of exchanged operators contains the twist-2 and twist-4 states below \eqref{eq106} and with dimensions \eqref{eq113} for every even spin, as expected from the OPE content of $\tr(\phi^\dagger_1(x_3)\phi^\dagger_1(x_4))$ for $x_3\to x_4$ \cite{Grabner:2017pgm,Gromov:2018hut}. Nevertheless, our calculation acknowledges the presence of an infinite tower of protected states with dimension $\Delta=S+k+3$ (with $k=1,3,\dots$) and even spin $S$. The existence of these states traces back to the coupling-independent poles of the pseudo-eigenvalue (see \eqref{eq79}) and thus to the ``gluing'' of only one basis graph-building operator $\hat{\mathcal{H}}_{\rm basis}$, as opposed to a geometric series of them \footnote{A similar remark was made in the context of the SYK model, see section 2 in \cite{Gross:2017aos}.}, to the chain graph in \eqref{eq59}.

We attempt to understand this phenomenon from a physical standpoint. The basis represents the leading contribution to the OPE states in \eqref{eq80} (see footnote \ref{foot}). As the loop corrections $\hat{\mathcal{H}}_B$ are added to the chain graph, the persistence of the basis causes some states existing at Born level to remain in the OPE and without developing quantum corrections. For twists $6,8,\dots$ (or $k=3,5,\dots$) the possible candidates are the protected primaries $\textrm{tr}(\phi^1 \square^{n} \phi^1)$ with $n=(k+1)/2\geq2$ of \cite{Gromov:2019bsj}.
For twist 4 (or $k=1$) the situation may be complicated by the fact that the (Born level) operators, like $\textrm{tr}(\phi^1 \square \phi^1)$, mix with each other.
In general it should be possible to list several $S$-tensor operators with the correct set of quantum numbers and solve the operatorial mixing \footnote{See section 7.2 in \cite{Gromov:2017cja} and appendix F in \cite{Kazakov:2018gcy} for examples.}. A solid interpretation of the protected states would greatly clarify the completeness of the form of the four-point function presented in this appendix.

A better understanding would benefit from concentrating on the technical steps of the derivation, for example in connection with the caveat below \eqref{eq31} and subtleties in the pole analysis \footnote{See section 4 in \cite{Gross:2017aos} for an occurrence in SYK.}. Further indications can come from weak/strong coupling. In the former regime, it would strengthen the perturbative test mentioned at end of section \ref{sec4ptweak}. In the latter regime, in analogy with \cite{Gromov:2018hut} the four-point function should exhibit the scaling behaviour in agreement with a semi-classical description of branes. Once these questions are settled on a firm ground, it would be interesting to look at the OPE content of the cross-channels and consider external operators with spin.

\section{Proof of the shift relation}
\label{secShift}

We give the technical details of the derivation of \eqref{eq57}. In the left-hand side of the equation we plug \eqref{eq56} and \eqref{eq65} with $\Delta_{\mathcal{O}_1}=\Delta_{\mathcal{O}_2}=1$:
\begin{gather}
\label{eq62}
\frac{1}{\left(4\pi^{2}\right)^{5}}\int\frac{d^{4}x_{1'}d^{4}x_{2'}\:\left(\frac{2n\cdot x_{02'}}{x_{02'}^{2}}-\frac{2n\cdot x_{01'}}{x_{01'}^{2}}\right)^{S}}{x_{11'}^{2}x_{12'}^{2}x_{21'}^{2}x_{22'}^{2}\left(x_{1'2'}^{2}\right)^{2-p}\left(x_{1'0}^{2}x_{2'0}^{2}\right)^{p}}\,.
\end{gather}
The integrand simplifies by making an inversion around $x_0$ (with $a,b=1,1',2,2'$)
\begin{gather}
x_{0a}^\mu =\frac{x_{0\bar a}^\mu}{x_{0\bar a}^2}\,,\qquad x_{0a}=\frac{1}{x_{0\bar a}}\,,\qquad x_{ab}=\frac{x_{\bar a\bar b}}{x_{0\bar a}x_{0\bar b}}\,,\qquad d^4 x_a = \frac{d^4 x_{\bar a}}{x_{0\bar a}^{8}}\,.
\end{gather}
This results in the integral
\begin{gather}
\label{eq63}
\frac{x_{\bar{1}0}^{4}x_{\bar{2}0}^{4}}{\left(4\pi^{2}\right)^{5}}\int\frac{d^{4}x_{\bar{1}'}\,d^{4}x_{\bar{2'}}\left(2n\cdot x_{\bar{1'}\bar{2'}}\right)^{S}}{x_{\bar{1}\bar{1'}}^{2}x_{\bar{1}\bar{2'}}^{2}x_{\bar{2}\bar{1'}}^{2}x_{\bar{2}\bar{2'}}^{2}\left(x_{\bar{1'}\bar{2'}}^{2}\right)^{2-p}}\,.
\end{gather}
We rewrite it in the dual coordinates $p^{\mu}=x_{\bar{1}\bar{2}}^{\mu}$, $k_{1}^{\mu}=x_{\bar{1'}\bar{1}}^{\mu}$ and $k_{2}^{\mu}=x_{\bar{2'}\bar{1}}^{\mu}$:
\begin{gather}
\label{eq66}
\frac{x_{\bar{1}0}^{4}x_{\bar{2}0}^{4}}{\left(4\pi^{2}\right)^{5}}\int\frac{d^{4}k_{1}\,d^{4}k_{2}\left[2n\cdot\left(k_{1}-k_{2}\right)\right]^{S}}{k_{1}^{2}k_{2}^{2}\left(p+k_{1}\right)^{2}\left(p+k_{2}\right)^{2}\left[\left(k_{1}-k_{2}\right)^{2}\right]^{2-p}}\,.\end{gather}
This is proportional to the two-loop Feynman integral (C.15) of \cite{Gromov:2018hut}. In this formula, the points $x_0$, $x_1$ and $x_2$ are chosen such that $p^{2}=n\cdot p=1$. If we relax this assumption, we can use dimensional analysis to reintroduce the correct powers of
\begin{gather}
p^{2}=\frac{x_{12}^{2}}{x_{10}^{2}x_{20}^{2}}
\,,
\qquad
n\cdot p=\left(\frac{n\cdot x_{02}}{x_{02}^{2}}-\frac{n\cdot x_{01}}{x_{01}^{2}}\right)^{S}
\end{gather}
and conclude that \eqref{eq66} equals
\begin{gather}
\label{eq64}
\frac{x_{10}^{4}x_{20}^{4}}{\left(4\pi^{2}\right)^{5}}\frac{\left(2n\cdot p\right)^{S}}{\left(p^{2}\right)^{2-p}}\pi^{4}I\left(\nu,S\right)\,,
\end{gather}
where the integral (C.26) of \cite{Gromov:2018hut} evaluates to
\begin{gather}
I\left(\nu,S\right)=\frac{1+\left(-1\right)^{S}}{2}i\left[\frac{\psi^{(1)}\left(\frac{\frac{S}{2}+i\nu+1}{2}\right)-\psi^{(1)}\left(\frac{\frac{S}{2}+i\nu+2}{2}\right)}{2\left(S+1\right)\nu}+\left(\nu\to-\nu\right)\right]\,.
\end{gather}
The result vanishes for odd spin because the integrand of \eqref{eq62} acquires the factor $(-1)^S$ under the exchange of the integration points. It is easy to recognise that \eqref{eq64} is equal to the right-hand side of \eqref{eq57} once we recall the definition \eqref{eq65} with $\Delta_{\mathcal{O}_1}=\Delta_{\mathcal{O}_2}=2$.

The connection between this derivation and the integrals worked out in \cite{Gromov:2018hut} is not incidental: the shift relation \eqref{eq57} is a rewriting of the eigenvalue equation \footnote{The eigenvalue equation (3.4) of \cite{Gromov:2018hut} is compatible with our definition \eqref{eq114} due to the symmetries of the kernel \eqref{eq85}.}
\begin{gather}
\left[\hat{H}_{\color{brown}\mathbbm{2}}\Phi_{\Delta,S,x_{0}}^{2,2}\right]\left(x_{1},x_{2}\right)
=E_{\color{brown}\mathbbm{2}}(\Delta,S)\,\Phi_{\Delta,S,x_{0}}^{2,2}\left(x_{1},x_{2}\right)
\end{gather}
of the two-magnon operator (see section 4.3 therein)
\begin{gather}
\label{eq85}
\left\langle x_{1},x_{2}|\hat{H}_{\color{brown}\mathbbm{2}}|x_{3},x_{4}\right\rangle =\frac{1}{\left(4\pi^{2}\right)^{4}x_{13}^{2}x_{14}^{2}x_{23}^{2}x_{24}^{2}}\,.
\end{gather}
The equivalence becomes transparent once we take note of the proportionality relations
\begin{gather}
\label{eq82}
\left\langle x_{1},x_{2}|\hat{H}_{\color{brown}\mathbbm{2}}|x_{3},x_{4}\right\rangle
=
4\pi^2 x_{34}^2 \left\langle x_{1},x_{2}|\hat{\mathcal{H}}_{\rm basis}|x_{3},x_{4}\right\rangle
\,,
\\
\Phi_{\Delta,S,x_{0}}^{1,1}\left(x_{1},x_{2}\right)
=
x_{12}^2\,\Phi_{\Delta,S,x_{0}}^{2,2}\left(x_{1},x_{2}\right)
\nonumber
\end{gather}
and consistently identify
\begin{gather}
\label{eq83}
E_{\color{brown}\mathbbm{2}}(\Delta,S)=4\pi^2 h_{\Delta,S}\,.
\end{gather}
The expression of the pseudo-eigenvalue \eqref{eq58} follows trivially from that of the eigenvalue $E_{\color{brown}\mathbbm{2}}(\Delta,S)$ given in (4.47) of \cite{Gromov:2018hut}. Further corollaries of \eqref{eq83} are the symmetry properties
\begin{gather}
\label{eq84}
h_{\Delta,S}=h_{4-\Delta,S}\,,
\qquad
\qquad
h_{S+3+k,S}=h_{S+3,S+k}
~~~{\rm with}~~~S,k=0,2,\dots\,.
\end{gather}

\acknowledgments
We thank Andrea Cavagli\`a, Nikolay Gromov and Brett Oertel for participation at the initial stage of this project.
We are grateful to them and Konstantin Zarembo for valuable discussions and comments on the draft. We thank Michelangelo Preti for countless discussions on fishnets and help on the Mathematica code. We also thank Amit Sever for valuable discussions. The work of OS is supported by the 2020 Undergraduate Research Opportunities Programme bursary of the Theoretical Physics Group at Imperial College London and the EPSRC Mathematical Sciences Doctoral Training Partnership 2021-22, grant number EP/W524025/1. The work of EV is supported by the European Union's Horizon 2020 research and innovation programme under the Marie Sklodowska-Curie grant agreement No 895958. Nordita is supported in part by NordForsk.

\bibliographystyle{nb}
\bibliography{LetterRef}

\begin{thebibliography}{10}
\ifx\href\asklfhas\newcommand{\href}[2]{#2}\fi
\ifx\arxivref\asklfhas\newcommand{\arxivref}[2]{\href{http://arxiv.org/abs/#1}{#2}}\fi
\ifx\doiref\asklfhas\newcommand{\doiref}[2]{\href{http://dx.doi.org/#1}{#2}}\fi
\raggedright
\small
\parskip 0pt

\bibitem{Gurdogan:2015csr}
O.~G\"urdo\u{g}an and V.~Kazakov,
\textit{``{New Integrable 4D Quantum Field Theories from Strongly Deformed
  Planar $\mathcal N = $ 4 Supersymmetric Yang-Mills Theory}''},
\textsf{\doiref{10.1103/PhysRevLett.117.201602}{Phys.~Rev.~Lett.~117,~201602~(2016)}},
\texttt{\arxivref{1512.06704}{arxiv:1512.06704}},
[Addendum: Phys.Rev.Lett. 117, 259903 (2016)].

\bibitem{Frolov:2005dj}
S.~Frolov,
\textit{``{Lax pair for strings in Lunin-Maldacena background}''},
\textsf{\doiref{10.1088/1126-6708/2005/05/069}{JHEP~0505,~069~(2005)}},
\texttt{\arxivref{hep-th/0503201}{hep-th/0503201}}.

\bibitem{Caetano:2016ydc}
J.~a.~Caetano, O.~G\"urdo\u{g}an and V.~Kazakov,
\textit{``{Chiral limit of $ \mathcal{N} $ = 4 SYM and ABJM and integrable
  Feynman graphs}''},
\textsf{\doiref{10.1007/JHEP03(2018)077}{JHEP~1803,~077~(2018)}},
\texttt{\arxivref{1612.05895}{arxiv:1612.05895}}.

\bibitem{Jiang:2019xdz}
Y.~Jiang, S.~Komatsu and E.~Vescovi,
\textit{``{Structure constants in $ \mathcal{N} $ = 4 SYM at finite coupling as
  worldsheet g-function}''},
\textsf{\doiref{10.1007/JHEP07(2020)037}{JHEP~2007,~037~(2020)}},
\texttt{\arxivref{1906.07733}{arxiv:1906.07733}}.

\bibitem{Jiang:2019zig}
Y.~Jiang, S.~Komatsu and E.~Vescovi,
\textit{``{Exact Three-Point Functions of Determinant Operators in Planar $N=4$
  Supersymmetric Yang-Mills Theory}''},
\textsf{\doiref{10.1103/PhysRevLett.123.191601}{Phys.~Rev.~Lett.~123,~191601~(2019)}},
\texttt{\arxivref{1907.11242}{arxiv:1907.11242}}.

\bibitem{Ghoshal:1993tm}
S.~Ghoshal and A.~B.~Zamolodchikov,
\textit{``{Boundary S matrix and boundary state in two-dimensional integrable
  quantum field theory}''},
\textsf{\doiref{10.1142/S0217751X94001552}{Int.~J.~Mod.~Phys.~A~9,~3841~(1994)}},
\texttt{\arxivref{hep-th/9306002}{hep-th/9306002}},
[Erratum: Int.J.Mod.Phys.A 9, 4353 (1994)].

\bibitem{Piroli:2017sei}
L.~Piroli, B.~Pozsgay and E.~Vernier,
\textit{``{What is an integrable quench?}''},
\textsf{\doiref{10.1016/j.nuclphysb.2017.10.012}{Nucl.~Phys.~B~925,~362~(2017)}},
\texttt{\arxivref{1709.04796}{arxiv:1709.04796}}.

\bibitem{Affleck:1991tk}
I.~Affleck and A.~W.~W.~Ludwig,
\textit{``{Universal noninteger 'ground state degeneracy' in critical quantum
  systems}''},
\textsf{\doiref{10.1103/PhysRevLett.67.161}{Phys.~Rev.~Lett.~67,~161~(1991)}}.

\bibitem{Linardopoulos:2020jck}
G.~Linardopoulos,
\textit{``{Solving holographic defects}''},
\textsf{\doiref{10.22323/1.376.0141}{PoS~CORFU2019,~141~(2020)}},
\texttt{\arxivref{2005.02117}{arxiv:2005.02117}}.

\bibitem{Yang:2021hrl}
P.~Yang, Y.~Jiang, S.~Komatsu and J.-B.~Wu,
\textit{``{Three-Point Functions in ABJM and Bethe Ansatz}''},
\texttt{\arxivref{2103.15840}{arxiv:2103.15840}}.

\bibitem{komatsutoappear}
P.~Yang, Y.~Jiang, S.~Komatsu and J.-B.~Wu,
\textit{``{Structure Constants in ABJM and Integrable Bootstrap}''},
{to appear}.

\bibitem{Komatsu:2020sup}
S.~Komatsu and Y.~Wang,
\textit{``{Non-perturbative defect one-point functions in planar
  $\mathcal{N}=4$ super-Yang-Mills}''},
\textsf{\doiref{10.1016/j.nuclphysb.2020.115120}{Nucl.~Phys.~B~958,~115120~(2020)}},
\texttt{\arxivref{2004.09514}{arxiv:2004.09514}}.

\bibitem{Bajnok:2013wsa}
Z.~Bajnok, N.~Drukker, A.~Heged\"us, R.~I.~Nepomechie, L.~Palla, C.~Sieg and
  R.~Suzuki,
\textit{``{The spectrum of tachyons in AdS/CFT}''},
\textsf{\doiref{10.1007/JHEP03(2014)055}{JHEP~1403,~055~(2014)}},
\texttt{\arxivref{1312.3900}{arxiv:1312.3900}}.

\bibitem{Caetano:2020dyp}
J.~a.~Caetano and S.~Komatsu,
\textit{``{Functional equations and separation of variables for exact
  $g$-function}''},
\textsf{\doiref{10.1007/JHEP09(2020)180}{JHEP~2009,~180~(2020)}},
\texttt{\arxivref{2004.05071}{arxiv:2004.05071}}.

\bibitem{Giombi:2018qox}
S.~Giombi and S.~Komatsu,
\textit{``{Exact Correlators on the Wilson Loop in $\mathcal{N}=4$ SYM:
  Localization, Defect CFT, and Integrability}''},
\textsf{\doiref{10.1007/JHEP05(2018)109}{JHEP~1805,~109~(2018)}},
\texttt{\arxivref{1802.05201}{arxiv:1802.05201}},
[Erratum: JHEP 11, 123 (2018)].

\bibitem{Giombi:2018hsx}
S.~Giombi and S.~Komatsu,
\textit{``{More Exact Results in the Wilson Loop Defect CFT: Bulk-Defect OPE,
  Nonplanar Corrections and Quantum Spectral Curve}''},
\textsf{\doiref{10.1088/1751-8121/ab046c}{J.~Phys.~A~52,~125401~(2019)}},
\texttt{\arxivref{1811.02369}{arxiv:1811.02369}}.

\bibitem{Cavaglia:2018lxi}
A.~Cavagli\`a, N.~Gromov and F.~Levkovich-Maslyuk,
\textit{``{Quantum spectral curve and structure constants in $ \mathcal{N}=4 $
  SYM: cusps in the ladder limit}''},
\textsf{\doiref{10.1007/JHEP10(2018)060}{JHEP~1810,~060~(2018)}},
\texttt{\arxivref{1802.04237}{arxiv:1802.04237}}.

\bibitem{McGovern:2019sdd}
J.~McGovern,
\textit{``{Scalar insertions in cusped Wilson loops in the ladders limit of
  planar $ \mathcal{N} $ = 4 SYM}''},
\textsf{\doiref{10.1007/JHEP05(2020)062}{JHEP~2005,~062~(2020)}},
\texttt{\arxivref{1912.00499}{arxiv:1912.00499}}.

\bibitem{Gromov:2013pga}
N.~Gromov, V.~Kazakov, S.~Leurent and D.~Volin,
\textit{``{Quantum Spectral Curve for Planar $\mathcal{N} = 4$ Super-Yang-Mills
  Theory}''},
\textsf{\doiref{10.1103/PhysRevLett.112.011602}{Phys.~Rev.~Lett.~112,~011602~(2014)}},
\texttt{\arxivref{1305.1939}{arxiv:1305.1939}}.

\bibitem{Gromov:2014caa}
N.~Gromov, V.~Kazakov, S.~Leurent and D.~Volin,
\textit{``{Quantum spectral curve for arbitrary state/operator in
  AdS$_{5}$/CFT$_{4}$}''},
\textsf{\doiref{10.1007/JHEP09(2015)187}{JHEP~1509,~187~(2015)}},
\texttt{\arxivref{1405.4857}{arxiv:1405.4857}}.

\bibitem{Cavaglia:2019pow}
A.~Cavagli\`a, N.~Gromov and F.~Levkovich-Maslyuk,
\textit{``{Separation of variables and scalar products at any rank}''},
\textsf{\doiref{10.1007/JHEP09(2019)052}{JHEP~1909,~052~(2019)}},
\texttt{\arxivref{1907.03788}{arxiv:1907.03788}}.

\bibitem{Gromov:2019wmz}
N.~Gromov, F.~Levkovich-Maslyuk, P.~Ryan and D.~Volin,
\textit{``{Dual Separated Variables and Scalar Products}''},
\textsf{\doiref{10.1016/j.physletb.2020.135494}{Phys.~Lett.~B~806,~135494~(2020)}},
\texttt{\arxivref{1910.13442}{arxiv:1910.13442}}.

\bibitem{Gromov:2020fwh}
N.~Gromov, F.~Levkovich-Maslyuk and P.~Ryan,
\textit{``{Determinant form of correlators in high rank integrable spin chains
  via separation of variables}''},
\textsf{\doiref{10.1007/JHEP05(2021)169}{JHEP~2105,~169~(2021)}},
\texttt{\arxivref{2011.08229}{arxiv:2011.08229}}.

\bibitem{Cavaglia:2021mft}
A.~Cavagli\`a, N.~Gromov and F.~Levkovich-Maslyuk,
\textit{``{Separation of variables in AdS/CFT: functional approach for the
  fishnet CFT}''},
\textsf{\doiref{10.1007/JHEP06(2021)131}{JHEP~2106,~131~(2021)}},
\texttt{\arxivref{2103.15800}{arxiv:2103.15800}}.

\bibitem{Widen:2018nnu}
E.~Wid\'en,
\textit{``{One-point functions in $\beta$-deformed $ \mathcal{N}=4 $ SYM with
  defect}''},
\textsf{\doiref{10.1007/JHEP11(2018)114}{JHEP~1811,~114~(2018)}},
\texttt{\arxivref{1804.09514}{arxiv:1804.09514}}.

\bibitem{Gromov:2017cja}
N.~Gromov, V.~Kazakov, G.~Korchemsky, S.~Negro and G.~Sizov,
\textit{``{Integrability of Conformal Fishnet Theory}''},
\textsf{\doiref{10.1007/JHEP01(2018)095}{JHEP~1801,~095~(2018)}},
\texttt{\arxivref{1706.04167}{arxiv:1706.04167}}.

\bibitem{Ahn:2011xq}
C.~Ahn, Z.~Bajnok, D.~Bombardelli and R.~I.~Nepomechie,
\textit{``{TBA, NLO Luscher correction, and double wrapping in twisted
  AdS/CFT}''},
\textsf{\doiref{10.1007/JHEP12(2011)059}{JHEP~1112,~059~(2011)}},
\texttt{\arxivref{1108.4914}{arxiv:1108.4914}}.

\bibitem{Basso:2019xay}
B.~Basso, G.~Ferrando, V.~Kazakov and D.-l.~Zhong,
\textit{``{Thermodynamic Bethe Ansatz for Biscalar Conformal Field Theories in
  any Dimension}''},
\textsf{\doiref{10.1103/PhysRevLett.125.091601}{Phys.~Rev.~Lett.~125,~091601~(2020)}},
\texttt{\arxivref{1911.10213}{arxiv:1911.10213}}.

\bibitem{Zamolodchikov:1980mb}
A.~B.~Zamolodchikov,
\textit{``{'FISHNET' DIAGRAMS AS A COMPLETELY INTEGRABLE SYSTEM}''},
\textsf{\doiref{10.1016/0370-2693(80)90547-X}{Phys.~Lett.~B~97,~63~(1980)}}.

\bibitem{Kazakov:2018gcy}
V.~Kazakov, E.~Olivucci and M.~Preti,
\textit{``{Generalized fishnets and exact four-point correlators in chiral
  CFT$_{4}$}''},
\textsf{\doiref{10.1007/JHEP06(2019)078}{JHEP~1906,~078~(2019)}},
\texttt{\arxivref{1901.00011}{arxiv:1901.00011}}.

\bibitem{Grabner:2017pgm}
D.~Grabner, N.~Gromov, V.~Kazakov and G.~Korchemsky,
\textit{``{Strongly $\gamma$-Deformed $\mathcal{N}=4$ Supersymmetric Yang-Mills
  Theory as an Integrable Conformal Field Theory}''},
\textsf{\doiref{10.1103/PhysRevLett.120.111601}{Phys.~Rev.~Lett.~120,~111601~(2018)}},
\texttt{\arxivref{1711.04786}{arxiv:1711.04786}}.

\bibitem{Gromov:2018hut}
N.~Gromov, V.~Kazakov and G.~Korchemsky,
\textit{``{Exact Correlation Functions in Conformal Fishnet Theory}''},
\textsf{\doiref{10.1007/JHEP08(2019)123}{JHEP~1908,~123~(2019)}},
\texttt{\arxivref{1808.02688}{arxiv:1808.02688}}.

\bibitem{Leigh:1995ep}
R.~G.~Leigh and M.~J.~Strassler,
\textit{``{Exactly marginal operators and duality in four-dimensional N=1
  supersymmetric gauge theory}''},
\textsf{\doiref{10.1016/0550-3213(95)00261-P}{Nucl.~Phys.~B~447,~95~(1995)}},
\texttt{\arxivref{hep-th/9503121}{hep-th/9503121}}.

\bibitem{Lunin:2005jy}
O.~Lunin and J.~M.~Maldacena,
\textit{``{Deforming field theories with U(1) x U(1) global symmetry and their
  gravity duals}''},
\textsf{\doiref{10.1088/1126-6708/2005/05/033}{JHEP~0505,~033~(2005)}},
\texttt{\arxivref{hep-th/0502086}{hep-th/0502086}}.

\bibitem{Fokken:2014soa}
J.~Fokken, C.~Sieg and M.~Wilhelm,
\textit{``{A piece of cake: the ground-state energies in $\gamma_{i}$ -deformed
  $ \mathcal{N} $ = 4 SYM theory at leading wrapping order}''},
\textsf{\doiref{10.1007/JHEP09(2014)078}{JHEP~1409,~078~(2014)}},
\texttt{\arxivref{1405.6712}{arxiv:1405.6712}}.

\bibitem{Sieg:2016vap}
C.~Sieg and M.~Wilhelm,
\textit{``{On a CFT limit of planar $\gamma_i$-deformed $\mathcal{N}=4$ SYM
  theory}''},
\textsf{\doiref{10.1016/j.physletb.2016.03.004}{Phys.~Lett.~B~756,~118~(2016)}},
\texttt{\arxivref{1602.05817}{arxiv:1602.05817}}.

\bibitem{Balasubramanian:2001nh}
V.~Balasubramanian, M.~Berkooz, A.~Naqvi and M.~J.~Strassler,
\textit{``{Giant gravitons in conformal field theory}''},
\textsf{\doiref{10.1088/1126-6708/2002/04/034}{JHEP~0204,~034~(2002)}},
\texttt{\arxivref{hep-th/0107119}{hep-th/0107119}}.

\bibitem{Corley:2001zk}
S.~Corley, A.~Jevicki and S.~Ramgoolam,
\textit{``{Exact correlators of giant gravitons from dual N=4 SYM theory}''},
\textsf{\doiref{10.4310/ATMP.2001.v5.n4.a6}{Adv.~Theor.~Math.~Phys.~5,~809~(2002)}},
\texttt{\arxivref{hep-th/0111222}{hep-th/0111222}}.

\bibitem{McGreevy:2000cw}
J.~McGreevy, L.~Susskind and N.~Toumbas,
\textit{``{Invasion of the giant gravitons from Anti-de Sitter space}''},
\textsf{\doiref{10.1088/1126-6708/2000/06/008}{JHEP~0006,~008~(2000)}},
\texttt{\arxivref{hep-th/0003075}{hep-th/0003075}}.

\bibitem{Witten:1998xy}
E.~Witten,
\textit{``{Baryons and branes in anti-de Sitter space}''},
\textsf{\doiref{10.1088/1126-6708/1998/07/006}{JHEP~9807,~006~(1998)}},
\texttt{\arxivref{hep-th/9805112}{hep-th/9805112}}.

\bibitem{Vescovi:2021fjf}
E.~Vescovi,
\textit{``{Four-point function of determinant operators in $\mathcal{N}=4$
  SYM}''},
\textsf{\doiref{10.1103/PhysRevD.103.106001}{Phys.~Rev.~D~103,~106001~(2021)}},
\texttt{\arxivref{2101.05117}{arxiv:2101.05117}}.

\bibitem{Chen:2019gsb}
G.~Chen, R.~de~Mello~Koch, M.~Kim and H.~J.~R.~Van~Zyl,
\textit{``{Absorption of closed strings by giant gravitons}''},
\textsf{\doiref{10.1007/JHEP10(2019)133}{JHEP~1910,~133~(2019)}},
\texttt{\arxivref{1908.03553}{arxiv:1908.03553}}.

\bibitem{Chen:2019kgc}
G.~Chen, R.~De~Mello~Koch, M.~Kim and H.~J.~R.~Van~Zyl,
\textit{``{Structure constants of heavy operators in ABJM and ABJ theory}''},
\textsf{\doiref{10.1103/PhysRevD.100.086019}{Phys.~Rev.~D~100,~086019~(2019)}},
\texttt{\arxivref{1909.03215}{arxiv:1909.03215}}.

\bibitem{Gromov:2019aku}
N.~Gromov and A.~Sever,
\textit{``{Derivation of the Holographic Dual of a Planar Conformal Field
  Theory in 4D}''},
\textsf{\doiref{10.1103/PhysRevLett.123.081602}{Phys.~Rev.~Lett.~123,~081602~(2019)}},
\texttt{\arxivref{1903.10508}{arxiv:1903.10508}}.

\bibitem{Gromov:2019bsj}
N.~Gromov and A.~Sever,
\textit{``{Quantum fishchain in AdS$_{5}$}''},
\textsf{\doiref{10.1007/JHEP10(2019)085}{JHEP~1910,~085~(2019)}},
\texttt{\arxivref{1907.01001}{arxiv:1907.01001}}.

\bibitem{Gromov:2019jfh}
N.~Gromov and A.~Sever,
\textit{``{The holographic dual of strongly $\gamma$-deformed $ \mathcal{N} $ =
  4 SYM theory: derivation, generalization, integrability and discrete
  reparametrization symmetry}''},
\textsf{\doiref{10.1007/JHEP02(2020)035}{JHEP~2002,~035~(2020)}},
\texttt{\arxivref{1908.10379}{arxiv:1908.10379}}.

\bibitem{Gromov:2021ahm}
N.~Gromov, J.~Julius and N.~Primi,
\textit{``{Open Fishchain in N=4 Supersymmetric Yang-Mills Theory}''},
\texttt{\arxivref{2101.01232}{arxiv:2101.01232}}.

\bibitem{Basso:2021omx}
B.~Basso, L.~J.~Dixon, D.~A.~Kosower, A.~Krajenbrink and D.-l.~Zhong,
\textit{``{Fishnet four-point integrals: integrable representations and
  thermodynamic limits}''},
\textsf{\doiref{10.1007/JHEP07(2021)168}{JHEP~2107,~168~(2021)}},
\texttt{\arxivref{2105.10514}{arxiv:2105.10514}}.

\bibitem{Berenstein:2002ke}
D.~Berenstein, C.~P.~Herzog and I.~R.~Klebanov,
\textit{``{Baryon spectra and AdS /CFT correspondence}''},
\textsf{\doiref{10.1088/1126-6708/2002/06/047}{JHEP~0206,~047~(2002)}},
\texttt{\arxivref{hep-th/0202150}{hep-th/0202150}}.

\bibitem{Balasubramanian:2002sa}
V.~Balasubramanian, M.-x.~Huang, T.~S.~Levi and A.~Naqvi,
\textit{``{Open strings from N=4 superYang-Mills}''},
\textsf{\doiref{10.1088/1126-6708/2002/08/037}{JHEP~0208,~037~(2002)}},
\texttt{\arxivref{hep-th/0204196}{hep-th/0204196}}.

\bibitem{Das:2000st}
S.~R.~Das, A.~Jevicki and S.~D.~Mathur,
\textit{``{Vibration modes of giant gravitons}''},
\textsf{\doiref{10.1103/PhysRevD.63.024013}{Phys.~Rev.~D~63,~024013~(2001)}},
\texttt{\arxivref{hep-th/0009019}{hep-th/0009019}}.

\bibitem{Berenstein:2003ah}
D.~Berenstein,
\textit{``{Shape and holography: Studies of dual operators to giant
  gravitons}''},
\textsf{\doiref{10.1016/j.nuclphysb.2003.10.004}{Nucl.~Phys.~B~675,~179~(2003)}},
\texttt{\arxivref{hep-th/0306090}{hep-th/0306090}}.

\bibitem{Berenstein:2004kk}
D.~Berenstein,
\textit{``{A Toy model for the AdS / CFT correspondence}''},
\textsf{\doiref{10.1088/1126-6708/2004/07/018}{JHEP~0407,~018~(2004)}},
\texttt{\arxivref{hep-th/0403110}{hep-th/0403110}}.

\bibitem{Balasubramanian:2004nb}
V.~Balasubramanian, D.~Berenstein, B.~Feng and M.-x.~Huang,
\textit{``{D-branes in Yang-Mills theory and emergent gauge symmetry}''},
\textsf{\doiref{10.1088/1126-6708/2005/03/006}{JHEP~0503,~006~(2005)}},
\texttt{\arxivref{hep-th/0411205}{hep-th/0411205}}.

\bibitem{Berenstein:2005vf}
D.~Berenstein and S.~E.~Vazquez,
\textit{``{Integrable open spin chains from giant gravitons}''},
\textsf{\doiref{10.1088/1126-6708/2005/06/059}{JHEP~0506,~059~(2005)}},
\texttt{\arxivref{hep-th/0501078}{hep-th/0501078}}.

\bibitem{deMelloKoch:2007rqf}
R.~de~Mello~Koch, J.~Smolic and M.~Smolic,
\textit{``{Giant Gravitons - with Strings Attached (I)}''},
\textsf{\doiref{10.1088/1126-6708/2007/06/074}{JHEP~0706,~074~(2007)}},
\texttt{\arxivref{hep-th/0701066}{hep-th/0701066}}.

\bibitem{deMelloKoch:2007nbd}
R.~de~Mello~Koch, J.~Smolic and M.~Smolic,
\textit{``{Giant Gravitons - with Strings Attached (II)}''},
\textsf{\doiref{10.1088/1126-6708/2007/09/049}{JHEP~0709,~049~(2007)}},
\texttt{\arxivref{hep-th/0701067}{hep-th/0701067}}.

\bibitem{deMelloKoch:2010zrl}
R.~de~Mello~Koch, G.~Mashile and N.~Park,
\textit{``{Emergent Threebrane Lattices}''},
\textsf{\doiref{10.1103/PhysRevD.81.106009}{Phys.~Rev.~D~81,~106009~(2010)}},
\texttt{\arxivref{1004.1108}{arxiv:1004.1108}}.

\bibitem{DeComarmond:2010ie}
V.~De~Comarmond, R.~de~Mello~Koch and K.~Jefferies,
\textit{``{Surprisingly Simple Spectra}''},
\textsf{\doiref{10.1007/JHEP02(2011)006}{JHEP~1102,~006~(2011)}},
\texttt{\arxivref{1012.3884}{arxiv:1012.3884}}.

\bibitem{Carlson:2011hy}
W.~Carlson, R.~de~Mello~Koch and H.~Lin,
\textit{``{Nonplanar Integrability}''},
\textsf{\doiref{10.1007/JHEP03(2011)105}{JHEP~1103,~105~(2011)}},
\texttt{\arxivref{1101.5404}{arxiv:1101.5404}}.

\bibitem{Hofman:2007xp}
D.~M.~Hofman and J.~M.~Maldacena,
\textit{``{Reflecting magnons}''},
\textsf{\doiref{10.1088/1126-6708/2007/11/063}{JHEP~0711,~063~(2007)}},
\texttt{\arxivref{0708.2272}{arxiv:0708.2272}}.

\bibitem{Mann:2006rh}
N.~Mann and S.~E.~Vazquez,
\textit{``{Classical Open String Integrability}''},
\textsf{\doiref{10.1088/1126-6708/2007/04/065}{JHEP~0704,~065~(2007)}},
\texttt{\arxivref{hep-th/0612038}{hep-th/0612038}}.

\bibitem{Berenstein:2005fa}
D.~Berenstein, D.~H.~Correa and S.~E.~Vazquez,
\textit{``{Quantizing open spin chains with variable length: An Example from
  giant gravitons}''},
\textsf{\doiref{10.1103/PhysRevLett.95.191601}{Phys.~Rev.~Lett.~95,~191601~(2005)}},
\texttt{\arxivref{hep-th/0502172}{hep-th/0502172}}.

\bibitem{Berenstein:2006qk}
D.~Berenstein, D.~H.~Correa and S.~E.~Vazquez,
\textit{``{A Study of open strings ending on giant gravitons, spin chains and
  integrability}''},
\textsf{\doiref{10.1088/1126-6708/2006/09/065}{JHEP~0609,~065~(2006)}},
\texttt{\arxivref{hep-th/0604123}{hep-th/0604123}}.

\bibitem{Basso:2017khq}
B.~Basso, F.~Coronado, S.~Komatsu, H.~T.~Lam, P.~Vieira and D.-l.~Zhong,
\textit{``{Asymptotic Four Point Functions}''},
\textsf{\doiref{10.1007/JHEP07(2019)082}{JHEP~1907,~082~(2019)}},
\texttt{\arxivref{1701.04462}{arxiv:1701.04462}}.

\bibitem{Basso:2018cvy}
B.~Basso, J.~a.~Caetano and T.~Fleury,
\textit{``{Hexagons and Correlators in the Fishnet Theory}''},
\textsf{\doiref{10.1007/JHEP11(2019)172}{JHEP~1911,~172~(2019)}},
\texttt{\arxivref{1812.09794}{arxiv:1812.09794}}.

\bibitem{Derkachov:2020zvv}
S.~Derkachov and E.~Olivucci,
\textit{``{Exactly solvable single-trace four point correlators in
  $\chi$CFT$_4$}''},
\textsf{\doiref{10.1007/JHEP02(2021)146}{JHEP~2102,~146~(2021)}},
\texttt{\arxivref{2007.15049}{arxiv:2007.15049}}.

\bibitem{Benvenuti:2006qr}
S.~Benvenuti, B.~Feng, A.~Hanany and Y.-H.~He,
\textit{``{Counting BPS Operators in Gauge Theories: Quivers, Syzygies and
  Plethystics}''},
\textsf{\doiref{10.1088/1126-6708/2007/11/050}{JHEP~0711,~050~(2007)}},
\texttt{\arxivref{hep-th/0608050}{hep-th/0608050}}.

\bibitem{Usyukina:1992jd}
N.~I.~Usyukina and A.~I.~Davydychev,
\textit{``{An Approach to the evaluation of three and four point ladder
  diagrams}''},
\textsf{\doiref{10.1016/0370-2693(93)91834-A}{Phys.~Lett.~B~298,~363~(1993)}}.

\bibitem{Drukker:2008pi}
N.~Drukker and J.~Plefka,
\textit{``{The Structure of n-point functions of chiral primary operators in
  N=4 super Yang-Mills at one-loop}''},
\textsf{\doiref{10.1088/1126-6708/2009/04/001}{JHEP~0904,~001~(2009)}},
\texttt{\arxivref{0812.3341}{arxiv:0812.3341}}.

\bibitem{Dolan:2000ut}
F.~A.~Dolan and H.~Osborn,
\textit{``{Conformal four point functions and the operator product
  expansion}''},
\textsf{\doiref{10.1016/S0550-3213(01)00013-X}{Nucl.~Phys.~B~599,~459~(2001)}},
\texttt{\arxivref{hep-th/0011040}{hep-th/0011040}}.

\bibitem{Derkachov:2019tzo}
S.~Derkachov and E.~Olivucci,
\textit{``{Exactly solvable magnet of conformal spins in four dimensions}''},
\textsf{\doiref{10.1103/PhysRevLett.125.031603}{Phys.~Rev.~Lett.~125,~031603~(2020)}},
\texttt{\arxivref{1912.07588}{arxiv:1912.07588}}.

\bibitem{Derkachov:2021rrf}
S.~Derkachov and E.~Olivucci,
\textit{``{Conformal quantum mechanics $\&$ the integrable spinning
  Fishnet}''},
\texttt{\arxivref{2103.01940}{arxiv:2103.01940}}.

\bibitem{Grisaru:2000zn}
M.~T.~Grisaru, R.~C.~Myers and O.~Tafjord,
\textit{``{SUSY and goliath}''},
\textsf{\doiref{10.1088/1126-6708/2000/08/040}{JHEP~0008,~040~(2000)}},
\texttt{\arxivref{hep-th/0008015}{hep-th/0008015}}.

\bibitem{Hashimoto:2000zp}
A.~Hashimoto, S.~Hirano and N.~Itzhaki,
\textit{``{Large branes in AdS and their field theory dual}''},
\textsf{\doiref{10.1088/1126-6708/2000/08/051}{JHEP~0008,~051~(2000)}},
\texttt{\arxivref{hep-th/0008016}{hep-th/0008016}}.

\bibitem{Bekker:2007ea}
D.~Bekker, R.~de~Mello~Koch and M.~Stephanou,
\textit{``{Giant Gravitons - with Strings Attached. III.}''},
\textsf{\doiref{10.1088/1126-6708/2008/02/029}{JHEP~0802,~029~(2008)}},
\texttt{\arxivref{0710.5372}{arxiv:0710.5372}}.

\bibitem{Ciavarella:2010tp}
A.~Ciavarella,
\textit{``{Giant magnons and non-maximal giant gravitons}''},
\textsf{\doiref{10.1007/JHEP01(2011)040}{JHEP~1101,~040~(2011)}},
\texttt{\arxivref{1011.1440}{arxiv:1011.1440}}.

\bibitem{deMelloKoch:2016mhc}
R.~de~Mello~Koch and H.~J.~R.~van~Zyl,
\textit{``{Inelastic Magnon Scattering}''},
\textsf{\doiref{10.1016/j.physletb.2017.02.056}{Phys.~Lett.~B~768,~187~(2017)}},
\texttt{\arxivref{1603.06414}{arxiv:1603.06414}}.

\bibitem{deMelloKoch:2018tlb}
R.~de~Mello~Koch, M.~Kim and H.~J.~R.~Zyl,
\textit{``{Integrable Subsectors from Holography}''},
\textsf{\doiref{10.1007/JHEP05(2018)198}{JHEP~1805,~198~(2018)}},
\texttt{\arxivref{1802.01367}{arxiv:1802.01367}}.

\bibitem{Chen:2018sbp}
H.-H.~Chen, H.~Ouyang and J.-B.~Wu,
\textit{``{Open Spin Chains from Determinant Like Operators in ABJM Theory}''},
\textsf{\doiref{10.1103/PhysRevD.98.106012}{Phys.~Rev.~D~98,~106012~(2018)}},
\texttt{\arxivref{1809.09941}{arxiv:1809.09941}}.

\bibitem{Bai:2019soy}
N.~Bai, H.-H.~Chen, H.~Ouyang and J.-B.~Wu,
\textit{``{Two-Loop Integrability of ABJM Open Spin Chain from Giant
  Graviton}''},
\textsf{\doiref{10.1007/JHEP03(2019)193}{JHEP~1903,~193~(2019)}},
\texttt{\arxivref{1901.03949}{arxiv:1901.03949}}.

\bibitem{Pittelli:2019ceq}
A.~Pittelli and M.~Preti,
\textit{``{Integrable fishnet from $\gamma$-deformed $\mathcal{N}=2$
  quivers}''},
\textsf{\doiref{10.1016/j.physletb.2019.134971}{Phys.~Lett.~B~798,~134971~(2019)}},
\texttt{\arxivref{1906.03680}{arxiv:1906.03680}}.

\bibitem{Gopakumar}
R.~Gopakumar,
\textit{``{Open-closed-open string duality}''},
{ talk at Second Joburg Workshop on String Theory},
\href{{http://neo.phys.wits.ac.za/workshop\_2/pdfs/rajesh.pdf}}{\texttt{{http://neo.phys.wits.ac.za/workshop\_2/pdfs/rajesh.pdf}}}.

\bibitem{Dobrev:1977qv}
V.~K.~Dobrev, G.~Mack, V.~B.~Petkova, S.~G.~Petrova and I.~T.~Todorov,
\textit{``{Harmonic Analysis on the n-Dimensional Lorentz Group and Its
  Application to Conformal Quantum Field Theory}''}.

\bibitem{DEramo:1971hnd}
M.~D'Eramo, G.~Parisi and L.~Peliti,
\textit{``{THEORETICAL PREDICTIONS FOR CRITICAL EXPONENTS AT THE lambda POINT
  OF BOSE LIQUIDS}''},
\textsf{\doiref{10.1007/BF02774121}{Lett.~Nuovo~Cim.~2,~878~(1971)}}.

\bibitem{Vasiliev:1981yc}
A.~N.~Vasiliev, Y.~M.~Pismak and Y.~R.~Khonkonen,
\textit{``{Simple Method of Calculating the Critical Indices in the 1/$N$
  Expansion}''},
\textsf{\doiref{10.1007/BF01030844}{Theor.~Math.~Phys.~46,~104~(1981)}}.

\bibitem{Dolan:2011dv}
F.~A.~Dolan and H.~Osborn,
\textit{``{Conformal Partial Waves: Further Mathematical Results}''},
\texttt{\arxivref{1108.6194}{arxiv:1108.6194}}.

\bibitem{Remiddi:1999ew}
E.~Remiddi and J.~A.~M.~Vermaseren,
\textit{``{Harmonic polylogarithms}''},
\textsf{\doiref{10.1142/S0217751X00000367}{Int.~J.~Mod.~Phys.~A~15,~725~(2000)}},
\texttt{\arxivref{hep-ph/9905237}{hep-ph/9905237}}.

\bibitem{Maitre:2005uu}
D.~Maitre,
\textit{``{HPL, a mathematica implementation of the harmonic
  polylogarithms}''},
\textsf{\doiref{10.1016/j.cpc.2005.10.008}{Comput.~Phys.~Commun.~174,~222~(2006)}},
\texttt{\arxivref{hep-ph/0507152}{hep-ph/0507152}}.

\bibitem{HPL}
D.~Maitre,
\textit{``{HPL, a mathematica implementation of the harmonic
  polylogarithms}''},
\href{{https://www.physik.uzh.ch/data/HPL/}}{\texttt{{https://www.physik.uzh.ch/data/HPL/}}}.

\bibitem{Gross:2017aos}
D.~J.~Gross and V.~Rosenhaus,
\textit{``{All point correlation functions in SYK}''},
\textsf{\doiref{10.1007/JHEP12(2017)148}{JHEP~1712,~148~(2017)}},
\texttt{\arxivref{1710.08113}{arxiv:1710.08113}}.

\end{thebibliography}
\end{document}